\documentclass[longauth]{aa}  

\usepackage{placeins}
\usepackage{graphicx}
\usepackage{array,multirow,makecell}
\usepackage{txfonts}
\usepackage{hyperref}
\hypersetup{colorlinks,linkcolor={blue},citecolor={blue},urlcolor={red}} 
\usepackage[capitalise]{cleveref}
\usepackage{natbib}
\bibpunct{(}{)}{;}{a}{}{,} 
\usepackage[normal]{threeparttable}

\newcommand{\lsun}{\mbox{$L_\odot$}}
\newcommand{\msun}{\mbox{$M_\odot$}}
\newcommand{\lbol}{\mbox{$L_{\rm bol}$}}
\newcommand{\tdust}{\mbox{$T_{\rm dust}$}}

\newcommand{\hii}{H\mbox{\sc ~ii} }

\newcommand{\bsens}{\texttt{bsens}\xspace}
\newcommand{\cleanest}{\texttt{cleanest}\xspace}
\newcommand{\getsf}{\textsl{getsf}\xspace}
\newcommand{\PPMAP}{\textsl{PPMAP}\xspace}
\newcommand{\summation}[2]{\sum\limits^{#1}_{#2}}

\begin{document} 

\title{ALMA-IMF XVI: Mass-averaged temperature of cores and protostellar luminosities in the ALMA-IMF protoclusters}
\titlerunning{ALMA-IMF XVI: Mass-averaged temperature of cores and protostellar luminosities}

\author{  F. Motte \inst{1} \and
          Y. Pouteau \inst{1} \and
          T. Nony \inst{2, 3} \and
          P. Dell'Ova \inst{4, 5, 6} \and
          A. Gusdorf \inst{4} \and
          N. Brouillet \inst{7} \and
          A. M. Stutz \inst{8, 9} \and
          S. Bontemps \inst{7}    \and
          A. Ginsburg \inst{10} \and
          T. Csengeri \inst{7} \and
          A. Men’shchikov \inst{11} \and
          M. Valeille-Manet \inst{7} \and
          F. Louvet \inst{1} \and
          M. Bonfand \inst{12} \and
          R. Galv\'an-Madrid \inst{2} \and 
          R. H. \'Alvarez-Guti\'errez \inst{8} \and
          M. Armante \inst{4, 5} \and
          L. Bronfman \inst{13} \and
          H.-R. V. Chen \inst{14} \and
          N. Cunningham \inst{1, 15} \and
          D. D\'iaz-Gonz\'alez \inst{2} \and 
          P. Didelon \inst{11} \and
          M. Fern\'andez-L\'opez \inst{16}
          F. Herpin \inst{7} \and 
          N. Kessler \inst{7} \and
          A. Koley \inst{8} \and
          B. Lefloch \inst{7} \and
          N. Le Nestour \inst{1}  \and
          H.-L. Liu \inst{17}
          E. Moraux \inst{1} \and
          Q. Nguyen Luong \inst{11, 18, 19} \and
          F. Olguin \inst{14} \and 
          J. Salinas \inst{8} \and
          N. A. Sandoval-Garrido \inst{8} \and
          P. Sanhueza \inst{20, 21} \and
          R. Veyry  \inst{1} \and
          T. Yoo \inst{10}
          }

\institute{Univ. Grenoble Alpes, CNRS, IPAG, 38000 Grenoble, France             
    \and Instituto de Radioastronom\'ia y Astrof\'isica, Universidad Nacional Aut\'onoma de M\'exico, Morelia, Michoac\'an 58089, M\'exico      
    \and INAF, Osservatorio Astrofisico di Arcetri, Largo Enrico Fermi 5, 50125 Firenze, Italy        
    \and Laboratoire de Physique de l'\'Ecole Normale Sup\'erieure, ENS, Universit\'e PSL, CNRS, Sorbonne Universit\'e, Universit\'e de Paris, Paris, France  
    \and Observatoire de Paris, PSL University, Sorbonne Universit\'e, LERMA, 75014, Paris, France 
    \and Universit\'e Paris-Saclay, CNRS, Institut d’Astrophysique Spatiale, 91405 Orsay, France 
    \and Laboratoire d'astrophysique de Bordeaux, Univ. Bordeaux, CNRS, B18N, all\'ee Geoffroy Saint-Hilaire, 33615 Pessac, France 
    \and Departamento de Astronom\'{i}a, Universidad de Concepci\'{o}n, Casilla 160-C, Concepci\'{o}n, Chile 
    \and Franco-Chilean Laboratory for Astronomy, IRL 3386, CNRS and Universidad de Chile, Santiago, Chile 
    \and Department of Astronomy, University of Florida, PO Box 112055, USA 
    \and Universit{\'e} Paris-Saclay, Universit{\'e} Paris Cit{\'e}, CEA, CNRS, AIM, 91191 Gif-sur-Yvette, France 
    \and Departments of Astronomy and Chemistry, University of Virginia, Charlottesville, VA 22904, USA 
    \and Departamento de Astronomía, Universidad de Chile, Casilla 36-D, Santiago, Chile   
    \and Institute of Astronomy, National Tsing Hua University, Hsinchu 30013, Taiwan 
    \and SKA Observatory, Jodrell Bank, Lower Withington, Macclesfield, SK11 9FT, United Kingdom 
    \and Instituto Argentino de Radioastronom\'\i a (CCT-La Plata, CONICET; CICPBA), C.C. No. 5, 1894, Villa Elisa, Buenos Aires, Argentina 
    \and Department of Astronomy, Yunnan University, Kunming 650091, People's Republic of China 
    \and CSMES, The American University of Paris, PL111, 2 bis, passage Landrieu, 75007, Paris, France 
    \and TNU Observatory, Tay Nguyen University, 567 Le Duan, Ea Tam, Buon Ma Thuot City, Dak Lak, 630000, Vietnam 
    \and National Astronomical Observatory of Japan, National Institutes of Natural Sciences, 2-21-1 Osawa, Mitaka, Tokyo 181-8588, Japan 
    \and Department of Astronomical Science, SOKENDAI (The Graduate University for Advanced Studies), 2-21-1 Osawa, Mitaka, Tokyo 181-8588, Japan   
}

\date{Received August 20, 2024; accepted November 11, 2024}
 
\abstract
{
The ALMA-IMF Large Program imaged 15 massive protoclusters down to a resolution of $\sim$2~kau scales, 
identifying about $10^3$ star-forming cores. The mass and luminosity of these cores, which are fundamental physical characteristics, are difficult to determine, a problem greatly exacerbated at the distances $\ge$2~kpc of ALMA-IMF protoclusters.
}
%
{
We combined new datasets and radiative transfer modeling to characterize these cores. We estimated their mass-averaged temperature and the masses these estimates imply. For one-sixth of the sample, we measured the bolometric luminosities, implementing deblending corrections when necessary.
}
%
{
We used spectral energy distribution (SED) analysis obtained with the point process mapping (\PPMAP) Bayesian procedure, which aims to preserve the best angular resolution of the input data. We extrapolated the luminosity and dust temperature images provided by \PPMAP at $2.5\arcsec$ resolution to estimate those of individual cores, which were identified at higher angular resolution. To do this, we applied approximate radiative transfer relationships between the luminosity of a protostar and the temperature of its surrounding envelope and between the external heating of prestellar cores and their temperatures. 
}
%
{
For the first time, we provide data-informed estimates of dust temperatures for 883 cores identified with ALMA-IMF: 17--31~K and 28--79~K (5th and 95th percentiles, up to 127~K) for the 617 prestellar and 266 protostellar cores, respectively. We also measured protostellar luminosities spanning $20-80\,000~\lsun$.
} 
%
{Dust temperatures previously estimated from SED-based analyses at a comparatively lower resolution validate our method. For hot cores, on the other hand, we estimated systematically lower temperatures than studies based on complex organic molecules. We established a mass-luminosity evolutionary diagram, for the first time at the core spatial resolution and for a large sample of intermediate- to high-mass protostellar cores. The ALMA-IMF data favor a scenario in which protostars accrete their mass from a larger mass reservoir than their host cores.}

\keywords{stars: formation -- stars: IMF -- stars: massive -- ISM: dust -- ISM: molecules}

\maketitle
\section{Introduction}
\label{s:intro}

Cores are considered to be the gas mass reservoir used for the formation of individual stars or small stellar systems \citep[e.g.,][]{mckeeOstriker2007, andre2014}. Their exact definition is subject to debate \citep[e.g.,][]{louvet2021simu}, but we assume in this work that cores are dense cloud structures on scales of $\sim$2000~au that form or will soon form stars via gravitational collapse. To distinguish the two cases mentioned above, we refer to prestellar cores as those that are about to collapse under their own gravity. In contrast, protostellar cores are those that, in the midst of their gravitational collapse, contain a central stellar embryo, that is, a protostar. Irrespective of the stage of collapse, the core mass is the most fundamental parameter, while the protostellar luminosity is used to evaluate the evolutionary stage of protostellar cores \citep[e.g.,][]{andre2000, molinari2008, duarte2013}.

Core mass and protostellar luminosity are two essential physical characteristics to constrain models of star and star cluster formation \citep[e.g.,][]{motte2018a, vazquez2019, molinari2019, peretto2020}. However, the masses of cores depend crucially on the dust temperature and emissivity used to convert their continuum flux, often measured at (sub)millimeter wavelengths, into the total mass of gas and dust. The mass-averaged dust temperature of cores is estimated (when possible) by fitting modified blackbody models to the cold component of their spectral energy distributions (SEDs), whose fluxes are measured at far-infrared, submillimeter, and millimeter wavelengths \citep[e.g.,][]{bontemps2010SI, konyves2015, furlan2016}. As for the luminosity of nearby protostars, they have been estimated by integrating their SEDs, that is by adding a warm component traced in the near-infrared to mid-infrared regimes to the cold component mentioned above \citep[e.g.,][]{bontemps1996, dunham2013}.

If measuring these parameters of protostars located in nearby star-forming regions, including the Gould Belt clouds, is a challenge, it is an even more Herculean task regarding protostars located beyond 1~kpc and worse still in galaxies other than the Milky Way. Regardless, when cores are identified with (sub)millimeter interferometers at $\sim$$0.1\arcsec-1\arcsec$ resolution, the lack of resolution-matched measurements in the far-infrared regime makes it difficult to build complete, meaningful SEDs and therefore to measure luminosities \citep[e.g.,][]{nguyen2011a, duarte2013}. This is the case for many (sub)millimeter interferometric studies and is particularly true for the ALMA-IMF\footnote{
    ALMA project \#2017.1.01355.L; see \url{http://www.almaimf.com} (PIs: Motte, Ginsburg, Louvet, Sanhueza).}
Large Program \citep[see Paper~I,][]{motte2022}, which aims to obtain statistically meaningful results on the origin of the initial mass function (IMF) of stars (see \citealt{motte2018b} and Papers~III, V, VI, X, XV, \citealt{pouteau2022, pouteau2023, nony2023, armante2024, louvet2024}). As ALMA-IMF has captured a large core sample, there is a pressing need for observationally driven constraints on their masses and luminosities. This is the central focus of this paper.

The ALMA-IMF program surveyed 15 massive ($2.5-33 \times 10^3~\msun$) relatively nearby ($2-5.5$~kpc) protoclusters that cover a wide variety of Galactic environments and three evolutionary stages: Young, Intermediate, and Evolved \citep[see][]{motte2022}. A total noncontiguous area of $\sim$53~pc$^2$ was imaged at 1.3~mm with a matched spatial resolution of $\sim$2000~au across the sample, corresponding to $0.3\arcsec$ to $0.9\arcsec$, depending on the distance of the protoclusters. ALMA-IMF discovered about a thousand cores with masses spanning four decades \citep{motte2018b, pouteau2022,louvet2024}. Of the entire ALMA-IMF core sample, around one-fourth are protostellar in nature \citep{nony2023, armante2024}. 

The Bayesian point process mapping (\PPMAP) procedure proposes a partial solution to the issue of measuring the luminosity and temperature of individual cores since it preserves at best the angular resolution of the best resolved datasets \citep{marsh2015, marsh2017}. This procedure was pushed to its extreme by combining mosaics with $\sim$$0.5\arcsec-0.9\arcsec$ resolution, which were obtained with the ALMA interferometer at 1.3~mm \citep{motte2018b, pouteau2022, armante2024}, with $5.6\arcsec-35\arcsec$ resolution images of \textit{Herschel} at $70\,\mu{\rm m}$ to $500\,\mu{\rm m}$ \citep[e.g.,][]{nguyen2013}.\footnote{
    \cite{motte2018b} also used in the \PPMAP procedure  $350\,\mu{\rm m}$ and $870\,\mu{\rm m}$ images obtained at the APEX telescope, as done in \cite{dellova2024}, and 3~mm images from the NOEMA interferometer \citep{louvet2014}. Their angular resolution, $2\arcsec$, $8\arcsec$, and $18\arcsec$, is intermediate between those of the ALMA and \textit{Herschel} data.}
Four dust temperature images were created at $2.5\arcsec$ angular resolution by the SED fits performed by  \PPMAP on the ALMA-IMF protoclusters W43-MM1, W43-MM2, W43-MM3, and G012.80 \citep{motte2018b, pouteau2022, armante2024}. More recently, \cite{dellova2024} applied the \PPMAP procedure to the entire ALMA-IMF. They produced column density, luminosity, and dust temperature maps at the unprecedented angular resolution of $2.5\arcsec$, corresponding to $5000-14\,000$~au at the ALMA-IMF cloud distances. Extrapolation to a higher angular resolution in order to estimate the dust temperature averaged over the mass of the cores, which are $\sim$2000~au in size, is therefore necessary. 

Alternatively, molecular lines are powerful tools to determine kinetic temperatures that can then be used to estimate the mass-averaged dust temperature of cores. These generally require follow-up observations at, for instance, the Very Large Array of the NH$_3$ thermometer \citep[e.g.,][Svoboda et al. in prep.]{battersby2014} or ALMA observations of multiple transitions of, for example, H$_2$CO, CH$_3$OH, CH$_3$CN, or other complex organic molecules \citep[COMs; e.g.,][]{giannetti2017, pols2018, molet2019, jeff2024, izumi2024}. These measurements are hampered by the difficulty of selecting lines tracing exclusively and entirely the core volume. For instance, as part of the ALMA-IMF survey, \cite{brouillet2022} studied the W43-MM1 protocluster and found that the H$_2$CO lines trace filaments rather than cores and that the filling factors of CH$_3$CHO, CH$_3$OCHO, and CH$_3$OCH$_3$ lines detected toward cores are poorly constrained. A recent study, however, showed that the methyl formate (CH$_3$OCHO, MF) emission is sometimes extended beyond the core size, suggesting that the excitation temperature could be a good proxy, or even a lower limit, for the mass-averaged dust temperature of cores \citep{bonfand2024}.

The present study proposes a methodology for estimating the mass-averaged temperature of cores detected with ALMA as well as the luminosity of individual high-mass protostars. We combine information from recently published catalogs of ALMA-IMF cores with dust temperature and luminosity images of their cloud environment derived from a new SED analysis (see Sects.~\ref{s:data-cat}--\ref{s:data-temp}). In Sects.~\ref{s:Tproto}--\ref{s:Tpre}, we describe in detail the proposed methodology, which based on approximate radiative-transfer relationships allows us to estimate mass-averaged temperatures of protostellar and prestellar cores and protostellar luminosities. We then discuss in Sect.~\ref{s:discussion} the consistency and differences of our results with previously published estimates, including those based on COM lines, and the limitations of our methodology. Finally, we summarize the paper and present our conclusions in Sect.~\ref{s:conc}.

\section{Catalogs of cores and luminosity peaks}  
\label{s:data-cat}

\begin{figure*}[htbp!]
\centering
\begin{minipage}{1\textwidth}
    \centering
    \includegraphics[width=0.495\textwidth]{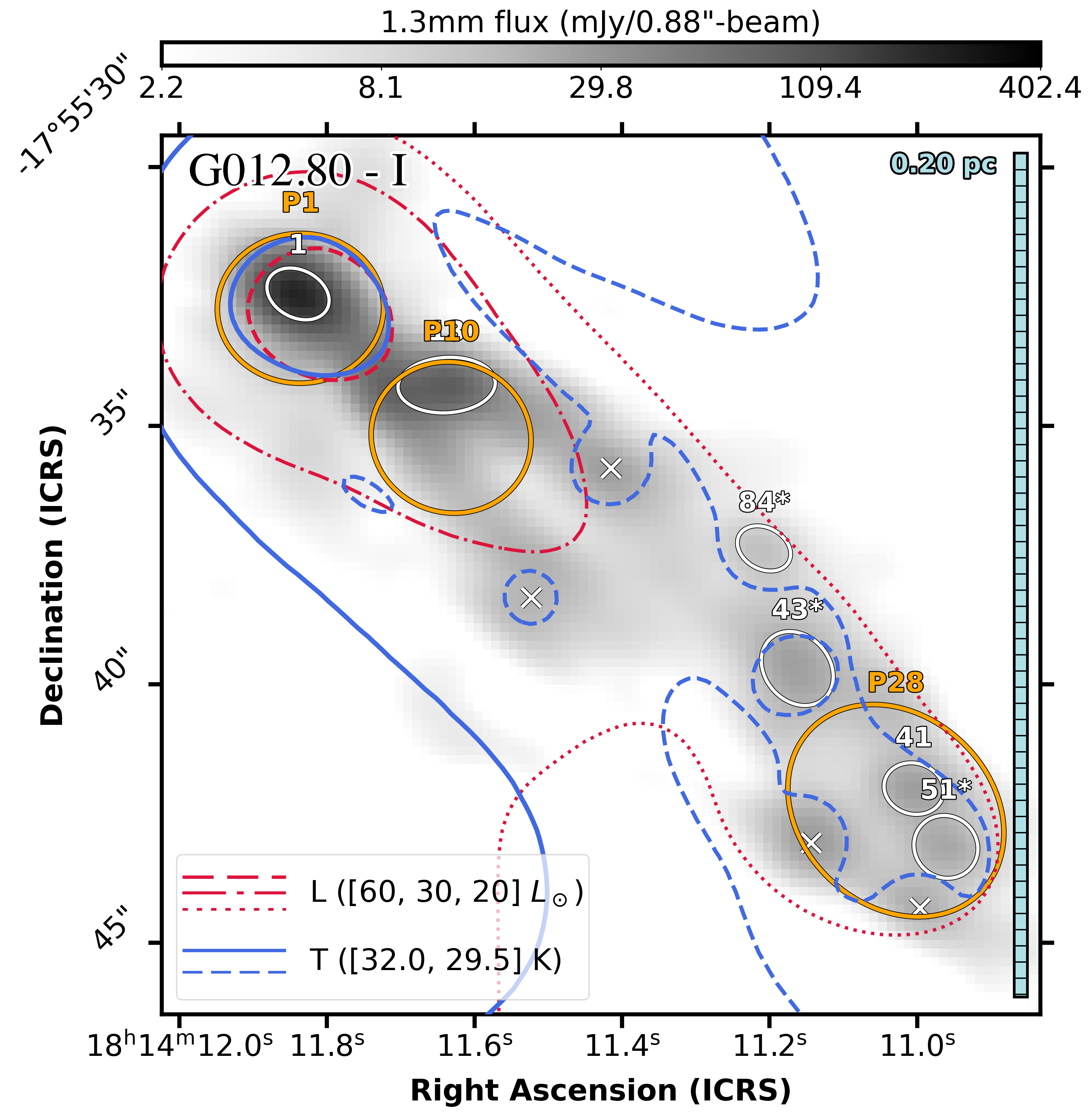} 
    \includegraphics[width=0.49\textwidth]{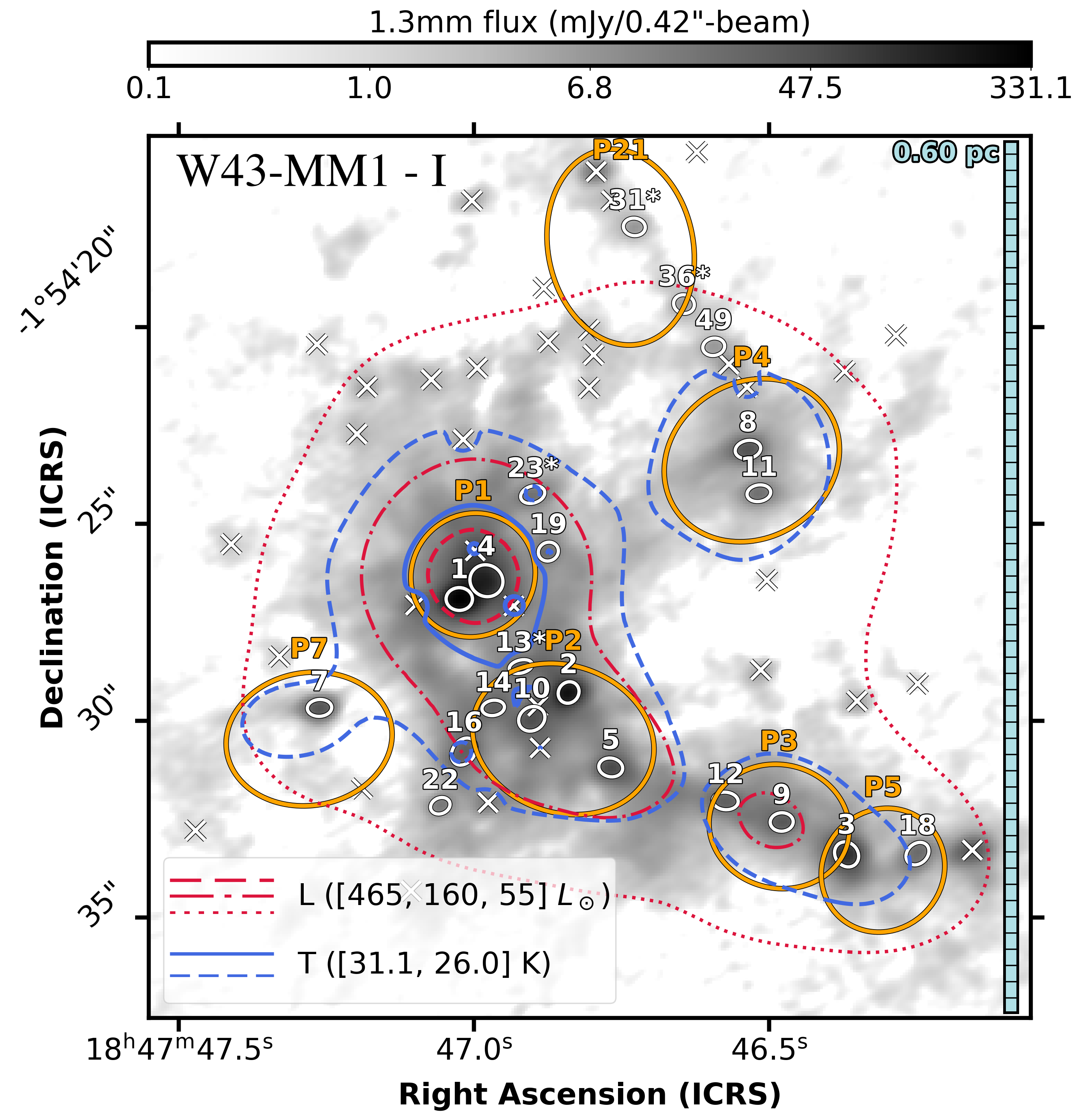}
    \vskip 15pt
    \includegraphics[width=0.494\textwidth]{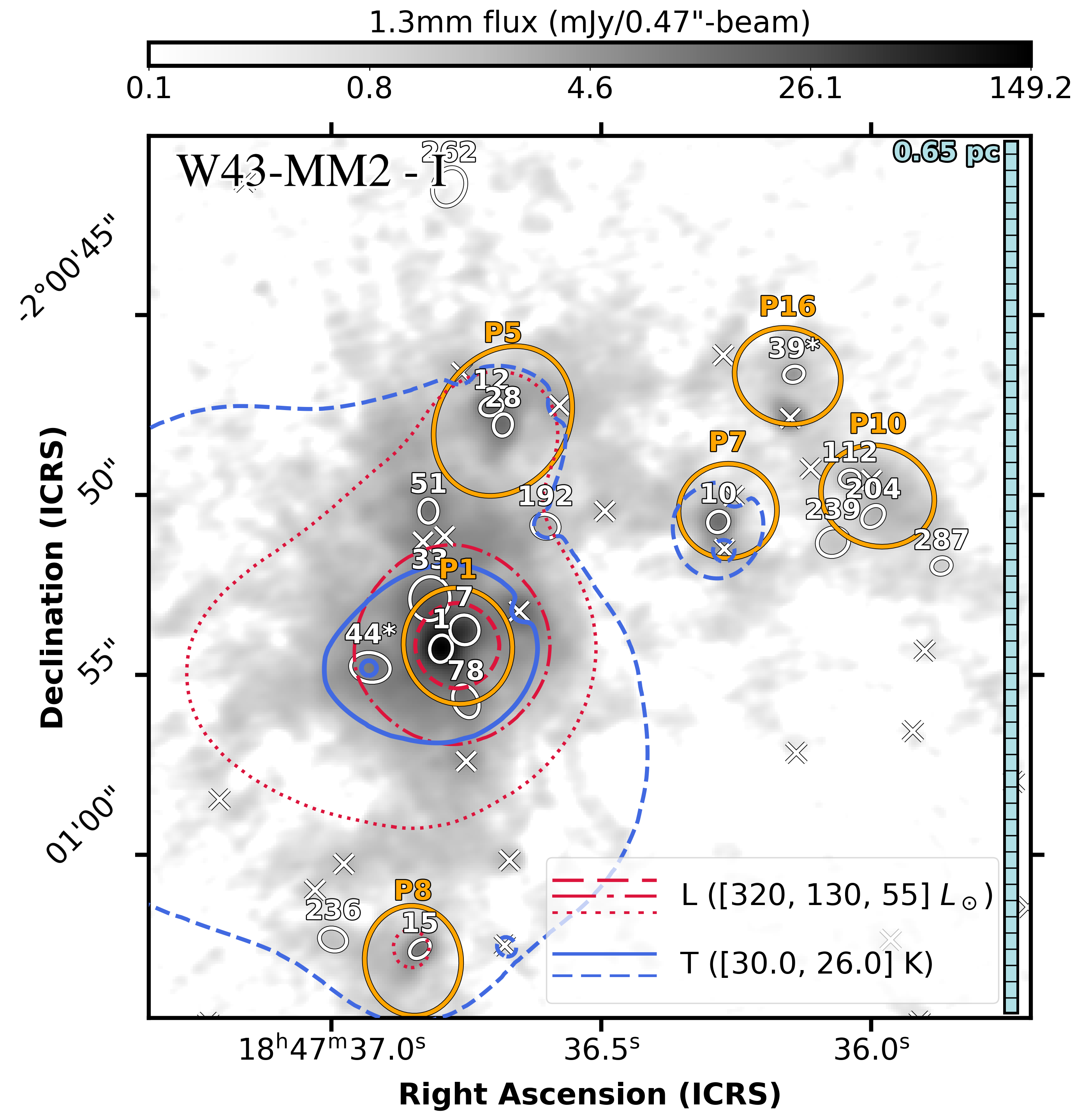}
    \includegraphics[width=0.49\textwidth]{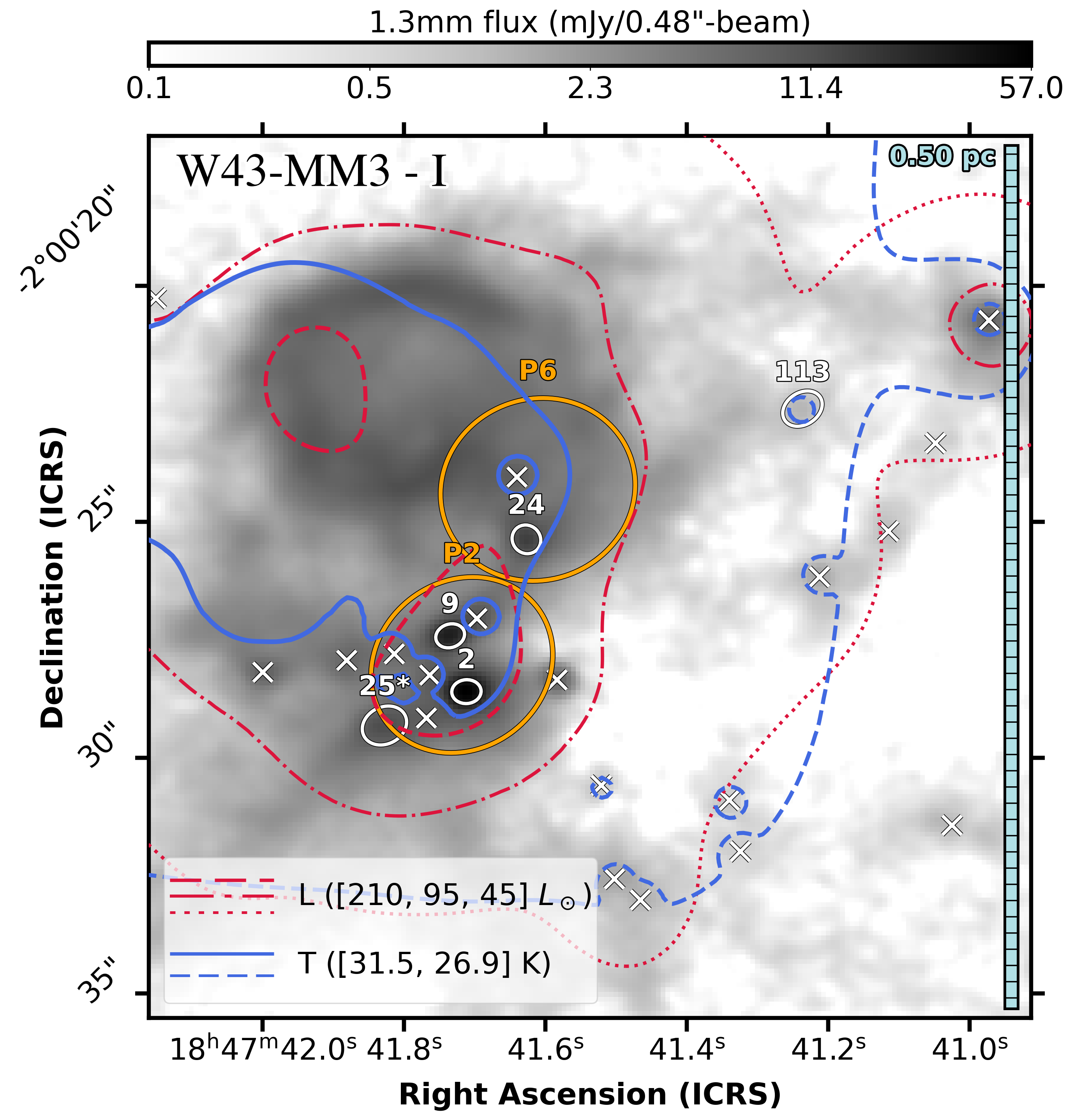}
\end{minipage}
\caption{Zoom into the luminosity and temperature peaks, characterized at $2.5\arcsec$, to identify their associated protostellar and prestellar cores, detected at $0.3\arcsec-0.9\arcsec$ resolution. Example zoom-in regions in our test-case ALMA-IMF protoclusters (the location of zoomed images is shown in \cref{appendixfig:zoom positions}): \textit{a}) the Evolved G012.80/W33-Main, \textit{b}) the Young W43-MM1, \textit{c}) the Young W43-MM2, and \textit{d}) the Intermediate W43-MM3. Red and blue contours display the \PPMAP luminosity and dust temperature map values, respectively, overlaid on the 1.3~mm continuum map shown in the grayscale background. Orange ellipses outline the FWHM size of the luminosity peaks associated with at least one protostellar core \citep[see][and \cref{tab:measures table evolved}]{dellova2024}. White ellipses and crosses locate the protostellar and prestellar cores, respectively, identified by \cite[][in prep.]{nony2020, nony2023}, \cite{pouteau2022}, and \cite{armante2024}. A scale bar is shown in the right-hand side of each panel. Some luminosity (and temperature) peaks host two and up to four protostellar cores  of 1900~au typical size (see, e.g., P1--5 in \textit{panel b} and P1 and P5 in \textit{panel c}).}
\label{fig:lum-temp peaks G012-W43}
\end{figure*}

The present study makes extensive use of the ALMA-IMF databases: the core catalogs (see Sect.~\ref{s:corecat}), which are taken from companion papers \citep[Papers~III, V, X, and XV,][]{pouteau2022, nony2023, armante2024, louvet2024}, and their association with outflows \citep[including Papers~V and X,][Nony et al. in prep.]{nony2020, nony2023, armante2024, valeille2024} and hot cores \citep[Papers~IV and XI,][]{brouillet2022, bonfand2024}. This study also uses luminosity peak catalogs, provided by Paper~XII \citep{dellova2024}, and properties derived for some hot cores by Paper~XI (see Sects.~\ref{s:lumcat}--\ref{s:extremehotcore}). 
In what follows, each of the above sources of information are integrated to classify the population of ALMA-IMF cores, as:
\begin{enumerate}
    \item[1] luminous protostellar cores: cores associated with a luminosity peak,  outflows, and sometimes a hot core. Several protostellar cores can be associated with a single luminosity peak (see Figs.~\ref{fig:lum-temp peaks G012-W43} and \ref{appendixfig:lum and temp peaks}). 
    \item[2] low-luminosity protostellar cores: cores associated with outflows but no luminosity peak, and rarely associated with a hot core.
    \item[3] prestellar cores: cores neither associated with a detected outflow nor with a hot core. 
\end{enumerate}
We outline below all the catalogs and databases used for this key core classification.

\subsection{ALMA-IMF prestellar and protostellar cores}
\label{s:corecat}

Our aim here is to determine the characteristics of as many ALMA-IMF cores as possible. We therefore used the best existing catalogs of cores for each ALMA-IMF protocluster, the most complete studies on the nature of these cores, and also excluded continuum sources whose nature is still too poorly determined. 

We used the homogeneous sample of sources obtained by \cite{louvet2024} for each of the 15 ALMA-IMF protoclusters, but replaced it for the W43-MM1, W43-MM2, W43-MM3, and G012.80 protoclusters by the more complete catalogs of \cite{pouteau2022}, \cite{nony2023}, and \cite{armante2024}. All catalogs are produced by the \getsf method \citep{men2021getsf} that identified and characterized cores in the 1.3~mm and 3~mm images, that are kept at their original angular resolutions ($0.3-0.9$\arcsec). While \cite{louvet2024} used the \cleanest images, which are computed to be free of contamination by line emission, the other three studies \citep{pouteau2022, nony2023, armante2024} complemented the \cleanest continuum images with the most sensitive, \bsens, continuum images of ALMA-IMF \citep[see definition in][]{ginsburg2022}. Moreover, the surveys of the W43 protoclusters used images whose noise has been reduced using the Multi-resolution non-Gaussian Segmentation software \citep[\textsl{MnGSeg,}][]{robitaille2019}. In total, the published ALMA-IMF catalogs, at the original resolution, contain $\sim$900 sources with typical deconvolved full width at half maximum (FWHM) sizes of 2300~au and fluxes spanning 3 decades.

The flux of these $\sim$900 sources at 1.3~mm generally corresponds to thermal dust emission, with the notable exception of $\sim$$10\%$ of these sources that present a 1.3~mm to 3~mm flux ratio suggesting their 1.3~mm flux is contaminated by free-free emission. \cite{louvet2024} labeled these sources as potentially contaminated by free-free emission, while \cite{pouteau2023}, \cite{armante2024}, and \cite{bonfand2024} performed a first-order correction of the 1.3~mm flux of some of these sources. They found that most of them have a 1.3~mm emission, mainly of thermal origin, requiring no correction or a minor correction of $10\%$ to $50\%$. These sources are cores that either simply lay on the line-of-sight of free-free emission from a nearby, compact or developed, \hii region or could be protostars hosting a hyper-compact (HC) \hii region \citep{hoare2007}. In addition, three cores driving outflows (see below) have their 1.3~mm flux corrected by larger factors and could thus represent high-mass protostars hosting an ultra-compact (UC) \hii region while still accreting. Defining the exact nature of these sources is out of the scope of the present paper but will be discussed in Nony et al. in (prep.). Cores whose 1.3~mm flux is corrected for free-free emission are preserved in our core catalog, but subsequent estimates for these objects are more uncertain.

We used studies searching for molecular outflows and hot cores to characterize the ALMA-IMF cores as either prestellar or protostellar in nature. With the exception of the SiO survey of companion paper, Paper~IX \citep{towner2024}, all ALMA-IMF outflow studies used both the CO(2-1) and SiO(5-4) lines, investigated the shape of spectral lines toward and around the cores, and the spatial distribution of their blue-shifted and red-shifted wings (\citealt{nony2020, nony2023}, in prep.; \citealt{armante2024, valeille2024}). Conducted over the entire ALMA-IMF protocluster sample, the outflow survey of Nony et al. (in prep.) found several hundreds of protostellar cores driving outflows, among which $\sim$$35\%$ are quoted tentative. Tentative protostellar cores, with potential indications of outflowing gas, are found either close to other protostellar cores \citep[see, e.g.,][]{nony2020} or within \hii regions where the ionized gas, entrained by the protostellar jet, escapes our detection with molecular tracers \citep[see, e.g.,][]{towner2024,armante2024}. Two thirds of the hot core candidates identified by the MF (CH$_3$OCHO) line survey of \cite{bonfand2024} are associated with a protostellar core driving an outflow. The MF sources associated with neither CO nor SiO outflow could correspond to protostars located in an \hii region, or to HC\hii sources, whose gas ejection is not detected by molecular lines \citep{bonfand2024}. They could also pinpoint shock locations \citep[e.g.,][]{lefloch2017, csengeri2017b} or correspond to protostars in their early warm-up phase that drive a weak outflow \citep{bouscasse2024}. The survey of high-velocity outflows driven by high-mass protostellar cores, performed by \cite{valeille2024}, refer to some of the protostars from Nony et al. (in prep.) as high-mass prestellar core candidates (HMPreSCs). They could be high-mass cores that only host low- to intermediate-mass protostars driving low-velocity outflows.

In total, the ALMA-IMF survey has identified 266 protostellar cores, 617 prestellar core candidates, and 81  sources associated with neither outflows nor hot cores and indicated as potentially contaminated by, but not yet corrected for, free-free emission. As long as the prestellar nature of the last group is not assured, we exclude these ALMA-IMF sources from the present study.
The masses of ALMA-IMF cores, computed from their thermal dust emission at 1.3~mm and initial temperature estimates, range from $\sim$$0.1~\msun$ to $\sim$$200~\msun$ \citep{motte2022}.

\subsection{Luminosity peaks in ALMA-IMF protoclusters}
\label{s:lumcat}

The luminosity maps of the ALMA-IMF protoclusters were created by \cite{dellova2024}, with each $2.5\arcsec$ resolution element of the maps corresponding to the luminosity integrated along the line of sight and over a wide wavelength range. Each SED covers the $3.6\,\mu{\rm m}$ to $24\,\mu{\rm m}$ and the $70\,\mu{\rm m}$ to 1.3~mm ranges that correspond to what are generally referred to as the warm and cold components of protostellar SEDs 
\citep[see][and in particular their Fig.~2 for details]{dellova2024}. The resulting luminosity images are made up of pixels whose value is the sum of the integration of a spline function to describe the warm SED component and a modified blackbody fit to the cold SED component, fitted by the Bayesian \PPMAP technique \citep{marsh2015, marsh2017}. The luminosities measured by this procedure are therefore the closest to a bolometric luminosity measurement. Unlike for their dust temperature and column density estimates, \cite{dellova2024} used ALMA-IMF 1.3~mm images that have not been corrected for free-free emission to create their luminosity images. As a result, the luminosity measured on certain pixels of their images does not correspond to a purely thermal luminosity.

\cite{dellova2024} used the \getsf method \citep{men2021getsf} to identify and characterize peaks in their luminosity images. The \getsf method spatially decomposes the observed image to isolate and separate relatively round sources from elongated structures and a larger-scale background. As a first step, the source detection provides a first-order estimate of the source sizes and luminosities. Improved measurements are then performed on background-subtracted images and provided in a catalog. \getsf identified 313 peaks, of which a one-third are associated with one or more protostellar cores that we will discuss below (see, e.g., \cref{fig:lum-temp peaks G012-W43}). Their median size is $\sim$$2.8\arcsec$, leading to a deconvolved physical size range from 2500~au to $7\,000$~au, depending to first order on the distance of the ALMA-IMF protocluster. Their luminosity varies from $2~\lsun$ to $75\,000~\lsun$ with a median of $\sim$360~$\lsun$. Given the noise level and cloud structure measured in the luminosity images, \cite{dellova2024} estimate that their census of protostellar groups is complete down to $\sim$$30~\lsun$ in Young protoclusters but only $\sim$$100~\lsun$ in Evolved ones. The present study will therefore have difficulty characterizing low-mass protostars, nor will it be complete for intermediate-mass protostars of the ALMA-IMF protoclusters, which are expected to have luminosities of $1-100~\lsun$ \citep[e.g.,][]{bontemps2010SI, duarte2013}.

For the ALMA-IMF protoclusters that were classified as Young or Intermediate by \cite{motte2022}, we directly extracted luminosity peaks from the catalog of \cite{dellova2024}. In the case of the four ALMA-IMF protoclusters classified as Evolved \citep{motte2022}, we performed a novel extraction to improve the measurement of luminosity peaks associated with protostars located at the edge or within \hii regions. Following \cite{dellova2024}, we used the \getsf algorithm version 230712 \citep{men2021getsf} to extract luminosity peaks with sizes ranging from one to two times the $2.5\arcsec$ beam of luminosity images. In contrast to \cite{dellova2024}, here we applied \getsf on luminosity images built using the 1.3~mm images corrected from free-free emission \citep{diazgonzalez2023, galvanmadrid2024}. At the periphery of \hii regions, our extraction favors dense structures over photo-dissociation region (PDR) structures, which are associated with free-free emission from these \hii regions. This process identifies fewer luminosity peaks, in total 255, and the same 111 luminosity peaks associated with protostellar cores ($90\%$ overlap) as the extraction performed on luminosity images constructed with 1.3~mm maps uncorrected for free-free emission \citep{dellova2024}. Our extraction does, however, better define the size and luminosity of peaks associated with protostellar groups that are located in PDRs because they are too spatially extended and therefore too bright in the catalog of \cite{dellova2024}.

We characterize here the luminosity peaks, that are associated with one or several protostellar cores. This association is positive when the center of at least a protostellar core, taken from the core catalogs described in Sect.~\ref{s:corecat}, lies within the ellipse describing the FWHM of the luminosity peaks. Figure~\ref{fig:lum-temp peaks G012-W43} presents coincidences of protostellar and prestellar cores identified in the G012.80, W43-MM1, W43-MM2, and W43-MM3  protoclusters (see \citealt{nony2020, nony2023}, in prep.; \citealt{armante2024}) with luminosity peaks. Luminosity peaks generally, but not univocally, correspond to temperature peaks (see Sect.~\ref{s:Tpeak}), because it is easier for the \PPMAP technique to produce luminosity peaks than temperature peaks. All the other coincidences, observed for the 14 studied ALMA-IMF protoclusters\footnote{
     The G351.77 protocluster is excluded from the present study because the quality of its CO(2-1) datacube precluded investigating outflows in this region \citep{cunningham2023, valeille2024}.}, 
are presented in \cref{appendixfig:lum and temp peaks}. Each luminosity peak of the ALMA-IMF protoclusters corresponds to one protostar or a protostellar group, composed of up to four protostars and three prestellar cores. Prestellar cores that lie within these luminosity peaks (see, e.g., P1 and P2 in W43-MM1 in \cref{fig:lum-temp peaks G012-W43}b) should only account for a negligible part of the measured luminosity. Protostellar cores associated with a luminosity peak are called "luminous protostellar cores", while others are called "low-luminosity protostellar cores".

Tables~\ref{tab:measures table}--\ref{appendixtab:measures table evolved} list, for each luminosity peak associated with protostars, their integrated luminosity, $\lbol(r<\theta_{L_{\rm bol}})$, their FWHM angular size, $\theta_{L_{\rm bol}}$, and their FWHM spatial size, FWHM$_{L_{\rm bol}}$. Tables~\ref{tab:measures table evolved} and \ref{appendixtab:measures table evolved} also provides the coordinates, RA and Dec, of the luminosity peaks extracted in the present paper. The luminosity peaks associated with protostars have FWHM sizes ranging from one to two times the $2.5\arcsec$ beam of \PPMAP images. For these protostellar groups, which have a size of $6000-29\,000$~au, the measured luminosity range from $\sim$$20~\lsun$ to $\sim$$7.5\times 10^4~\lsun$ (see Tables~\ref{tab:measures table}--\ref{appendixtab:measures table evolved}). They correspond to the brightest sources in the luminosity peak catalog of \cite{dellova2024} (see their Table~B.1). These values are in agreement with the luminosity of high-mass protostars, with the lowest luminosities for their youngest phase, qualified as infrared (IR)-quiet \citep[see definition in][]{motte2007, csengeri2017b}, and the brightest for their evolved, HC\hii phase \citep[see definition by][]{hoare2007}. For each luminosity peak, Tables~\ref{tab:measures table}--\ref{appendixtab:measures table evolved} also indicate if a temperature peak is detected and list all the protostellar cores associated. It gives the protostellar core numbers in the published catalogs of Sect.~\ref{s:corecat}, their FWHM angular size, $\theta_{\rm core}$, and their distance to the luminosity peak center, $d_{\rm L_{\rm bol}}$, notably used to characterize the robustness of their association.

\subsection{Highest-luminosity protostars in ALMA-IMF protoclusters}
\label{s:extremehotcore}
 
The brightest ALMA-IMF protostars heat cloud areas with diameters of $0.4-1$~pc (see \cref{appendixsect:ppmap} and, e.g., \cref{appendixfig:ppmap}-lower row). Hence, their protostellar luminosity, which must be measured over a wide area around the protostar while excluding the luminosity of its surrounding cloud, is poorly defined by a simple metric such as luminosity peak, $\lbol(r<\theta_{L_{\rm bol}})$ (see Sect.~\ref{s:lumcat}). The highest-luminosity protostars are also expected to be associated with bright, chemically rich hot cores if we consider that COMs are released into the gas by thermal desorption, after having formed onto grains \citep[e.g.,][]{garrod2006}. In fact, \cite{bonfand2024} have discovered six very extended, $4000-13\,400$~au, MF sources, and argue that they correspond to extreme hot cores heated, to over $\sim$100~K, by $10^5-10^6~\lsun$ protostars. In qualitative agreement with this interpretation, five of these six extreme hot cores are associated with the three brightest luminosity peaks of ALMA-IMF: W51-E-MF1 in W51-E\#P2, W51-E-MF2 and MF3 in W51-E\#P1, and W51-IRS2-MF1 and MF3 in W51-IRS2\#P1. However, these peak luminosities, $\lbol(r<\theta_{L_{\rm bol}}) = L_{\rm MF}(\rm peak)=3-7.5 \times 10^4~\lsun$, only allow us to estimate lower limits to the luminosities of their host protostars (see Tables~\ref{appendixtab:measures table}--\ref{tab:luminosity peaks from radial profile}).

Therefore, integration over larger areas and decomposition of different emission components are required for the protostellar cores associated with the six extreme MF sources of \cite{bonfand2024}. We investigated the luminosity images around four positions, encompassing, due to the lower resolution of the luminosity maps, these extreme MF sources: two toward G327.29-MF1 and W51-E-MF1, one toward the middle location of W51-E-MF2 and -MF3, encompassing both, and the last toward the middle location encompassing W51-IRS2-MF1 and -MF3. Figure~\ref{appendixfig:luminosity profile} presents their pixel luminosity profile as a function of angular radius, $L(\theta)$, which is made up of the median luminosity of pixels contained within annular rings, taking their medians in order to attenuate the effect of nearby luminosity peaks on the profile. As expected, the spheres of influence of these highest-luminosity protostars display no clear outer radii.  

Here, we model the pixel luminosity profiles by the sum of a power-law and a Gaussian, both centered at angular radius $\theta=0$. As shown below, we associate the power-law model, which dominates at radii smaller than $0.05-0.1$~pc, with the cold emission component of protostellar envelopes, and the Gaussian with a typical FWHM of $\sim$0.5~pc with the sum of the hot emission component of protostars and the cold emission of the cloud structures surrounding protostellar cores. This approach requires us to define two additional measures of luminosity, $L_{\rm MF}(\rm pow)$ and $L_{\rm MF}(\rm tot)$ (see definition below). \cref{tab:luminosity peaks from radial profile} lists the eight highest-luminosity ALMA-IMF protostars that require this special treatment, along with the three estimates of the luminosity discussed here.

The spherical power-law luminosity profile model is expected to correspond to the cold emission component of protostellar cores observed from far-infrared to submillimeter wavelengths. As a matter of fact, a protostellar envelope with infinite spherical symmetry, which has a density profile $\rho(r)\propto r^{-p}$, has a column density profile $N_{\rm H_2}(r) \propto r \times \rho(r) \propto r^{-p+1}$. In the Rayleigh-Jeans regime and with a radial temperature profile of $T(r) \propto r^{-q}$, the intensity profile of this protostellar envelope is $I_{\rm cold}(r) \propto N_{\rm H_2}(r) \times T(r) \propto r^{-m}$ with $m=p+q-1$ \citep[e.g.,][]{motte2001}. It follows that the cold component of a protostellar core, with $\rho(r)\propto r^{-2}$ and $\tdust(r)\propto r^{-0.36}$ profiles (see Sect.~\ref{s:Tproto-eq}), has a radial luminosity profile reasonably described by a power-law $L(\theta)\propto \theta^{-m}$ with an index of $m\simeq 2+0.36-1=1.36$. With the exception of W51-E-MF1, the best-fit power-laws have indices ranging from $1.2$ to $1.9$ (see \cref{appendixfig:luminosity profile}), close to the theoretical value of $m\simeq 1.36$. We thus argue that the pixel luminosity, integrated below this power-law component, $L_{\rm MF}(\rm pow)$, represents a good estimate of the cold SED component of the underlying protostars.

In addition, the Gaussian component completing the fit of the profiles of pixel luminosity could correspond to the sum of the warm SED component of protostars and the cold SED component of their parental clumps, heated by nearby \hii regions, internal shocks, and the interstellar radiation field (ISRF). The latter has no reason to be spherically symmetric. Similarly, the warm component of protostar SED is far more affected by geometric effects such as disk inclination and outflow cavities than its cold component \citep[e.g.,][]{furlan2016}. Hence, this Gaussian component is complex in nature and its three-dimensional modeling remains less well constrained than the power-law component. Therefore, we provided in \cref{tab:luminosity peaks from radial profile} a third estimate of the luminosity of protostars or protostellar groups driving the brightest hot cores of \cite{bonfand2024}: that integrated under the total luminosity profile, $L_{\rm MF}({\rm tot})$. The two $L_{\rm MF}({\rm pow})$ and $L_{\rm MF}({\rm tot})$ estimates of the protostellar luminosities should contain most of the radiation escaping along the outflow cavities, since they are integrated up to the map edges, at $\sim$0.6--1.2~pc from the protostars studied, whereas protostellar outflows are generally smaller in size \citep[e.g.,][]{nony2023}. 

For each of the four locations encompassing the six brightest hot cores of \cite{bonfand2024}, \cref{tab:luminosity peaks from radial profile} lists their name and coordinates along with the three estimates of the underlying  luminosity discussed above. We measured hot core luminosities ranging from $L_{\rm MF}({\rm pow})=9\times 10^3~\lsun$ to $2\times 10^5~\lsun$, increased by 3\% to more than a factor of 6 with respect to their peak luminosities. The total luminosities of these extreme hot cores, $L_{\rm MF}({\rm tot})$, are $2.5-9$ times higher, but should be considered as upper limit values because they include the luminosity of the hot core background. In the following, we use $L_{\rm MF}({\rm pow})$ to compute the contribution of each protostar and use it as the best estimate of protostellar luminosities.

\section{Database of temperature images}
\label{s:data-temp}

The mean temperature of clouds is not identical either across our Galaxy or throughout the cloud's evolution, but depends directly on the intensity of the cloud's interactions with sources of irradiation, cosmic rays, and shocks. While it is obvious that the densest parts of clouds self-shield more easily from the external ISRF, filaments and cores nevertheless form in a medium whose temperature could be cold or warm depending on their parental cloud.

From the dust temperature images provided by Paper~XII \citep{dellova2024}, we measure in Sect.~\ref{s:Tbck} the background-diluted temperature of the ALMA-IMF cores, that is the dust temperature measured toward the core but diluted in the \PPMAP beam, which is larger than the size of the core.
We then investigate in Sect.~\ref{s:Tpeak} variations of the cloud temperature and identified temperature peaks.

\subsection{Background-diluted dust temperature of cores}
\label{s:Tbck}

The ALMA-IMF database includes dust temperature images for all 15 protoclusters, at an angular resolution of $2.5\arcsec$ \citep{dellova2024}. The \PPMAP procedure \citep{marsh2015, marsh2017} was used to fit modified blackbody models to the fluxes measured in the $70\,\mu{\rm m}$ to 1.3~mm wavelength range, using a dust opacity law of $\kappa_\nu \propto \nu^{\beta}$ with $\beta=1.8$ \citep{planck2011, koehler2015}. The resulting temperature images correspond to column density-weighted temperatures. Neglecting line-of-sight effects \citep[e.g,][]{malinen2011}, these maps provide a good approximation of the mass-averaged dust temperature for the densest cloud structures observed at $2.5\arcsec$ in the ALMA-IMF fields.

Since ALMA-IMF protoclusters have been selected to be clusters in their embedded phase \citep{motte2022}, the high column density of the protocluster cloud and dense structures \citep{diazgonzalez2023, dellova2024} partially extinguishes their $70\,\mu{\rm m}$ emission. As a result, the \PPMAP fits, which use the $70\,\mu{\rm m}$ fluxes, underestimate the dust temperature globally, over the whole protocluster, and even more, locally, toward protostars. \cite{dellova2024} corrected the dust temperature images for the $70\,\mu{\rm m}$ opacity, using the column density image computed by \PPMAP and a grid of temperature corrections. This grid, originally built for the pilot ALMA-IMF study \citep{motte2018b}, estimates temperature corrections resulting from the radiative transfer model of a luminous source, a protostar or an UC\hii region, located behind a cloud \citep[using COREFIT,][]{marsh2014}. This correction increases the global dust temperature of ALMA-IMF protoclusters by $1-3$~K (see \cref{appendixtab:ppmap table}). 
This correction also locally raises the dust temperature, from $20-30$~K to $40-80$~K, toward column density peaks hosting cores, revealing hot spots within cold filaments \citep{dellova2024}.

As stated in \cite{dellova2024} and shown in \cref{appendixsect:ppmap}, the best-suited image to estimate the background-diluted dust temperature of cores, that is the dust temperature measured at the location of ALMA-IMF cores but diluted in the $2.5\arcsec$ \PPMAP beam, is the dust temperature image corrected for the $70\,\mu{\rm m}$ opacity. However, at the location of cold cores such as prestellar cores and low-luminosity protostellar cores located in Evolved protoclusters, these opacity-corrected dust temperature images display local heating. The \PPMAP assumption that part of the $70\,\mu{\rm m}$ emission, associated with the heating by \hii regions, originates from a bright protostar embedded within prestellar cores or cold envelopes of low-luminosity protostars is indeed erroneous. Therefore, the temperature for prestellar cores and low-luminosity protostellar cores in Evolved protoclusters are overestimated. We propose that the least biased temperature estimate for the background-diluted temperature of these low-luminosity cores is the average between the temperatures measured in the original and in the opacity-corrected images (see \cref{appendixsect:ppmap}). This simple average takes into account both the underestimation of cloud heating of the original image and the overestimation of local heating of the opacity-corrected image. We then constructed combined dust temperature images, from the opacity-corrected and original temperature images of \cite{dellova2024}. Their default, opacity-corrected, temperature images are modified at the location of prestellar cores and protostars not bright enough to be seen in the luminosity images. For these cores, whose nature is defined by companion papers (\citealt{nony2020, nony2023}, in prep.; \citealt{armante2024}; \citealt{bonfand2024}, see Tables~\ref{tab:measures table}--\ref{appendixtab:measures table evolved}), we used the average between the original and the opacity-corrected images, as measured at the central position of the cores and smoothed by Gaussians representing the resolution elements of the 1.3~mm images.

\begin{figure*}[htbp!]
    \centering
    \includegraphics[width=1\linewidth]{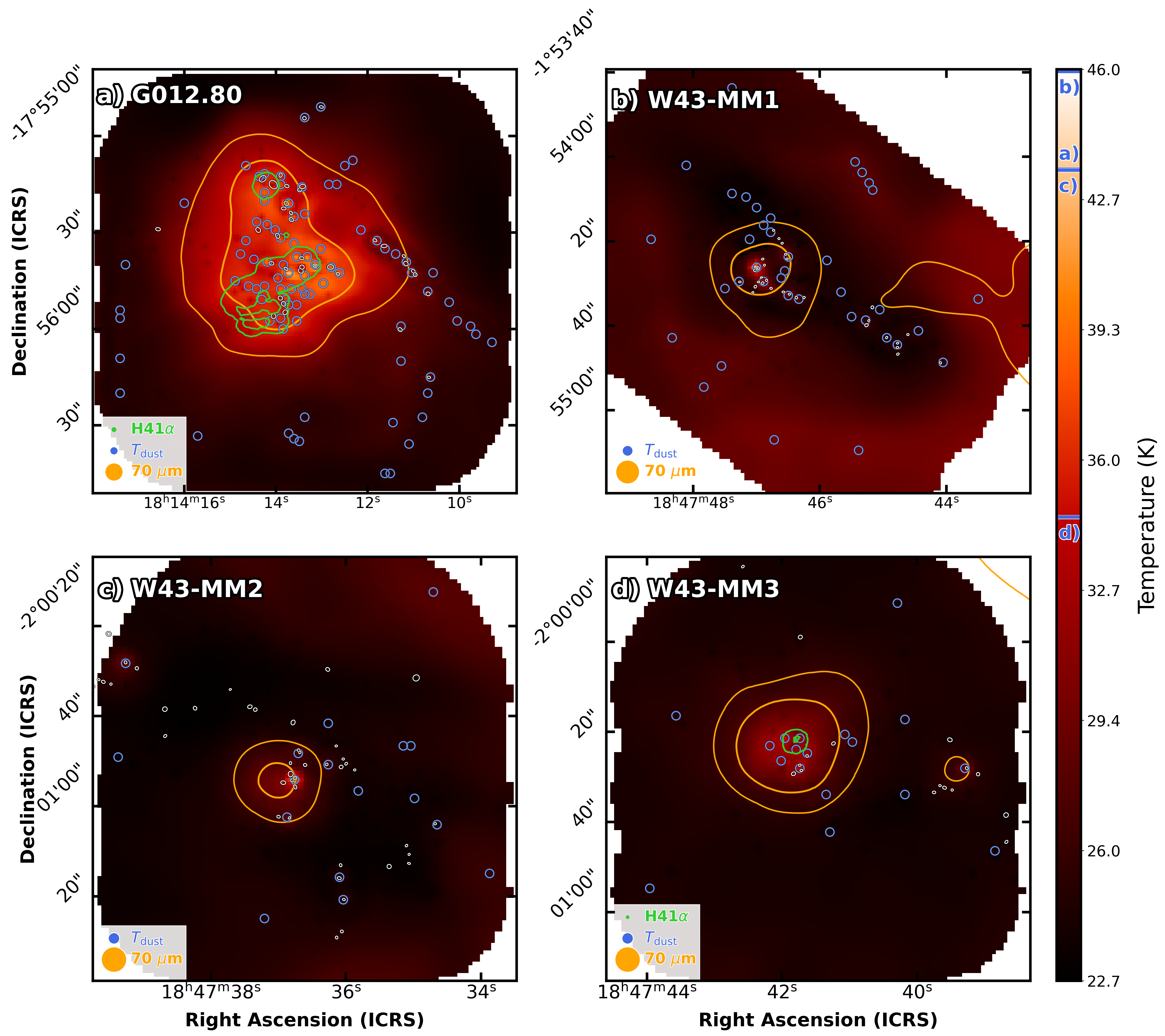}
    \vskip -0.cm
    \caption{Temperature maps used to estimate the background-diluted temperature of cores detected in \textit{a}) G012.80, \textit{b}) W43-MM1, \textit{c}) W43-MM2, and \textit{d)} W43-MM3. These images use the corrected \PPMAP images along with the original \PPMAP products (which ignore the $70~\mu$m opacity correction), revealing both heating at the location of the most luminous protostars and self-shielding toward prestellar cores and low-luminosity protostars (see Sect.~\ref{s:Tbck}, \cref{appendixsect:ppmap}, \cref{appendixfig:ppmap} and \citealt{dellova2024}).
    Green and orange contours outline the H41$\alpha$ emission (at 25$\sigma$ levels) and the $70\,\mu{\rm m}$ emission (at 3$\sigma$ and 7$\sigma$ levels), respectively. A common color bar for temperature is placed on the right-hand side of the figure and blue horizontal lines indicate the maximum temperature of each panel. Temperature peaks are indicated by blue circles, protostellar cores by white ellipses. The resolution of the ALMA H41$\alpha$ emission, $0.38\arcsec-2.1\arcsec$, the \PPMAP dust temperature image, $2.5\arcsec$, and the $70\,\mu{\rm m}$ emission, $5.6\arcsec$, are given in the lower left corner. The background-diluted core temperature in the Evolved G012.80 protocluster is globally $\sim$4~K hotter than in the three other, Young and Intermediate, protoclusters.}
    \label{fig:temperature maps}
\end{figure*}

We measured the background-diluted temperature of cores in the combined dust temperature image of each ALMA-IMF protocluster. This temperature corresponds to the one weighted by the column density measured with a $2.5\arcsec$ beam. Tables~\ref{tab:measures table}--\ref{appendixtab:measures table evolved} list the \PPMAP dust temperatures at core location, $\overline{T_{\rm dust}^{\rm PPMAP}}$[$1.25\arcsec$], which range from 17~K to 73~K. The uncertainty on the background-diluted core temperature is the quadratic sum of the uncertainty associated with the \PPMAP fitting procedure ($\sigma_{\rm PPMAP}\sim 16\%$) and that caused by the opacity correction. In practice, at core location the uncertainty covers the two extreme values provided by the original and opacity-corrected temperature maps.

\subsection{Variations of the dust temperature and hot spots}
\label{s:Tpeak}

The ALMA-IMF protoclusters are located in the inner part of the Galactic plane and are often located in the immediate vicinity of massive star clusters \citep[][see their Fig.~1]{motte2022}. They were also chosen to be active protoclusters forming high-mass stars, thus associated with clusters of outflows \citep{nony2020, nony2023, towner2024, valeille2024} and \hii regions \citep{diazgonzalez2023, galvanmadrid2024}. We therefore expect the cloud temperature of the ALMA-IMF protoclusters to vary from one protocluster to another and be higher in the Evolved protoclusters. As a matter of fact, \cite{dellova2024} measured spatially averaged, column-density weighted line-of-sight temperatures that range from 21~K to 29~K, with higher temperatures for Evolved protoclusters.

We illustrate the difference in the ALMA-IMF background-diluted core temperature using our four case-study regions: G012.80, W43-MM1, W43-MM2, and W43-MM3. Figure~\ref{fig:temperature maps} displays their combined dust temperatures images (see definition in Sect.~\ref{s:Tbck}), along with their \textit{Herschel} $70\,\mu{\rm m}$ and ALMA-IMF H41$\alpha$ emission \citep{galvanmadrid2024}. The cloud gas of the Evolved G012.80 protocluster is expected to be mainly heated by its three internal \hii regions  \citep{beilis2022, armante2024} with temperatures up to 43~K, 
in line with the spatially averaged temperature weighted by the column density of $\sim$28~K \citep{dellova2024}. Its $70\,\mu{\rm m}$ emission is indeed extended over $\sim$1~pc and very well correlated with both the large-scale structure of the temperature map and the \hii regions, outlined by their H41$\alpha$ emission (see \cref{fig:temperature maps}a). 
In contrast, the cloud gas of the Young W43-MM1 and W43-MM2 protoclusters, which are mostly heated by the external ISRF, are  colder than G012.80, $\sim$26~K on average, with $\sim$23~K at their center and temperature increases toward their periphery (see Figs.~\ref{fig:temperature maps}b--c). As for the Intermediate protocluster W43-MM3, it remains globally cold, at $\sim$25~K \citep{dellova2024}, with some moderate heating, $32$~K over an area of $\sim$0.35~pc diameter, which is associated with its UC\hii region (see \cref{fig:temperature maps}d). The $70\,\mu{\rm m}$ emission from Young and Intermediate protoclusters is more compact than in Evolved protoclusters because it originates from protostellar groups and compact \hii regions (see Figs.~\ref{fig:temperature maps}b--d). 

We used the \textsl{skimage/peak\_local\_max}\footnote{
    See documentation at \url{https://scikit-image.org/docs/stable/auto_examples/segmentation/plot_peak_local_max.html}} 
python function to identify dust temperature peaks, which are inflection points whose minimum size corresponds to the resolution element of the combined dust temperature images built in Sect.~\ref{s:Tbck}. Figures~\ref{fig:temperature maps}a--d locate the temperature peaks that we found in the G012.80, W43-MM1, W43-MM2, and W43-MM3 protoclusters. The large majority, $70\%$, of these hot gas pockets are low-density structures of PDRs, which are either associated with the \hii regions of Evolved protoclusters or with the outer parts of the Young or Intermediate clouds heated by the ISRF of nearby stellar clusters. This is the case throughout \cref{fig:temperature maps}a and particularly at the location of the G012.80 \hii regions \citep[see][]{armante2024}. We observe similar PDR features at the periphery of the W43-MM1 and W43-MM2 protoclusters, which are heated by the OB/WR stellar cluster located 5~pc away \citep[see Figs.~\ref{fig:temperature maps}b--c,][]{motte2003, nguyen2013}.

In addition to these PDR features, we identified in the entire ALMA-IMF survey, 83 temperature peaks, which coincide with a luminosity peak and at least one, and up to four, protostellar cores (see \cref{fig:lum-temp peaks G012-W43}). Coincidence occurs when the temperature peak lies within the \getsf ellipse describing the FWHM of the luminosity peaks \citep[see][and Tables~\ref{tab:measures table}--\ref{appendixtab:measures table evolved}]{dellova2024}. These peaks of both temperature and luminosity correspond to the heating imprints, ranging in size from 6000 to $29\,000$~au, of protostellar groups. In our case-study regions, these hot spots have line-of-sight dust temperatures ranging from 24~K to 42~K (see Tables~\ref{tab:measures table}--\ref{tab:measures table evolved} and \cref{fig:temperature maps}). In the Evolved G012.80 and Intermediate W43-MM3 protoclusters, three of these temperature and luminosity peaks are located at the periphery or within \hii regions: in G012.80 at the periphery of "4" and within "2-3" \citep[see the \hii regions outline definition in][and \cref{fig:temperature maps}a]{beilis2022, armante2024} and in W43-MM3 at the periphery of its UC\hii region \citep[see][and \cref{fig:temperature maps}d]{nguyen2017}.

In the following, Sects.~\ref{s:Tproto}--\ref{s:Tpre} describe our methodology, based on the temperature distributions in protostellar envelopes and prestellar cores. We extrapolate below the constraints provided by \PPMAP (see Sects.~\ref{s:lumcat} and \ref{s:Tbck}) toward even higher spatial resolution and estimate the mass-averaged temperature of the 883 ALMA-IMF cores.

\begin{figure*}[htbp!]
    \centering
    \includegraphics[width=0.33\linewidth]{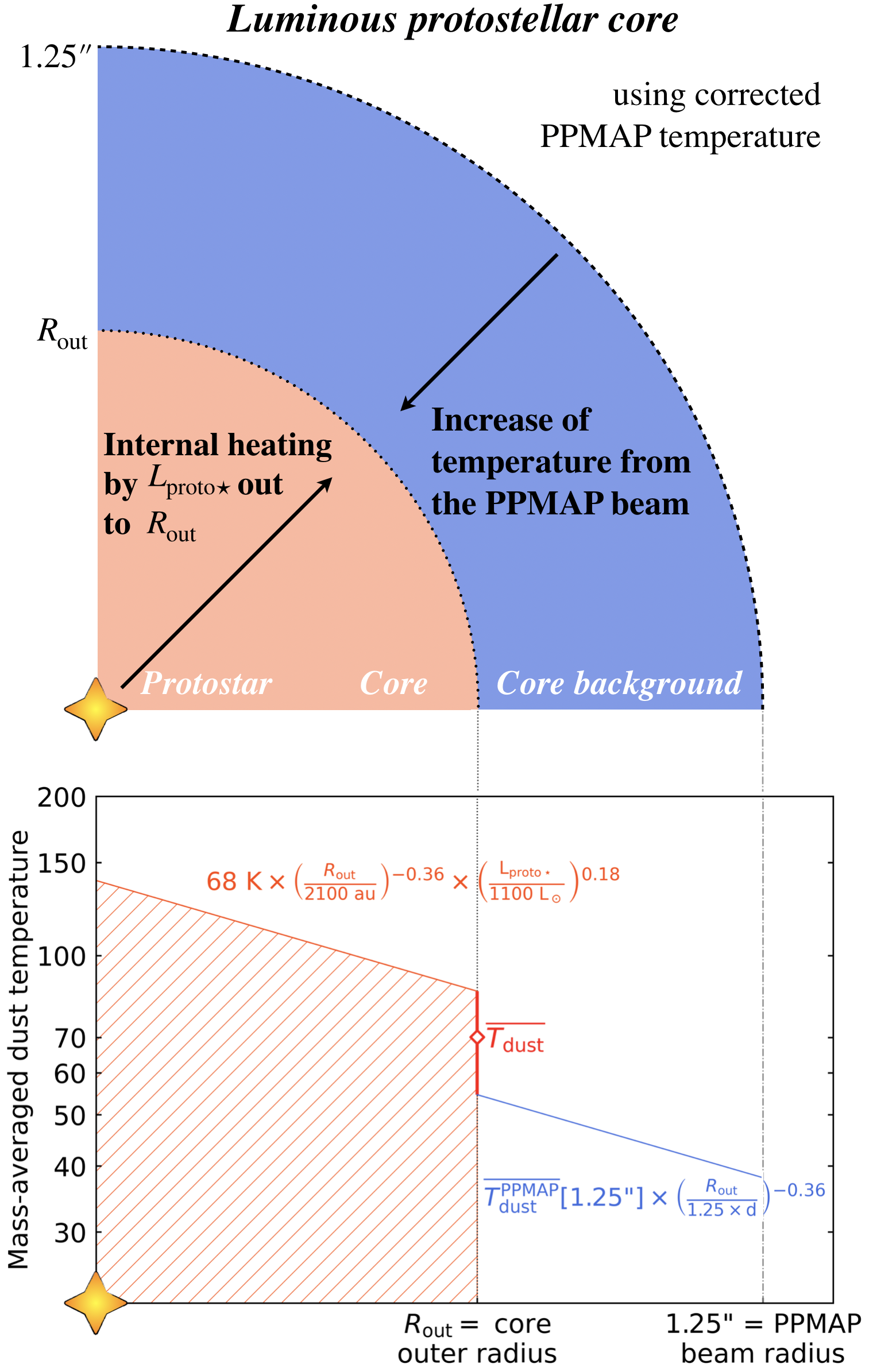}
    \includegraphics[width=0.33\linewidth]{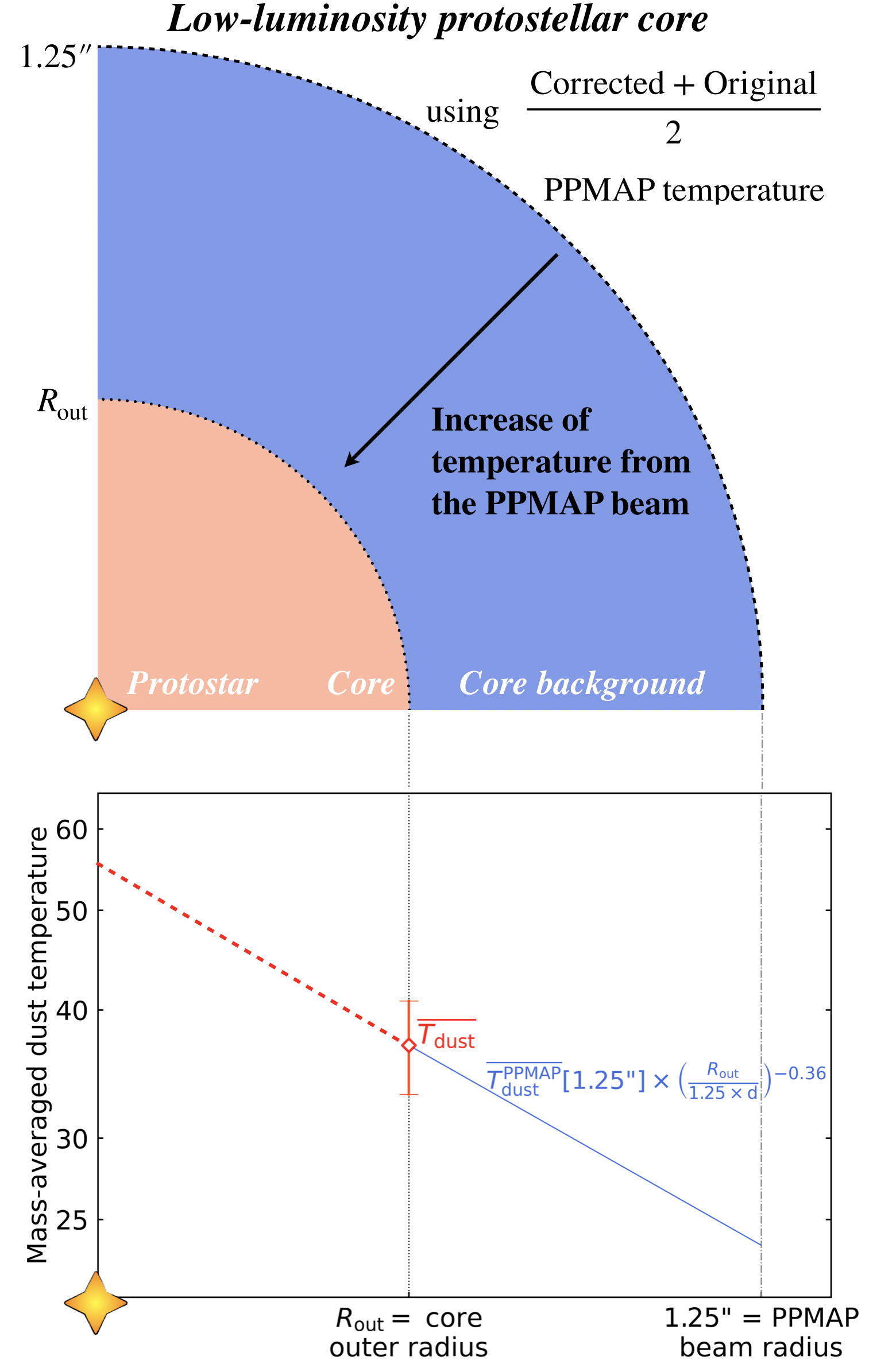}
    \includegraphics[width=0.33\linewidth]{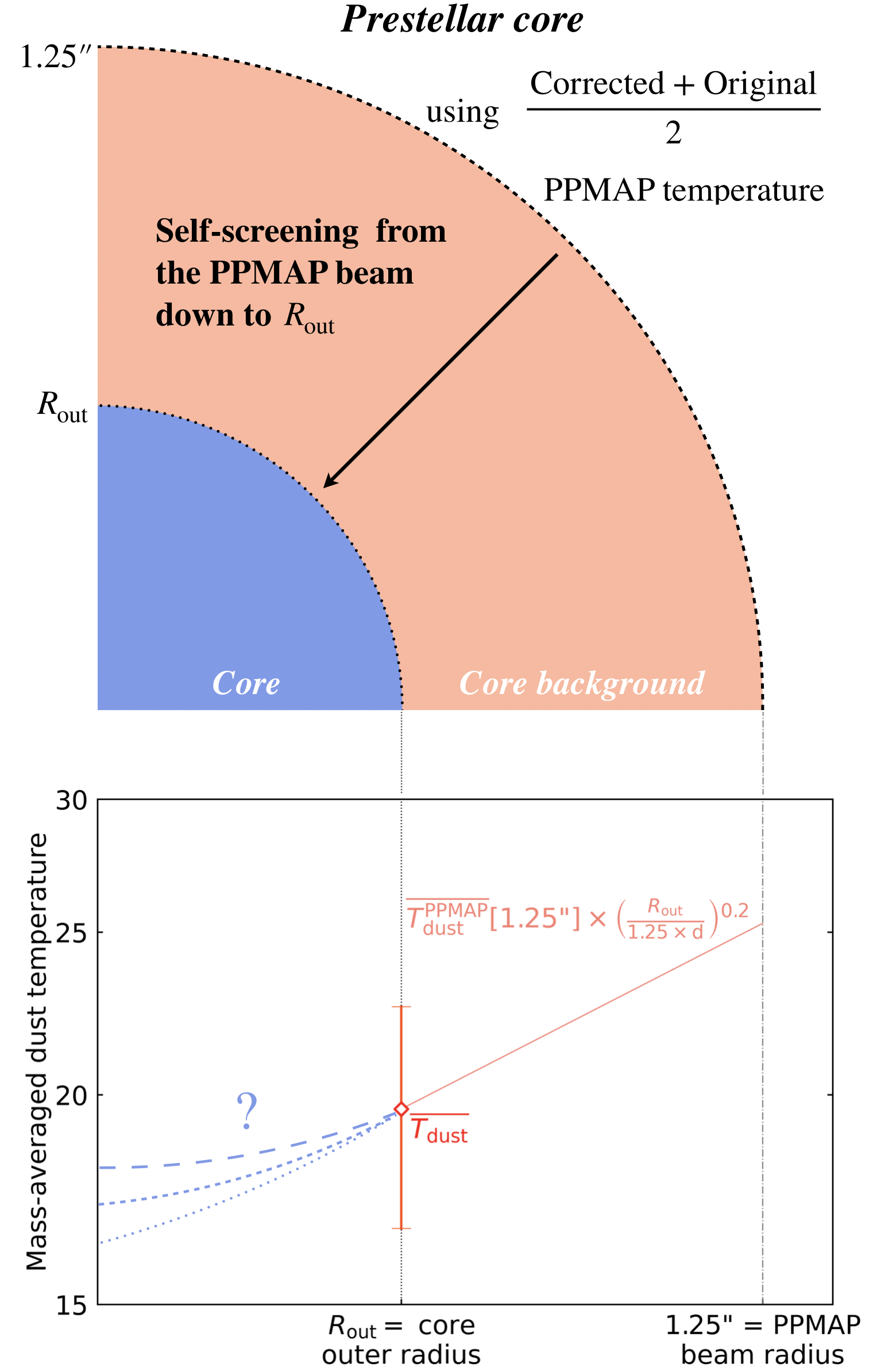}
    \vskip 0.1cm
    \caption{
    Methodology used to estimate the mass-averaged dust temperature of ALMA-IMF cores: luminous protostellar cores (\textit{left panels}), low-luminosity protostellar cores (\textit{central panels}), and prestellar cores (\textit{right panels}). Sketch (\textit{upper panels}) and radial profile (\textit{lower panels}) representing the mass-averaged dust temperature of a given core, as estimated by constraints obtained outside the core and, for the luminous protostars, within the core. The absolute values of the equations, the mean $\overline{T_{\rm dust}}$, and its uncertainty that are used for illustration are those computed for three cores of the W43-MM1 protocluster: core \#1 for luminous protostars,  core \#26 for low-luminosity protostars, and core \#6 for prestellar cores.
    In the left panels, for luminous protostellar cores, the mass-averaged dust temperature was computed with two approaches. The temperature was first measured in \PPMAP images at a radius of $1.25\arcsec$ and extrapolated within the \PPMAP beam to the outer radius of the core, $R_{\rm out}$ (blue area and blue equation). Its value at $R_{\rm out}$ was also computed from the internal heating of the protostar with a luminosity of $L_{\rm proto\star}$ (orange area and orange equation).
    In the central panel, for low-luminosity protostellar cores, the mass-averaged dust temperature measured in \PPMAP images at a radius of $1.25\arcsec$ is extrapolated within the \PPMAP beam to the outer radius of the core, $R_{\rm out}$ (blue area and blue equation).
    In the right panels, for prestellar cores, the mass-averaged dust temperature measured in \PPMAP images at a radius of $1.25\arcsec$ is extrapolated within the \PPMAP beam to the outer radius of the core, $R_{\rm out}$ (orange area and orange equation). The mathematical relationships describing the mass-averaged temperature expected for the inner parts of low-luminosity protostellar cores and prestellar cores are not used in the present study.}
    \label{fig:temperature three profiles}
\end{figure*}

\section{Mass-averaged dust temperature and luminosity of protostellar cores}
\label{s:Tproto}

Using the well-documented temperature model of protostellar envelopes (see Sect.~\ref{s:Tproto-eq}), we make two estimates for the mass-averaged temperature of single protostellar cores (see Sects.~\ref{s:Tproto-OutIn}--\ref{s:Tproto-InOut}). As illustrated by \cref{fig:temperature three profiles}, the first estimate uses the background-diluted core temperatures determined in Sect.~\ref{s:Tbck} and corresponds to an outside-in extrapolation. In addition and for the protostellar cores that are bright enough to be detected in the \PPMAP luminosity images, the second estimate is based on the luminosity peaks identified in Sect.~\ref{s:lumcat}, and is an inside-out extrapolation. The G012.80, W43-MM1, W43-MM2, and W43-MM3 protoclusters are used as case studies (see \cref{tab:computed table1}) and results for the entire protocluster sample of the ALMA-IMF Large Program are given in Appendix (see \cref{appendixtab:computed table}).

\subsection{Analytical description of dust temperature within protostellar cores}
\label{s:Tproto-eq}

When a protostellar envelope is centrally heated and is optically thin to far-infrared radiation, the temperature inside the envelope follows a power-law as a function of radius, with an index that depends on the dust emissivity index $\beta$ \citep{terebey1993, kauffmann2008, peretto2020}. When neglecting the background temperature, the absolute value of the temperature varies as a function of the luminosity of the protostar, $L_{\rm proto \star}$, once again with an index that depends on $\beta$:
\begin{eqnarray}
    \tdust(r, L_{\rm proto \star}) & = & T_0 \times \left(\frac{r}{r_0}\right)^{-q} \times \left(\frac{L_{\rm proto \star}}{L_0}\right)^{q/2}, \label{eq:theo-tdust}
\end{eqnarray}
with $q = 2/(4 + \beta)$ and the normalization constants, $T_0=25$~K, $r_0=6685$~au, and $L_0=520~\lsun$, taken from the model of \cite{terebey1993}. For a protostellar core described by a density profile of $\rho(r) \propto r^{-p}$ and a temperature law following \cref{eq:theo-tdust}, the mass-averaged temperature within its outer radius $R_{\rm out}$ is
\begin{eqnarray}
    \overline{\tdust(r, L_{\rm proto \star})}[R_{\rm out}]  & = & \frac{ \int_{r=0}^{r=R_{\rm out}} \tdust(r, L_{\rm proto \star}) \times 4\pi r^2 \rho(r) ~{\rm d}r } { \int_{r=0}^{r=R_{\rm out}} 4\pi r^2 \rho(r)~{\rm d}r} \nonumber\\ 
    & = & \frac{3-p}{3-p-q} \times \tdust(R_{\rm out}, L_{\rm proto \star}). 
\label{eq:theo-meantdust}
\end{eqnarray}

In the case of both low- and high-mass protostellar envelopes, the best suited power-law indices for their density and dust opacity laws are $p=2\pm 0.3$ (\citealt{motte2001, beuther2002, gieser2021}, in agreement with theoretical models by, e.g., \citealt{shu1977, gomez2021}) and $\beta=1.5\pm 0.3$ \citep{andre1993, rathborne2010, juvela2015, palau2021}, respectively. We note, however, that recent dust models and observations of low-mass protostellar disks suggest that flatter emissivity indices may be more appropriate in very dense cloud structures such as these disks, and potentially high-mass cores such as those characterized here \citep{ysard2019, galametz2019}. The emissivity index chosen above yields a power-law index for the temperature profile of $q\simeq 0.36$, which is close to the $q\approx 0.4$ index recently measured \citep{gieser2021} and adopted in initial studies \citep[e.g.,][]{motte2001, beuther2002}. We adapted the normalisation constants of \cref{eq:theo-tdust} to the characteristics of protostellar cores in ALMA-IMF protoclusters. We assumed that the spatial distribution of cores intensity is described by a Gaussian function\footnote{
    The protostellar cores extracted by \getsf have a size of $\sim$1.5 times the beam \cite[e.g.,][]{pouteau2022, nony2023, louvet2024}. A Gaussian therefore reasonably describes their intensity distribution, which corresponds to the convolution of a barely resolved power-law with the clean beam, which is a pure 2D Gaussian.}
and took their median deconvolved FWHM sizes \citep[$\sim$2100~au, ][]{motte2022} as a proxy for their outer radius. As for the typical protostellar luminosity, our only constraint at this stage is the median luminosity of protostellar groups: $\sim$$1100~\lsun$ \citep{dellova2024}. With these assumptions and typical outer radius and luminosity, Eqs.~\ref{eq:theo-tdust}--\ref{eq:theo-meantdust} become
\begin{eqnarray}
    \tdust(r, L_{\rm proto \star}) & \simeq & 43.5~{\rm K} \times \left(\frac{r}{\rm 2100~\text{au}}\right)^{-0.36} \times \left(\frac{L_{\rm proto \star}}{1100~\lsun}\right)^{0.18}
 \label{eq:tdust}
\end{eqnarray}
and 
\begin{eqnarray}
    \overline{\tdust(L_{\rm proto \star})}[R_{\rm out}] & \simeq & 
    68~{\rm  K} \times \left(\frac{R_{\rm out}}{2100~\text{au}}\right)^{-0.36} \times \left(\frac{L_{\rm proto \star}}{1100~\lsun}\right)^{0.18}.
\label{eq:tdust-fromL}
\end{eqnarray}

The relationship of \cref{eq:tdust-fromL} is found, with variations of $\sim$20\%, in grids of radiative transfer models describing protostars with a protostellar envelope and an outflow cavity \citep{robitaille2017}. It stands, but with greater dispersion, for protostellar envelope models with a disk.

\subsection{Outside-in estimate using \PPMAP dust temperatures}
\label{s:Tproto-OutIn}
A first estimate of the mass-averaged dust temperature of protostellar cores, $T_{\rm dust}^{\rm protostellar~core}$, can be obtained from the temperatures provided by \PPMAP (see \cref{fig:temperature three profiles}). We used \cref{eq:tdust-fromL}, to extrapolate the \PPMAP temperature averaged over a $1.25\arcsec\times d \in[2500; 7000]$~au radius to the core outer radius, taken to be equal to the deconvolved FWHM of the core,\footnote{
   FWHM$^{\rm dec}_{\rm core} =\sqrt{\theta_{\rm core}^2-\Theta_{\rm beam}^2} \times d$, where  $\theta_{\rm core}$ is listed in Tables~\ref{tab:measures table}--\ref{appendixtab:measures table evolved}, $\Theta_{\rm beam}$ can be found in published catalogs \citep{pouteau2022, nony2023, armante2024, louvet2024} 
   and $d$ in \cite{motte2022} (see their Table~1).
} $R_{\rm out}$:
\begin{eqnarray}
T_{\rm dust}^{\rm protostellar~core}  & \approx & \overline{\tdust(L_{\rm proto \star})}[R_{\rm out}] \nonumber
\end{eqnarray}
with 
\begin{eqnarray}
\overline{\tdust(L_{\rm proto \star})}[R_{\rm out}] & = & \overline{T_{\rm dust}^{\rm PPMAP}}[1.25\arcsec] 
\times \left(\frac{R_{\rm out}}{1.25\arcsec \times d}\right)^{-0.36}, \label{eq:tdust-fromT}
\end{eqnarray}
where $\overline{T_{\rm dust}^{\rm PPMAP}}[1.25\arcsec]$ is the \PPMAP dust temperature of cores diluted in their background, averaged over a radius of $1.25\arcsec$ (see Sect.~\ref{s:Tbck}), and $d$ is the distance of the ALMA-IMF protoclusters to the Sun, 5.5~kpc for the W43 regions, 2.4~kpc for G012.80, and 2.0-5.4~kpc for the others \citep[see Table~1 of][]{motte2022}. As explained in Sect.~\ref{s:Tbck} and \cref{appendixsect:ppmap}, protostars bright enough to be detected in the luminosity images have their dust temperature, $\overline{T_{\rm dust}^{\rm PPMAP}}[1.25\arcsec]$, measured in the \PPMAP image corrected for the opacity at $70\,\mu$m. For the other protostars, $\overline{T_{\rm dust}^{\rm PPMAP}}[1.25\arcsec]$ is taken to be the average of values measured in the opacity-corrected and original \PPMAP images (see Sect.~\ref{s:Tbck}). We then used \cref{eq:tdust-fromT} to compute a first estimate of the mean dust temperature of the ALMA-IMF protostellar cores. 

Tables~\ref{tab:computed table1}--\ref{appendixtab:computed table} characterize the protostellar cores of the G012.80, W43-MM1, W43-MM2, and W43-MM3 protoclusters and of the remaining ALMA-IMF protoclusters, respectively. They list their outer radius and their mass-averaged dust temperature computed, as shown in \cref{fig:temperature three profiles}, from \cref{eq:tdust-fromT}. The dust temperature of protostellar cores is observed to vary from $\sim$17~K to $\sim$127~K.

This extrapolation from the outside in, that is to say from $\frac{{\rm HPBW}_{\rm PPMAP}}{2}=1.25\arcsec$ ($2500-7000$~au) to $R_{\rm out}=0.35-0.95\arcsec$ ($1900$~au), should be correct for most protostellar cores but, in several cases, it corresponds to an upper limit. First, \cref{eq:tdust-fromT} is correct as far as the protostellar core remains optically thin to far-infrared radiation at the $R_{\rm out}$ radius. In the very few cases when this condition is not fulfilled, the extrapolated dust temperature is an upper limit. Then, for evolved protostellar cores such as bright hot cores, which, due to the propagation of the inside-out collapse, could have a slightly flatter density profile, $\rho(r)\propto r^{-1.5}$ instead of $\rho(r)\propto r^{-2}$ according to theoretical studies \citep[e.g.,][]{shu1977, gong2013} and some observations \citep{didelon2015, jeff2024}, \cref{eq:tdust-fromT} gives a dust temperature slightly, by $18\%$, overestimated. Moreover, when no protostellar heating is observed in the \PPMAP image as a hot spot associated with a luminosity peak (see Tables~\ref{tab:measures table}--\ref{appendixtab:measures table evolved}), the proposed equation also provides upper limit values. And finally when the background temperature is not negligible, the temperature profile would also be flatter and the protostellar core temperature at $R_{\rm out}$ would be overestimated as well. For a smaller dust emissivity index, for instance, $\beta=1$ \citep{galametz2019}, the dust temperature remains almost unchanged. In contrast, in the case of a core with a geometry far from being spherical and potentially subfragmenting in a few components, much larger variations are expected. For the purposes of the present statistical study, we will ignore this last case.

\subsection{Inside-out estimate using \PPMAP peak luminosities}
\label{s:Tproto-InOut}

\begin{figure}[htbp!]
    \centering
    \includegraphics[width=1\linewidth]{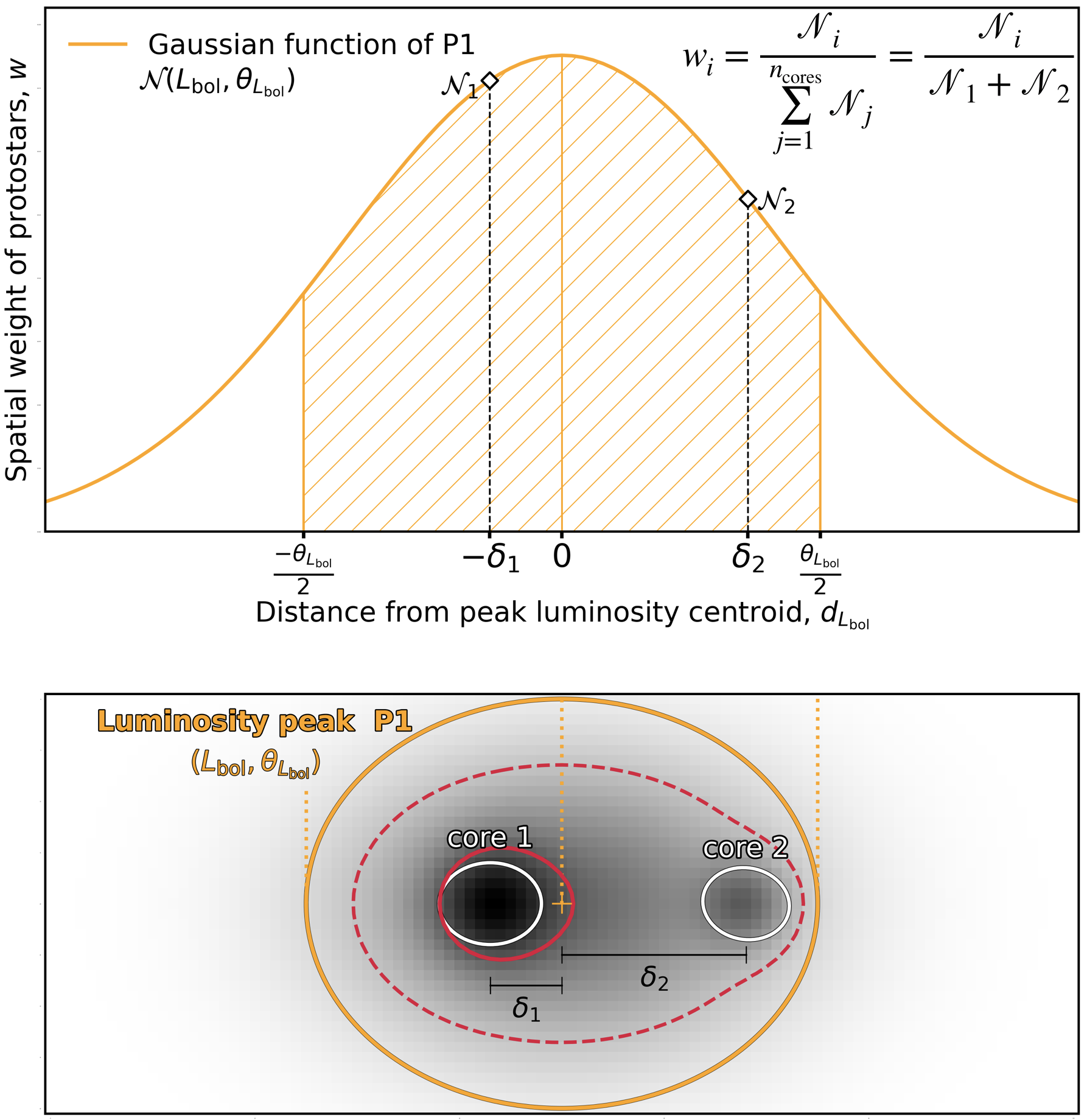}
    \vskip 0.2cm
    \caption{
    Gaussian distribution, $\mathcal{N}$(\lbol, RA, Dec, $\theta_{L_{\rm bol}}$) (orange curve in the \textit{upper panel}), which represents the luminosity peak P1 associated with two protostellar cores, and its intensity map (\textit{lower panel}). The spatial contribution of the protostellar cores, $w_1$ and $w_2$, to luminosity peak P1 is assumed to be proportional to the value of the Gaussian function at the cores' location, $\mathcal{N}_{1}$ and $\mathcal{N}_{2}$ at $\delta_1$ and $\delta_2$, respectively. The FWHM of the luminosity peak P1, $\theta_{L_{\rm bol}}$, is outlined by the dashed area in the \textit{upper panel} and the orange ellipse in the \textit{lower panel}. Red contours separate the luminosity peak into the intensity contribution of its two host cores.}
    \label{fig:spatial weight}
\end{figure}

As shown in \cref{eq:tdust-fromL}, a second estimate of the mass-averaged dust temperature of protostellar cores can be obtained using constraints on their protostellar luminosity (see scheme of \cref{fig:temperature three profiles}). It is clear from Figs.~\ref{fig:lum-temp peaks G012-W43} and \ref{appendixfig:lum and temp peaks} (see also Sect.~\ref{s:Tpeak}) that the luminosity peaks of Tables~\ref{tab:measures table}--\ref{appendixtab:measures table evolved} correspond to the luminosity imprints of protostellar groups, containing one to four protostars. Determining the contribution of each protostar to the luminosity of protostellar groups requires knowledge of their relative intensity at wavelengths ranging from near-infrared to millimeter. While ALMA provides a good estimate of the relative intensities of protostellar cores in the (sub)millimeter regime, \textit{Herschel} and \textit{Spitzer} observatories lack the angular resolution to do so \citep{dellova2024}. We developed a method, which uses the spatial distribution of luminosity peaks to distribute the luminosity of these groups to each of the individual protostellar cores that they host (see \cref{fig:spatial weight}). 

Tables~\ref{tab:measures table}--\ref{appendixtab:measures table evolved} show that about $30\%$ of the luminosity peaks have contributions from multiple protostellar cores. This ratio is highly variable throughout the ALMA-IMF protoclusters. It primarily depends on the distance of the protoclusters to the Sun; more cores are expected in a $2.5\arcsec$ area when they have typical sizes of $0.5\arcsec$, as in W43-MM1, instead of $0.9\arcsec$ sizes, as in G012.80 (compare Figs.~\ref{fig:lum-temp peaks G012-W43}a and \ref{fig:lum-temp peaks G012-W43}b). This ratio also depends on the number density of cores within protoclusters as observed when comparing, for instance, W43-MM1 and W43-MM2. While the most clustered, W43-MM1, has $3/4$ of its luminosity peaks associated with multiple protostellar cores, only $1/4$ are multiple in the least clustered of the two, W43-MM2. Moreover, some of the luminosity peaks hosting two or three cores display an elongated shape along the axis that connects them (e.g., P3 in \cref{fig:lum-temp peaks G012-W43}b, P1 of G010.62~I and P3 of G327.29~II in \cref{appendixfig:lum and temp peaks}), suggesting that these cores make a balanced and significant contribution to the luminosity peak. Therefore, the location of each core within their host luminosity peak could give a good estimate of their contribution to the luminosity. Making this assumption, we calculate below the luminosity of protostars listed in Tables~\ref{tab:computed table1}--\ref{appendixtab:computed table}.

As illustrated in \cref{fig:spatial weight}, the luminosity peaks from Sect.~\ref{s:lumcat} have been fitted by \getsf with Gaussian functions, $\mathcal{N}(\rm RA, Dec, \lbol, \theta_{L_{\rm bol}})$, whose parameters are their center coordinates, their total luminosities, $\lbol(r<\theta_{L_{\rm bol}})$, and their FWHM sizes, $\theta_{L_{\rm bol}}$. For each protostellar core $i$ contained in one luminosity peak, we then defined its contribution, $\mathcal{N}_{i}$, as the value of the luminosity peak's Gaussian function at the location of the core, $\delta_i$, corresponding to $d_{L_{\rm bol}}$ for core $i$.
The contribution of protostellar core $i$ among $n_{\rm cores}$ to a given luminosity peak is then considered to be
\begin{equation}
    w_{\rm spatial,~i}= \frac{\mathcal{N}_{i}}{\summation{n_{\rm cores}}{j=1} \mathcal{N}_{j}}.
        \label{eq:spatial contribution}
\end{equation}
The Gaussian parameters of each luminosity peak and the contribution to their luminosity of the host protostellar cores are tabulated in Tables~\ref{tab:measures table}--\ref{appendixtab:measures table evolved}.

Assuming the contribution computed in \cref{eq:spatial contribution}, the luminosity of protostar $i$ is
\begin{eqnarray}
    L_{\rm proto\star,~i}
    = w_{\rm spatial,~i}  
    \times [\lbol(r<\theta_{L_{\rm bol}}) - L_{\rm bck} ],
    \label{eq:spatial luminosity}
\end{eqnarray}
where $L_{\rm bck}$ represents the background luminosity of the $0.5\arcsec-0.9\arcsec$ cores within a given luminosity peak, typically $2.5\arcsec-5\arcsec$ in size. Neglecting $L_{\rm bck}$ is probably acceptable for luminous protostars in Young protoclusters. It however should lead to overestimating the luminosity of low-luminosity protostars, especially when they are located in the luminous background of Evolved protoclusters \citep[e.g.,][]{evans2009,kryukova2012} or close to the highest-luminosity protostars. As we have no information whatsoever on the value of this background (at scale between $2.5\arcsec$ and the core size), we are forced to set it here to zero. Beyond this source of uncertainty, the main bias of this method is linked to the chance alignment of cores fortuitously located along the line of sight of a luminosity peak. 

Finally, for each luminous protostellar core, thus associated with a luminosity peak, we plugged its protostellar luminosity computed from \cref{eq:spatial luminosity} into \cref{eq:tdust-fromL}, which gives a second estimate of its mass-averaged dust temperature. Tables~\ref{tab:computed table1}--\ref{appendixtab:computed table} list, for 151 (i.e., $55\%$) ALMA-IMF protostars, an estimate of their protostellar luminosity and associated mass-averaged dust temperature. This second estimate of the dust temperature provides values that range from $\sim$30~K to $\sim$143~K and are generally 20~K higher than the outside-in measurements of Sect.~\ref{s:Tproto-OutIn}.

For the eight cores associated with the brightest ALMA-IMF hot cores \citep{bonfand2024}, three different temperatures are calculated from the three luminosity estimates presented in Sect.~\ref{s:extremehotcore} (see \cref{tab:luminosity peaks from radial profile}).  For the eight highest-luminosity protostars located within these extreme hot cores,  given their sizes $1200-3200$~au, \cref{eq:tdust-fromL} gives a mass-averaged dust temperature of $80-140$~K, with an upper value of $100-170$~K (see \cref{tab:luminosity peaks from radial profile}). We compare these different derived temperatures in Sects.~\ref{s:finalTdust}
and \ref{s:consistency}.

\section{Mass-averaged dust temperature of prestellar cores}
\label{s:Tpre}

There is as yet no well-established temperature model to characterize the self-shielding of starless structures in the interstellar medium. Therefore, we examine the temperature profiles of prestellar cores and starless clumps from the literature (see Sect.~\ref{s:Tpre-litterature}), and we determine, in Sect.~\ref{s:Tpre estimate}, the expected temperature profile for the immediate surrounding of ALMA-IMF prestellar cores. The prestellar cores are indeed embedded in dense clumps, whose dust temperature measured toward the core are diluted in spheres of $5000-14\,000$~au radii (i.e., $2.5\arcsec$ at the protocluster distance) as defined in Sect.~\ref{s:Tbck}. The mass-averaged dust temperature of prestellar cores is estimated by extrapolating their clump-diluted temperature, by a factor of 1 to 4 in radii, down to the outer radius of the prestellar cores, ranging from 700 to 9900~au (median at 2900~au). Temperature extrapolation for prestellar cores is illustrated in \cref{fig:temperature three profiles} and performed in Sect.~\ref{s:Tpre estimate}.

\subsection{Current knowledge of the temperature profiles of prestellar cores}
\label{s:Tpre-litterature}

The density profile of prestellar cores is often assimilated to that of a Bonnor-Ebert sphere \citep{bonnor1956, alves2001b}. However, the existence of an outer edge is still much debated (see, e.g., references in Sect.~\ref{s:Tpre estimate}). On the other hand, it is well established that the outer part of their density profile is similar to that of protostars, that is close to $\rho(r)\propto r^{-2}$, and that their inner part, often reported at radii $r<R_{\rm flat}$, shows a definite flattening \citep{wardthompson1999, launhardt2013, roy2014}. 

In absence of internal heating from protostars, the gas in prestellar cores of typical Galactic disk environments is heated by the local ISRF, cosmic rays, and shocks, with the ISRF dominating the heating of isolated prestellar cores \citep{evans2001}. The dust temperature profile of prestellar cores, therefore controlled by their self-shielding against the ISRF, decreases with increasing density and thus decreasing radius toward their center. As already suggested before the \textit{Herschel} era \citep[e.g.][]{zucconi2001,evans2001,pagani2003, crapsi2007}, the temperature distribution of the Taurus prestellar cores and isolated Bok globules is made up of a central plateau of $\rho_{\rm C}$ density and cold, $\sim$7--8~K, temperature and a trend of increasing temperature toward the outside \citep{launhardt2013, marsh2014, roy2014}. Many efforts have been made in radiative transfer models to describe in detail the temperature profiles of a handful of iconic prestellar cores, notably B68 and L1544 \citep{crapsi2007, nielbock2012, juarez2017} or similarly isolated cores \citep{malinen2011, marsh2014}. Several parametrizations of temperature profiles have been proposed to interpret observations of samples of prestellar cores \citep[][see their Eq.~(8), Eq.~(8), Eq.~(3), and Eq.~(8), respectively]{kauffmann2008, launhardt2013, marsh2014, koumpia2020}. The parameters of these empirical temperature distributions are currently only constrained by observation, and could only be linked to a simple isothermal model such as Bonnor-Ebert spheres \citep{bonnor1956}. Radiative transfer models of observed and simulated cores, with more realistic density structure than that of Bonnor-Ebert spheres are therefore necessary to make major improvements.

Analytical solutions are therefore not straightforward and the outcome of Monte Carlo radiative transfer models strongly depends on the cooling functions and the  structure assumed for the synthetic prestellar cores  \citep[e.g.,][]{goncalves2004}. For instance, the inner temperature of prestellar cores crucially depends on its inner structure, such as its mean density and its density distribution, including its subfragmentation and deviation from spherical symmetry \citep{wilcock2012}. In addition, the environment of prestellar cores, through its ISRF and dust properties, obviously has a major impact on their temperature at their outer radius, but also on their mass-averaged temperature, since most of a core's mass is found in its outer parts \citep{goncalves2004}. Although the physical scales of the outer part of nearby prestellar cores are similar to those of the immediate surrounding of ALMA-IMF cores, that is $\sim$$1000-10\,000$~au, ALMA-IMF environments are probably more filamentary, denser, and subject to a higher ISRF.

Despite these many caveats, we give a first-order estimate of the self-shielding effect of prestellar cores. This is, most probably, better than just taking a single fixed temperature, the background-diluted temperature of cores, or even decreasing it by a few degrees like was done in the pilot and initial ALMA-IMF studies \citep{motte2018b, pouteau2022, armante2024}.

\subsection{Outside-in estimate for prestellar cores using \PPMAP dust temperatures}
\label{s:Tpre estimate}

The common and most robust part of the structure of prestellar cores remains the outer part of its density profile, which follows a $\rho(r)\propto r^{-2}$ power-law \citep[see, however,][]{nielbock2012}. A similar density power-law is observed for the outer part of cloud filaments, which displays a Plummer-like profile:
\begin{equation} \label{eq:plummer}
    \rho_{\rm p}(r) = \frac{\rho_{\rm C}}{[1+(r/R_{\rm flat})^2]^{p/2}}
\end{equation}
with $p$ often set to $2$ (\citealt{palmeirim2013, stutz2018, arzoumanian2019}, but also see \citealt{stutzGould2016, peretto2023}). At large radii, that is for $r \gg R_{\rm flat}$ and $r$ between $10\,000$ to $80\,000$~au, this density profile is reminiscent of the $\rho(r)\propto r^{-2}$ structure observed for the outer parts of prestellar cores \citep[e.g.,][]{wardthompson1999, launhardt2013}, but at radii ten times greater than these. The temperature distribution of high-density filaments has also been observed and predicted to decrease as one approaches their crest \cite[e.g.,][]{hill2012, anathpindika2015}. Besides, on similar spatial scales as the Plummer-like filaments above, \cite{wilcock2011} modeled a handful of infrared-darks clouds (IRDCs), which are cold elongated clumps forming intermediate- to high-mass stars, and found a decreasing temperature profile with decreasing radius. 

According to published radiative transfer models, those of \cite{wilcock2012} are the most adequate to model the gas surrounding ALMA-IMF cores. They  state that an exponential relationship with distance is a good description of the dust temperature profile produced by self-shielding effects. Given the small range of radii investigated here, from $\sim$2900~au (the core size) to $5000-14\,000$~au (the $2.5\arcsec$ \PPMAP beam), this exponential law is, over this range, approximated by a simple power-law, $\tdust(r)\propto r^{-q}$. We measured the power-law indices $q$ for the temperature profiles presented in the figures provided by many authors for the observed or modeled prestellar cores or IRDCs, focusing on the radius range where their density profile follows a power-law close or equal to $\rho(r)\propto r^{-2}$.

About twenty observed and/or modeled prestellar cores, which are low-mass and located nearby, display temperature profiles in their outer parts, before reaching their background temperature, of $\tdust(r) \propto r^{-q}$ with $-q=0.1-0.32$ with a median value of $-q\simeq0.2$ \citep{goncalves2004, crapsi2007, nielbock2012, launhardt2013, marsh2014, juarez2017}. IRDCs have been modeled assuming a non spherical shape and four times stronger ISRF than what is generally assumed for low-mass prestellar cores \citep{wilcock2011, wilcock2012}. Surprisingly, their median temperature profiles are very similar, with $-q=0.2$, with $\Delta q \simeq0.05$ variations when considering the distribution in the mid-plane of the elongated IRDC or perpendicular to it. Moreover, \cite{jorgensen2006} made radiative transfer modeling and showed that the temperature profile of a prestellar core should quantitatively vary with its surrounding ISRF. Most ALMA-IMF protoclusters should be under the influence of a strong ISRF \citep{motte2022}, stronger than that required to model the IRDCs of \cite{wilcock2012}. The lack of dedicated radiative transfer modeling for these extreme protoclusters nevertheless prevents us from predicting the effect of a strong ISRF on the index studied here. In the following, we therefore assumed the temperature distribution of the background of ALMA-IMF prestellar cores to be following a $\tdust(r)\propto r^{0.2}$ power-law in the range of radii considered for the extrapolation (see \cref{fig:temperature three profiles}).

When the core FWHM is smaller than the \PPMAP beam, we  extrapolated the mean dust temperature of prestellar cores from the temperature measured in the combined \PPMAP images, averaged over a radius of $1.25\arcsec$, $\overline{T_{\rm dust}^{\rm PPMAP}}[1.25\arcsec]$. Like protostellar cores, the core outer radius, $R_{\rm out}$, is taken to be equal to the deconvolved FWHM of the core (see Sect.~\ref{s:Tproto-OutIn}). We also assume that the core background follows the same density profile as the outer parts of prestellar cores, IRDCs or filaments: $\rho(r) \propto r^{-2}$. Given the temperature law defined above, the mass-averaged temperature within its outer radius $R_{\rm out}$ is
\begin{eqnarray}
    T_{\rm dust}^{\rm prestellar~core}  & \approx & \overline{\tdust({\rm prestellar~core})}[R_{\rm out}]  \nonumber
\end{eqnarray}
with 
\begin{eqnarray}
    \overline{\tdust({\rm prestellar~core})}[R_{\rm out}] & = & \overline{T_{\rm dust}^{\rm PPMAP}}[1.25\arcsec] 
    \times \left(\frac{R_{\rm out}}{1.25\arcsec \times d}\right)^{0.2}.
\label{eq:tdustPrestellar}
\end{eqnarray}

We used \cref{eq:tdustPrestellar} to estimate the mean dust temperature of the ALMA-IMF prestellar cores or directly measured it in the \PPMAP images when the core FWHM is larger than the \PPMAP beam. \cref{appendixtab:prestellar table} lists the prestellar cores\footnote{
    \cref{appendixtab:prestellar table} lists the ALMA-IMF prestellar cores more massive than $6.5~\msun$. The complete catalog of ALMA-IMF prestellar cores can be found at CDS.}
of the ALMA-IMF protoclusters, along with their estimated mass-averaged dust temperatures, which vary from $\sim$17~K to $\sim$31~K (5th and 95th percentiles). The highest temperatures are found in Evolved protoclusters, in particular in the PDR of \hii regions and under the influence of luminous protostars, in agreement with measurements done in the Rosette molecular cloud \citep{motte2010, boegner2022} and in clustered environments \citep{sanchezmonge2013}. With an uncertainty of $\pm0.1$ for the power law index of their dust temperature profile, the resulting mass-averaged temperature of prestellar cores could vary by $\pm1$~K in the closest ALMA-IMF protoclusters, and by $\pm2$~K in the furthest protoclusters.

\section{Discussion}
\label{s:discussion}

In Sects.~\ref{s:Tproto}--\ref{s:Tpre}, we estimated the mass-averaged dust temperature, one of the two essential parameters necessary to determine the mass of the 266 protostellar cores (1--2 independent estimates per core) and 617 prestellar cores (1 estimate per core) found in the ALMA-IMF protoclusters. Sections~\ref{s:finalTdust}--\ref{s:mass} define the adopted dust temperature of all cores and compute their mass. \cref{s:consistency} then compares these core temperatures with previously estimated values. Core masses and luminosities, two of the most fundamental characteristics of protostars, are finally used in a Mass-versus-Luminosity diagram that traces the evolution of protostars in protoclusters (see Sect.~\ref{s:M-L}).

\begin{figure*}[htbp!]
    \centering
    \includegraphics[width=1\linewidth]{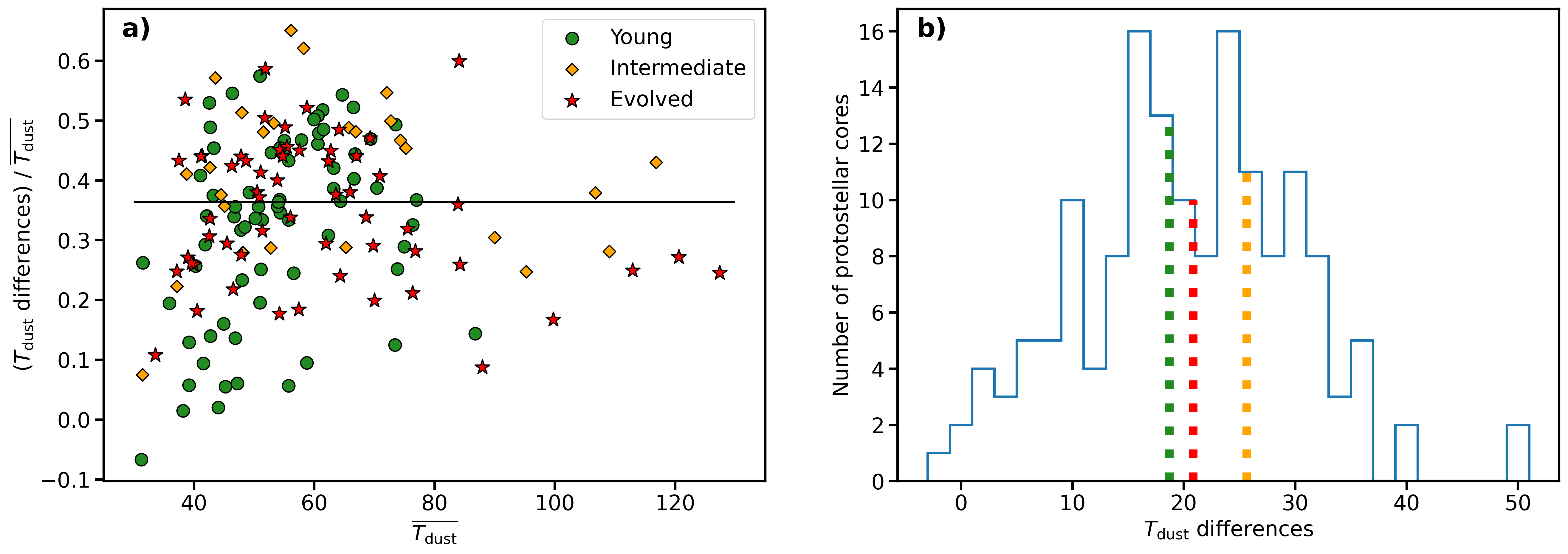}
       \vskip -0.cm
    \caption{Comparison of the two mass-averaged temperature estimates made for the 151 luminous protostellar cores of ALMA-IMF bright enough to be detected in the luminosity maps. \textit{(a)} Fractional difference between estimates from the protostellar core luminosity (Col.~5 of Tables~\ref{tab:computed table1}--\ref{appendixtab:computed table}) and the background-diluted temperature of cores (Col.~6 of Tables~\ref{tab:computed table1}--\ref{appendixtab:computed table}). Protostellar cores in Young, Intermediate, and Evolved protoclusters are indicated by green circles, orange diamonds, and red stars, respectively. The black line shows that the median fractional difference between these two estimates is $\sim$36\%. 
    \textit{(b)} Distribution of the differences between the two temperature estimates of Cols.~5--6 in Tables~\ref{tab:computed table1}--\ref{appendixtab:computed table} (blue histogram). Colored dashed lines locate the median of the differences: $\sim$19~K in Young and Evolved protoclusters (green and red, respectively) and $\sim$26~K in Intermediate protoclusters (orange). 
    }
    \label{fig:tdust variation}
\end{figure*}

\subsection{Adopted core temperatures}
\label{s:finalTdust}

For each ALMA-IMF core, we generally provide a single estimate of the mass-averaged dust temperature, which is extrapolated from the images of the background-diluted core temperatures (see Sects.~\ref{s:Tbck}, \ref{s:Tproto-OutIn}, and \ref{s:Tpre estimate}, and see \cref{fig:temperature three profiles}). In addition, for the 151 luminous protostellar cores that are bright enough to be detected in the \PPMAP luminosity images, we provide a second temperature estimate, derived from their protostellar luminosity (see Sect.~\ref{s:Tproto-InOut} and \cref{fig:temperature three profiles}).

Figure~\ref{fig:tdust variation} compares these two dust temperature estimates and shows that they are correlated. More precisely, their fractional difference is on average $36\%$, with a $1\sigma$ dispersion of about $15\%$ (see \cref{fig:tdust variation}a). Equivalently, the dust temperature estimated from the protostellar luminosity is on average $\sim$19~K higher than the dust temperature estimated from the extrapolation of the background-diluted core temperature (see \cref{fig:tdust variation}b). Given the assumptions necessary to make these temperature estimates (see Sects.~\ref{s:Tproto-OutIn}--\ref{s:Tproto-InOut}), there is no reason to favor one estimate over the other. We therefore simply averaged them and list their mean value in Tables~\ref{tab:computed table1}--\ref{appendixtab:computed table}, with an uncertainty that covers these two dust temperature estimates. 

Here we investigate the origin of the difference found for the two dust temperature estimates obtained for the luminous protostellar cores of ALMA-IMF. Differences range from $-2$~K to $+50$~K, with a median around $\sim$19~K and the fractional differences range from $-7\%$ to $+76\%$ with a median of $+36\%$ (see \cref{fig:tdust variation}). First, we could reconcile these estimates with a temperature versus radius relation, which has a steeper power-law index, $q>0.36$, than that assumed in \cref{eq:theo-tdust}. As a matter of fact, \cite{jeff2024} measured somewhat steeper temperature indices, $q\in[0.1;1]\simeq 0.45\pm 0.3$, for ten hot cores of the Galactic center that are partly optically thick. \cite{menshchikov2016}, which carried out radiative transfer modeling within protostellar envelopes, also revealed a similarly steeper temperature profile for their optically thick inner part: $\tdust(r)\propto r^{-0.88}$. This could thus apply for the eight cores associated with the brightest hot cores of the ALMA-IMF sample and a few other optically thick protostellar cores. For the other protostellar cores, a steeper temperature gradient would correspond to a higher index for the dust emissivity law. However, the opposite is now observed and predicted in hot, dense structures such as low-mass protostellar disks \citep{ysard2019, galametz2019}, somewhat reminiscent of the ALMA-IMF high-mass, small-scale ($>$$10~\msun$, $\sim$2400~au) protostellar cores. If the theoretical power-law index of \cref{eq:theo-meantdust} is not valid, the $q\approx 0.4$  values observed on the line-of-sight are probably underestimated and correspond to indices closer to $q = 0.5-0.6$ \citep{estalella2024}.

These differences may also arise from the assumptions used to compute these two separate estimates. On the one hand, the dust temperature extrapolated from outside, that is from the \PPMAP dust temperature image, could be underestimated. One reason could be that the \PPMAP mass-averaged temperature represents a line-of-sight temperature biased by the low temperature of the protocluster background and foreground. This marginally applies to Evolved protoclusters. The background-diluted temperature of luminous protostars could also be underestimated by the correction, itself underestimated, of the opacity of the emission at $70\,\mu$m as applied to the \PPMAP results \citep{dellova2024}. On the other hand, the dust temperature extrapolated from inside, that is from the protostellar luminosity, could alternatively be overestimated. This happens when the inner part of the protostellar core becomes optically thick. Moreover, the fact that we neglect the background luminosity of the cores in $2.5\arcsec$-beams in \cref{eq:spatial luminosity} explains why the fractional difference between these two temperature estimates is greater for protostellar cores in the Intermediate and Evolved protoclusters than for those in the Young protoclusters. Internal geometry and subfragmentation of cores could be another good reason for this disagreement. For example, the gas distribution is not correctly described by a single $\rho(r)\propto r^{-2}$ power-law when the inner part of the protostellar envelope free-falls, with a flatter density profile, or when subfragmentation distributes the gas over several sites, leading to a flatter mean density profile. For a $\rho(r)\propto r^{-1}$ law, \cref{eq:tdust-fromL} would give a temperature up to 22\% lower: $\sim$20~K and $\sim$10~K differences for $\tdust=100$~K and $\tdust=50$~K, respectively. The combination of high-resolution studies of  both molecular lines and infrared emission will be necessary to improve current estimates of the mass-averaged temperature of cores.

\begin{figure}[htbp!]
    \centering
    \includegraphics[width=1\linewidth]{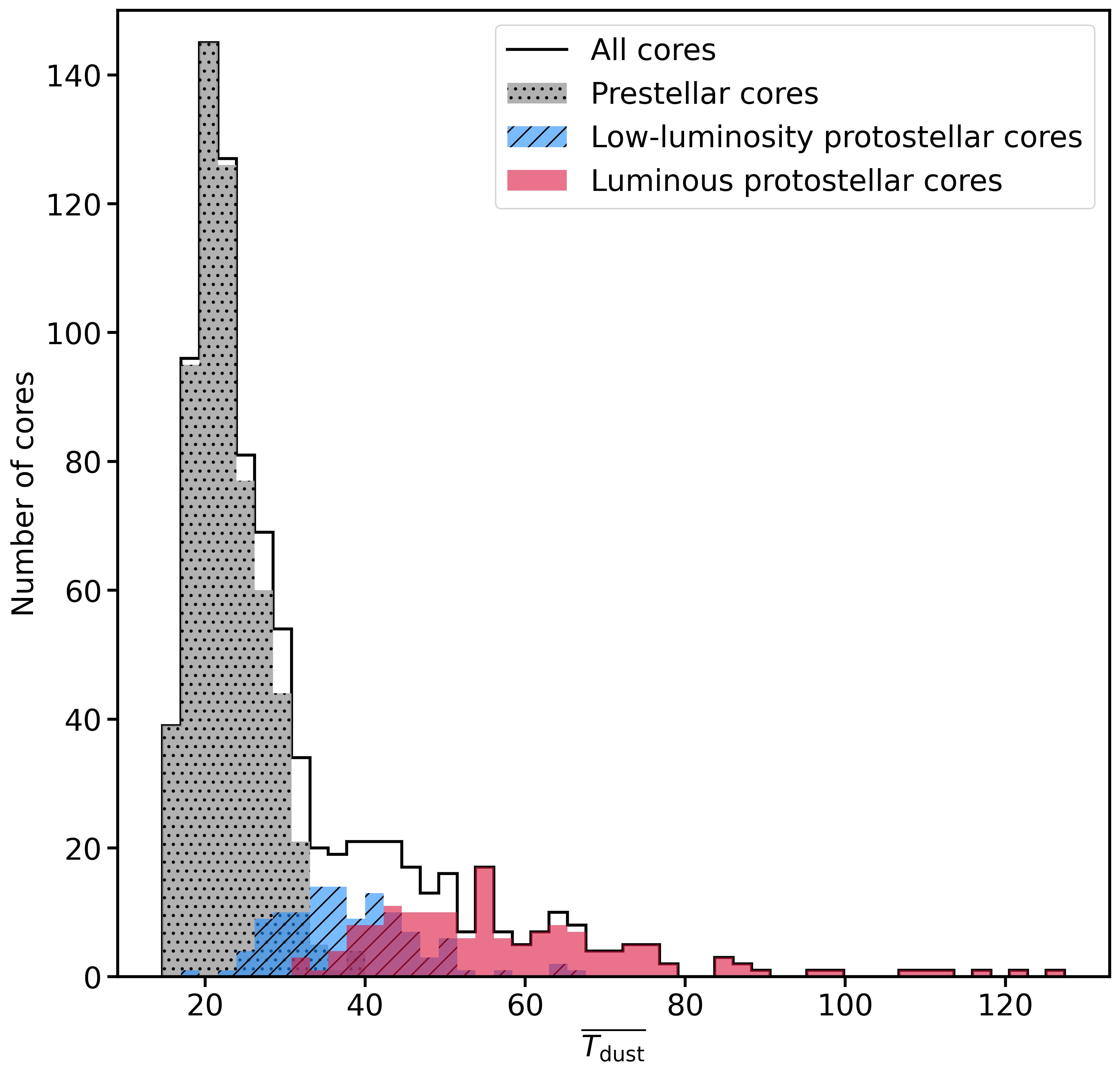}
    \vskip -0.cm
    \caption{Distribution of the mass-averaged dust temperature of prestellar cores (gray histogram), low-luminosity protostellar cores (blue histogram), and luminous protostellar cores (red histogram), as defined in Sect. ~\ref{s:corecat}. Despite the different assumptions we used to compute the dust temperatures of these three groups of cores (see Sects.~\ref{s:Tproto}--\ref{s:Tpre} and \cref{fig:temperature three profiles}), there is a clear continuity between their distributions, highlighted by the cumulated histogram of all cores (black outline).}
    \label{fig:tdust all cores}
\end{figure}

Figure~\ref{fig:tdust all cores} displays dust temperature histograms for the three groups of cores, which we treated differently according to their prestellar or protostellar nature, and according to their protostellar luminosity (see \cref{fig:temperature three profiles}). Their temperatures are complementary and have their 5th and 95th percentiles that range from 17~K to 31~K, 26~K to 51~K, and 38~K to 92~K for prestellar cores, low-luminosity and luminous protostellar cores, respectively. Interestingly, prestellar cores located in luminosity peaks and therefore in the immediate vicinity of luminous protostars have a higher dust temperature than cores located further away from them: $\sim$24~K versus $\sim$22~K median values (see \cref{appendixfig:tdust prestellar cores}).

\begin{figure}[htbp!]
    \centering
    \includegraphics[width=1\linewidth]{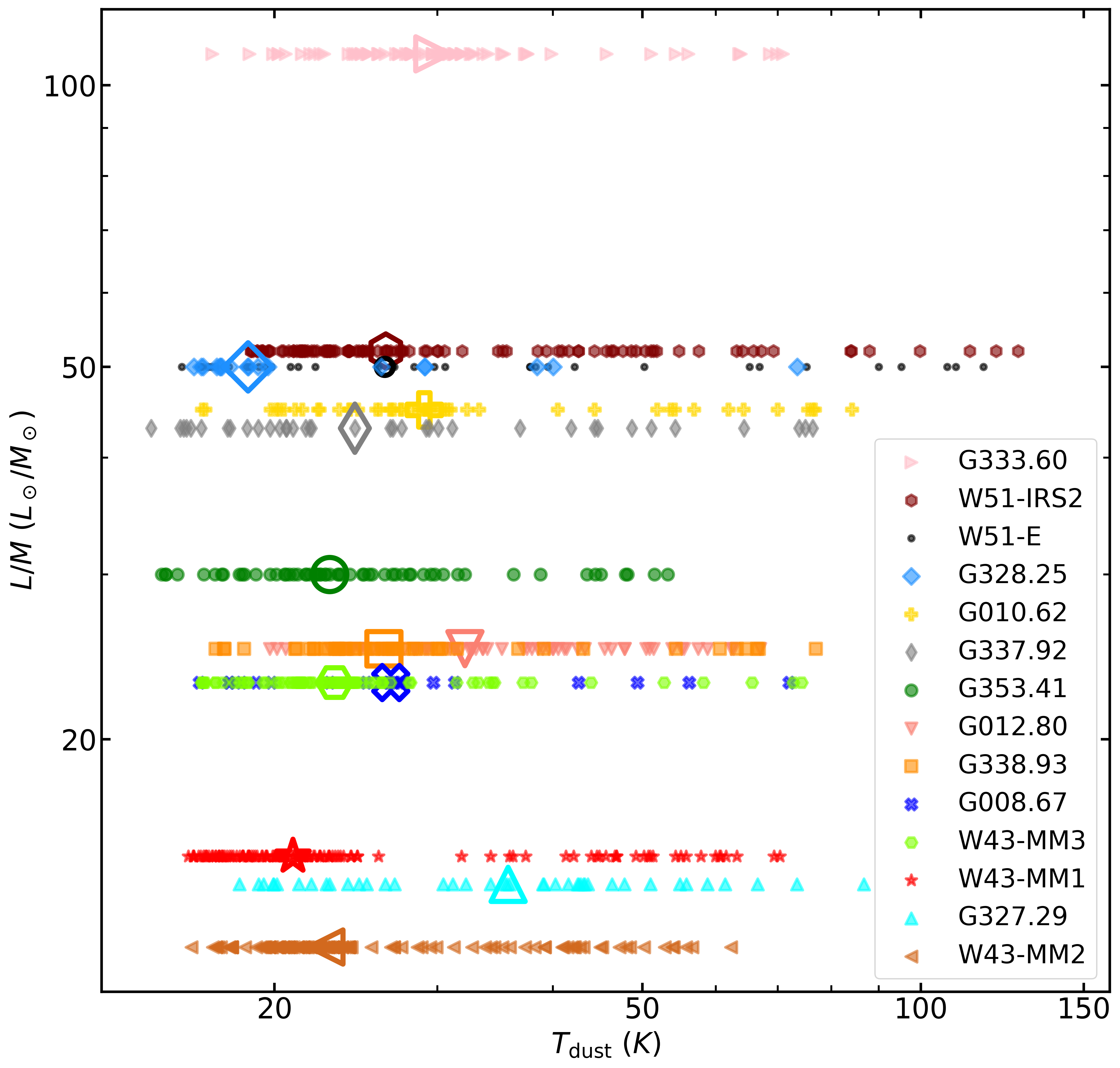}
    \vskip -0.cm
    \caption{Distribution of the mass-averaged temperature of ALMA-IMF cores (Col.~7 of Tables~\ref{tab:computed table1}--\ref{appendixtab:computed table}) versus  luminosity-to-mass ($L/M$) ratios of their protocluster \citep{dellova2024}. Neither the median temperature (large markers) measured for the cores of each protocluster nor their minimum or maximum values seem to correlate with the protocluster $L/M$ ratio.}
    \label{fig:tdust all regions}
\end{figure}

Figure~\ref{fig:tdust all regions} shows, for each ALMA-IMF protocluster, the distribution of the mass-averaged temperature of its cores, whether prestellar or protostellar. The core temperatures have been computed above and in Sect.~\ref{s:Tpre estimate} and the luminosity-to-mass ratio, $L/M$, of the ALMA-IMF protoclusters is taken from \cite{dellova2024}. As they stated, the $L/M$ ratio of ALMA-IMF protoclusters correlates with the evolutionary stage determined in \cite{motte2022} but also exhibits a noticeable variability within the groups of Young, Intermediate, and Evolved  protoclusters. Assuming that the $L/M$ ratio is a reasonable indicator for the evolutionary state of ALMA-IMF protoclusters, the vertical axis thus could roughly represent the evolutionary path of extreme protoclusters in our Milky Way. Figure~\ref{fig:tdust all regions} displays no obvious correlation between the $L/M$ ratio of ALMA-IMF protoclusters and the median, minimum, and maximum values of its core temperatures. This is surprising, as we would expect core temperatures to increase in more evolved protoclusters, and therefore potentially with $L/M$ \citep{molinari2016, giannetti2017}. If we instead use the ALMA-IMF classification that separates Young, Intermediate, and Evolved protoclusters, the median core temperature increases between these 3 groups: from $\sim$22~K (with G327.29 as the obvious outlier) to $\sim$26~K and finally $\sim$29~K, with continuity between protoclusters in these three groups. We thus argue that the amount of dense gas impacted by \hii regions, which is the criterion used by \cite{motte2022}, is a better evolutionary tracer of protoclusters than the $L/M$ ratio that can only be reliably used for individual protostellar cores. G328.25 and G327.29 host, on average, the coldest and hottest cores of the ALMA-IMF protoclusters, consistent with the relative lack of MF emission in G328.25 \citep{bonfand2024} and the impact on G327.29 of an external \hii region that was not previously taken into account in our classification \citep{motte2022, galvanmadrid2024}.

\subsection{Resulting core mass}
\label{s:mass}
Since the thermal dust emission of most ALMA-IMF cores is largely optically thin at 1.3~mm, we could have computed the mass of ALMA-IMF cores using the following equation: 
\begin{eqnarray}
    M_{\rm \tau\ll 1} \: & = \frac{S^{\rm int}_{\rm 1.3\,mm}\; d^2}{ \kappa_{\rm 1.3\,mm}\; B_{\rm 1.3\,mm}(T_{\rm dust})},
    \label{eq:optically thin mass}
\end{eqnarray}
where the dust $+$ gas mass opacity is set to $\kappa_{\rm 1.3mm} = \rm 0.01~cm^2\,g^{-1}$,  $S^{\rm int}_{1.3{\rm mm}}$ is the integrated flux of cores \citep[see core catalogs in][]{pouteau2022, nony2023, armante2024, louvet2024}, and $B_{1.3{\rm mm}}(\tdust)$ is the Planck function at the core temperature listed in Col.~7 of Tables~\ref{tab:computed table1}--\ref{appendixtab:computed table}.

However, several of the densest cores of ALMA-IMF are known to be partly optically thick \citep{motte2018b, pouteau2022, jeff2024}. We therefore used the equation proposed by \cite{pouteau2022} \citep[see also][]{motte2018b} to compute the mass of all the 883 ALMA-IMF cores, assuming an optical thickness that at most is close to one: 
\begin{equation} \label{eq:optically thick mass}
M_{\rm \tau\gtrsim 1} \: = -\, \frac{\Omega_{\rm beam} \;d^2} {\kappa_{1.3{\rm mm}}}\, \frac{S^{\rm int}_{1.3{\rm mm}}} {S^{\rm peak}_{1.3{\rm mm}}} \,
	\ln\left(1\,-\,\frac{S^{\rm peak}_{1.3{\rm mm}}}{\Omega_{\rm beam}\;B_{1.3{\rm mm}}(\tdust)}\right),
\end{equation}
where $\Omega_{\rm beam}$ is the solid angle of the 1.3~mm beam \citep[see Table~2 of][]{motte2022} and $S^{\rm peak}_{1.3{\rm mm}}$ is the peak flux of cores. The core optical thickness, measured for their peak flux, is applied to their total flux. This leads to an overestimate of the mass of the few cores whose emission is optically thick, but this overestimate is small because the size of these cores is close to the beam.

Integrated and peak fluxes of cores are corrected for line contamination over the entire ALMA-IMF sample \citep[see][]{pouteau2022, nony2023, louvet2024}. In contrast, the core continuum fluxes were not homogeneously corrected for contamination by free-free emission. As detailed in Sect.~\ref{s:corecat}, the complete core sample of the four study-case regions, W43-MM1, W43-MM2, W43-MM3, and G012.80, and the hot core candidates found in the other ALMA-IMF protoclusters had their core fluxes corrected when necessary \citep{pouteau2022, nony2023, armante2024, bonfand2024}. The remaining protostellar cores potentially contaminated by free-free emission have fluxes that correspond to upper limits, and consequently so do their masses.

For the vast majority, $\sim$$96\%$, of the ALMA-IMF cores, their 1.3~mm emission remains optically thin and \cref{eq:optically thick mass} gives the same mass value as \cref{eq:optically thin mass}. In contrast, for $\sim$30 cores, \cref{eq:optically thick mass} increases their mass by at least $10\%$. And in fact, four of these particular protostellar cores have a $50\%-90\%$ increase in mass, compared with the masses estimated in the optically thin hypothesis. The resulting core masses range from $\sim$0.1$~\msun$ to $\sim$260$~\msun$ (see Tables~\ref{tab:computed table1}--\ref{appendixtab:computed table}). As in companion papers \citep[e.g.,][]{pouteau2022, armante2024}, the uncertainty on the mass of cores takes into account errors on their fluxes and on the estimate of their mass-averaged dust temperature (see Sect.~\ref{s:finalTdust}). The temperature uncertainties are dominated, for luminous protostellar cores, by the differences between our two temperature estimates, and for low-luminosity protostellar cores and prestellar cores, by uncertainties on their background-diluted temperature. This leads to relative uncertainties of $20\%$, up to $\sim$$40\%$, for the masses of the entire ALMA-IMF core sample. Uncertainties on the dust emissivity of dense cloud structures lead to even greater uncertainties in core mass estimates, a factor of at least 5 according to \cite{koehler2015}. Preliminary studies of core mass functions (CMFs) in the W43 protoclusters show that the results of \cite{motte2018b}, \cite{pouteau2022}, and \cite{nony2023} hold with these new mass estimates. Forthcoming papers (Nony et al. in prep.; Louvet et al. in prep.) will be dedicated to this topic.

\subsection{Comparison with other estimates of core temperatures}
\label{s:consistency}

We here discuss the consistency of the mass-averaged dust temperature we propose for ALMA-IMF cores (see Sect.~\ref{s:finalTdust}) with other temperature estimates. These estimates are taken from papers already published by our consortium: dust temperature of the first ALMA-IMF studies (see Sect.~\ref{s:consistencyPilot}), COM line temperatures assumed for ALMA-IMF hot cores (see Sect.~\ref{s:consistencyCOM}), and mass-averaged temperatures of prestellar cores in massive protoclusters (see Sect.~\ref{s:consistencyPcores}).

\subsubsection{Dust temperatures in previous ALMA-IMF papers}
\label{s:consistencyPilot}

The pilot study and two first studies of the ALMA-IMF survey, \citep{motte2018b, pouteau2022, armante2024} estimated the mass-averaged dust temperature of protostellar and prestellar cores, following a methodology close to that presented in the present article. They indeed used \PPMAP products, which have a lower angular resolution than ALMA images, and extrapolated at higher resolution the mass-averaged dust temperature for each core. The differences in methodology are shown below, along with their effect on core temperature and mass estimates.

For the \cite{motte2018b} and \cite{pouteau2022} studies, the first difference stems from the dust temperature images they used. They were less well corrected for the opacity of the $70~\mu$m emission than the images produced by \cite{dellova2024}, and used in \cite{armante2024}, for which the weighting of the far-infrared images compared with the millimeter images has been improved. As a result, the dust temperature measured in the improved images of \cite{dellova2024} with a $2.5\arcsec$ beam are a few, up to 15, degrees higher toward the brightest protostars of the W43-MM1, W43-MM2, and W43-MM3 protoclusters. 

The second difference in methodology is that these first three ALMA-IMF studies \citep{motte2018b, pouteau2022, armante2024} did not have \PPMAP luminosity images to use. As the luminosity of these protoclusters was not constrained with the $2.5\arcsec$ resolution of \PPMAP, they used the luminosity integrated in areas $\sim$$15\arcsec$ in diameter \citep{motte2003, herpin2012}. They correspond to the sum of the individual luminosity of a handful or dozen protostellar cores and of their background luminosity (see Figs.~\ref{fig:lum-temp peaks G012-W43} and \ref{appendixfig:zoom positions}). In the G012.80 protocluster, the luminosity of their \hii regions dominates, precluding using integrated luminosites from the literature. Rather than using the relative flux at 1.3~mm of the protostars in W43-MM1 and W43-MM2\&MM3 to determine their contribution to the luminosity of their protocluster, \cite{motte2018b} and \cite{pouteau2022} used their COM line emission, thus focusing on the most luminous protostars associated with a hot core. In detail, \cite{motte2018b} assumed a proportional relationship between the contribution of protostars hosting hot cores to the protocluster luminosity, $L_{\rm proto\star}^{\rm COM}/L_{\rm bol}^{\rm protocluster}$, and the intensity ratio of COM line to continuum emission, $\gamma_{\rm COM}$, of the number of hot cores, $n_{\rm HC}$, of the protocluster:
\begin{eqnarray}
    \frac{L_{\rm proto\star}^{\rm COM}}{L_{\rm bol}^{\rm protocluster}} \sim 
    \frac{(\gamma_{\rm COM})^{3.1}}            
    {\summation{n_{\rm HC}}{j=1} (\gamma_{\rm COM,\,j})^{3.1}}    .   
    \label{eq:COM luminosity Nature}
\end{eqnarray}
This approximate relation and its power-law index of $3.1$ are obtained by the demonstration, and its hypotheses are provided in Appendix~\ref{appendixsect:gammaCOM} (in particular see \cref{eq:lproto-gammaCOM}). \cite{motte2018b} and \cite{pouteau2022} applied \cref{eq:COM luminosity Nature} to the nine W43-MM1 and four W43-MM2 protostellar cores, identified by \cite{brouillet2022} and \cite{bonfand2024} to host a hot core. The resulting luminosities are $\sim$5 times smaller than the protostellar luminosities estimated in Sect.~\ref{s:Tproto-InOut}, leading to a  $\sim$$33\%$ underestimate of the temperatures for the thirteen brightest protostellar cores of W43-MM1 and W43-MM2.

The biggest discrepancies with the dust temperatures estimated by \cite{motte2018b}, and similarly by \cite{louvet2024}, concern protostellar cores not associated with a hot core and prestellar cores. Their temperatures were measured in the images of the background-diluted core temperature, $\sim$23~K, leading to an underestimate by $\sim$70\% for protostellar cores and an overestimate by $\sim$20\% for prestellar cores. As a result, the masses of protostellar cores were overestimated by $\sim$40\% and those of prestellar cores were underestimated by $\sim$40\%. \cite{pouteau2022} and \cite{armante2024} went beyond the background-diluted temperature of cores provided by \PPMAP \citep{dellova2024}. To account for protostellar heating, they increased by $4\pm 4$~K the mass-averaged temperature of protostellar cores driving outflows but no hot core. In addition, they decreased by $2\pm2$~K the mass-averaged temperature of prestellar cores in order to account for their self-shielding. When compared to the present study, the masses resulting from these temperature estimates for these protostellar cores remain overestimated, but only by $\sim$30\%, and those of the prestellar cores are similar in \cite{pouteau2022} and \cite{armante2024}.

To conclude, on average, the dust temperature of \cite{motte2018b} is consistent with those of the present study to within $40\%$ and those of \cite{pouteau2022} and \cite{armante2024} to within $30\%$.

\subsubsection{COM temperatures of the ALMA-IMF survey}
\label{s:consistencyCOM}

Studies of ALMA-IMF molecular lines provide complementary estimates to the dust temperatures estimated by \PPMAP SED fits (see Sect.~\ref{s:lumcat}). These also  have a higher angular resolution than the dust based estimates. If the line emission is spatially resolved and assuming that the molecular abundance profile is known, the measured excitation temperatures of a given molecule can approximately represent the core temperature. Implicit in this are the assumptions that dust and gas must be well coupled over the full extent of the core, and that dust and gas emissions are both optically thin. Good coupling is expected for molecules excited in dense envelopes by the internal heating of protostars. As for the dust and line emissions, they should be optically thin in moderate-luminosity protostellar cores associated with hot cores.

\cite{bonfand2024} used the CH$_3$OCHO molecule and specifically its doublet line at 216.1~GHz, to study images of its integrated emission on the 15 massive protoclusters of ALMA-IMF. They discovered 76 sources and found that $74\%$ of these hot core candidates are slightly more extended than their host core detected in continuum: mean FWHM of $\sim$2300~au versus $\sim$1900~au. They interpreted this as evidence for a filling factor close to one. Besides, detailed studies of eight hot cores in the W43-MM1 protocluster have revealed excitation temperatures of $120-160$~K for the CH$_3$CN line emission \citep{brouillet2022}. Therefore and because these temperatures are also close to the thermal desorption temperature of CH$_3$OCHO, \cite{bonfand2024} set a dust temperature of $\tdust = 100\pm50$~K for the ALMA-IMF protostellar cores associated with candidate hot cores. For eight protostellar cores of their sample, however, \cite{bonfand2024} increased the temperature up to $\tdust = 300\pm100$~K, based on previous measurements \citep{gibb2000, ginsburg2017}. In particular, \citet{ginsburg2017} imaged three extended hot cores of the W51-E and W51-IRS2 protoclusters and found $T_{\rm rot}>200$ K out to radii of $\sim$$10^4$~au, using LTE modeling of CH$_3$OH. These eight protostellar cores are notably associated with very extended, $4000-13\,400$~au, CH$_3$OCHO emission \citep[][see also \cref{tab:luminosity peaks from radial profile}]{bonfand2024}.

Among the hot core candidates discovered by \cite{bonfand2024}, 51 are associated with a core in the 1.3~mm ALMA images of the 14 ALMA-IMF protoclusters studied here. Virtually all of these hot core candidates are associated with a luminosity peak in the \PPMAP images (see Tables~\ref{tab:computed table1}--\ref{appendixtab:computed table}). The core temperatures proposed by \cite{bonfand2024} and measured by \cite{gibb2000}, \cite{ginsburg2017}, and \cite{brouillet2022} are on average $1.7$ times higher than the dust-based estimates of Sect.~\ref{s:finalTdust}: $100\pm50$~K versus $35-100$~K and $300\pm100$~K versus $70-130$~K (see Tables~\ref{tab:luminosity peaks from radial profile}--\ref{appendixtab:computed table}).

If the relation between the mass-averaged temperature and protostellar luminosity given in \cref{eq:tdust-fromL} is correct, reaching mass-averaged dust temperatures within cores of 2100~au typical sizes of 100~K and 300~K requires protostars with a luminosity of $9.3\times 10^3~\lsun$ and $4\times 10^6~\lsun$, respectively. In W43-MM1, \cite{bonfand2024} found 14 hot core candidates associated with protostellar cores and propose that their mass-averaged temperature is $100\pm50$~K. The sum of their luminosities, predicted to be $14\times 9.3\times 10^3~\lsun \simeq 1.3\times 10^5~\lsun$, is 1.6 times larger than the luminosity of the W43-MM1 central dense clump \citep{koenig2017,dellova2024}. Assuming 50~K, which is the lower limit temperature proposed by  \cite{bonfand2024}, instead of 100~K would make these luminosity estimates more consistent, although this would require non-thermal desorption processes to bring CH$_3$OCHO to the gas phase. As for the three ALMA-IMF protoclusters that \cite{bonfand2024} propose to contain one to three cores of $300\pm100$~K, the sum of their predicted protostellar luminosity amounts to $4.1-12.5\times 10^6~\lsun$, values $10-80$ times greater than the luminosity of their central dense clump \citep{dellova2024}. These inconsistencies probably suggest that \cref{eq:tdust-fromL} is not valid for the most massive ALMA-IMF cores, which are expected to subfragment and be partly optically thick to infrared radiation. 

Alternatively, these inconsistencies could question the interpretation that the extended CH$_3$OCHO emission is evidence of intense protostellar irradiation alone. In agreement with this questioning, \cite{brouillet2022} found that the hot cores of W43-MM1 are not resolved when they are traced by the integrated emission of all COM lines detected in the 2~GHz-wide spectral band at 233~GHz. As already stated in \cite{bonfand2024}, the sum of several COM emission peaks associated with protostellar multiples and/or shocks created by protostellar outflows or accretion \citep{lefloch2017, palau2017, csengeri2019, busch2024} would mimic an extended emission. 

Moreover, we could question the thermal coupling between dust and gas, as was done by  studies of massive clumps in their earliest phase of evolution \citep{merello2019, urquhart2019, mininni2021}. Our results, however, correspond to scales ten to 100 times smaller and therefore to much denser cloud structures where thermal coupling will be considerably better than in massive clumps. Finally, the spatial coupling between dust emission and COM lines could be much weaker than expected, given a protostellar source geometry that is much more complex than a spherical envelope with a regular concentration of density toward the center. Follow-up studies, in particular on the core subfragmentation and the origin of CH$_3$OCHO emission, are necessary to reconcile SED-based and COM-based methods, or explain why they give different results, to estimate the protostellar luminosity and mass-averaged temperature of protostellar cores.

\subsubsection{Prestellar core temperatures}
\label{s:consistencyPcores}

In the solar neighborhood, cores of the quiet Taurus cloud and isolated globules have cold, 10--12~K, temperatures \citep{launhardt2013, marsh2014}. Cores located in low- to intermediate-mass protoclusters have slightly warmer, 12--15~K, temperatures \citep[e.g.][]{konyves2015, ladjelate2020, nony2021}. In agreement with this slight temperature increase, low- to intermediate-mass protoclusters are expected to have a stronger (external and internal) ISRF than Taurus \citep{jorgensen2006}. High-mass, nearby star-forming complexes, under the influence of massive star clusters and \hii regions, exhibit filaments and clumps of higher column density. Despite that, these \textit{Herschel}/HOBYS complexes have higher, $15-20$~K, temperatures when measured with SEDs \citep{motte2010, hill2012, tige2017} and ammonia \citep{wienen2012, boegner2022}. A handful of starless clumps are also observed to have temperatures of up to $\sim$40~K \citep{motte2010, tige2017, boegner2022}. Since the ALMA-IMF protoclusters are subject to a stronger ISRF, we expect their core environment to be hotter than in \textit{Herschel}/HOBYS clouds. \cite{dellova2024} measured spatially averaged temperature of $\sim$21~K to $\sim$29~K for the ALMA-IMF protoclusters, with similar variations in most protoclusters from their cold and dense filaments at $19-23$~K to areas heated up to $\sim$45~K by \hii regions and luminous protostars  (see, e.g., Figs.~\ref{fig:temperature maps} and \ref{appendixfig:ppmap}). We computed, for ALMA-IMF prestellar cores, mass-averaged temperatures of $15-40$~K, with a median value at 23~K and 5th and 95th percentiles at 17~K and 31~K. 

The companion paper \cite{valeille2024} used temperatures ranging from 20~K to 27~K for the most massive ALMA-IMF cores, which do not drive powerful outflows and which they call PreSCs. These assumed temperatures are in good agreement, to within $30\%$, with the measurements of the present paper (see \cref{appendixtab:prestellar table} and their Table~3), with the exception of seven of their 42 PreSCs. These PreSCs are either qualified here as protostellar cores (Nony et al. in prep., see Tables~\ref{tab:computed table1}--\ref{appendixtab:computed table}) or they are located at the immediate proximity of, and thus are heated by, a luminous protostar (see, e.g., Figs.~\ref{appendixfig:ppmap}c and \ref{appendixfig:zoom positions} G333.60~I). Three of these sources are quoted as tentative protostellar cores (Nony et al. in prep.); if we would have assumed them to be prestellar in nature their temperature would agree with the assumptions of \cite{valeille2024}. Therefore, for more than $80\%$ of the PreSCs of \cite{valeille2024}, the masses computed using the temperatures of the present study agree, to within $30\%$, with their mass estimates (see Tables~\ref{tab:computed table1}--\ref{appendixtab:prestellar table}).

\begin{figure*}
    \centering    \includegraphics[width=0.9\linewidth]{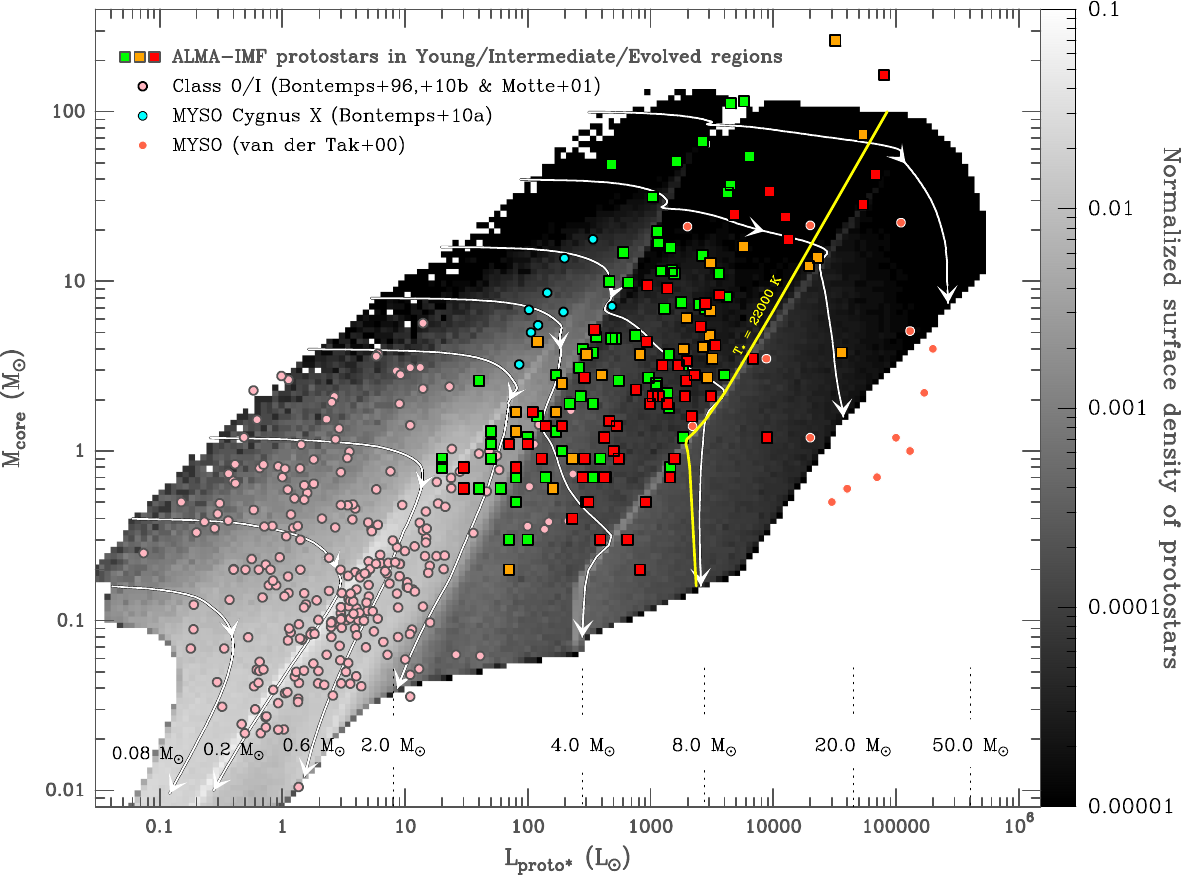}
    \caption{Mass versus luminosity diagram of the 151 protostellar cores discovered, with detectable luminosity, in the ALMA-IMF massive protoclusters (color-filled squares, green, orange, and red for Young, Intermediate, and Evolved protoclusters) compared to reference studies of low-mass protostars (pink circles), IR-quiet high-mass protostars (cyan circles), and clumps hosting UC\hii regions (red circles). The surface density of observed protostellar cores is to be compared to that predicted by a parameterized model (gray scale image), here with decreasing accretion rates and intermittent accretion \citep{duarte2013}. The final stellar mass of a protostar is predicted by the evolutionary tracks, for $0.08-50~\msun$ core masses, associated to this model (white curves). The yellow curve theoretically separates high-mass protostars from sources developing an \hii region. Current scenario, assuming an accretion from a gas reservoir decreasing with time, does not fit the protostellar number density observed for ALMA-IMF protoclusters: protostellar cores more massive than $2~\msun$ are clustered in areas where the normalized surface density is predicted to be very low ($\sim$$3\times10^{-5}$), while only a few of these protostellar cores are found in areas where the surface density is 30 times higher.}
    \label{fig:mass vs lum}
\end{figure*}

\subsection{Protostellar evolution in massive protoclusters}
\label{s:M-L}

Robust measurements of the basic properties of protostellar cores, such as luminosity and mass, are crucial to build the protostellar evolutionary diagrams necessary to constrain the accretion history of protostars. In particular, this was poorly achieved for protostars forming in massive protoclusters \citep[see, however, the first attempt of][]{duarte2013}. One of the most widely used evolutionary diagrams is that of the mass of protostellar cores as a function of the luminosity of their inner illuminating protostar, the so-called $M_{\rm core}-L_{\rm proto\star}$ diagram \citep[e.g.,][]{andre2000, molinari2008, duarte2013, peretto2020}. While $M_{\rm core}$ provides a measure of the mass reservoir, the protostar luminosity, $L_{\rm proto\star}$, represents, at early stages, the accretion luminosity and, at later stages, the stellar luminosity related to the mass of the central protostar.

Figure~\ref{fig:mass vs lum} presents the $M_{\rm core}-L_{\rm proto\star}$ diagram for the 151 luminous protostellar cores, found by the ALMA-IMF survey to be bright enough to have their protostellar luminosities characterized (see Sects.~\ref{s:Tproto-InOut} and \ref{s:mass}). Their masses and luminosities are taken from Tables~\ref{tab:computed table1}--\ref{appendixtab:computed table}. The present sample represents intermediate- to high-mass protostars forming in massive protocluster environments. Despite the apparent bias toward luminous sources, many of these protostellar cores are the equivalent of Class~0s (young low-mass protostars), with a core mass much larger than the protostar mass \citep{motte2018b, pouteau2022}. We expect the sample to be complete for protostars of $100~\lsun$ and $300~\lsun$ luminosities for the Young and Evolved protoclusters, respectively \citep{dellova2024}. 

We first use the $M_{\rm core}-L_{\rm proto\star}$ diagram to compare the location of ALMA-IMF protostellar cores with that of other samples of the young stellar objects. We present the result of surveys for protostellar cores in nearby clouds \citep{bontemps1996, bontemps2010SI, motte2001} and two studies dedicated to a handful of high-mass IR-quiet protostellar cores in the Cygnus~X star-forming complex \citep{duarte2013} and a dozen of massive clumps hosting UC\hii regions \citep{vandertak2000}. The size of protostellar cores is $6000-10\,000$~au in nearby protoclusters, set to $4000$~au for the protostellar cores of Cygnus~X, while UC\hii clumps have much larger sizes, $40\,000-140\,000$~au. Therefore, all of these young stellar objects are cloud structures of a larger size and thus potentially of greater mass than the ALMA-IMF luminous protostellar cores, which are $\sim$1900~au in size.

As shown in \cref{fig:mass vs lum}, ALMA-IMF protostellar cores appear clustered in between the location of the young high-mass protostellar cores of \cite{duarte2013} and the clumps harboring UC\hii region of \cite{vandertak2000}. There is a small location gradient between Young, Intermediate, and Evolved protoclusters, with the protostellar population of Evolved protoclusters twice as dispersed as that of Young protoclusters. This result suggests that Evolved protoclusters have formed (high-mass) stars for longer time than Young protoclusters, in line with their more evolved stage. With the exception of seven sources, ALMA-IMF protostellar cores are located in a distinct part of the diagram from UC\hii regions, separated by the $T=22\,000$~K line (see \cref{fig:mass vs lum}). As for the Cygnus~X protostellar cores, they look younger, that is more massive for a given luminosity, than most of the ALMA-IMF protostellar cores. A higher mass by a factor $\sim$2 is expected, as the mass of Cygnus~X protostellar cores were extrapolated by \cite{bontemps2010} to a radius of 4000~au, twice the measured size of ALMA-IMF protostellar cores. Dividing the core mass of Cygnus~X protostars by this factor of two, would provide a better match in the $M_{\rm core}-L_{\rm proto\star}$ diagram between the locations of high-mass protostellar cores in Cygnus~X and in ALMA-IMF (see \cref{fig:mass vs lum}). Besides, there is a substantial overlap between the samples of low- to intermediate-mass protostars in nearby star-forming regions and in the ALMA-IMF protoclusters (see \cref{fig:mass vs lum}). It will allow us soon to study the potential variation in accretion history for protostellar cores of the same mass but forming stars in different environments.

We then use the $M_{\rm core}-L_{\rm proto\star}$ diagram to compare the location of observed protostars with predictions of protostellar accretion models. \cite{duarte2013} proposed several parametrized models, adapted from \cite{bontemps1996} and \cite{andre2000}, with constant and decreasing accretion rates, and with continuous or intermittent accretion. They computed the associated probability density distribution of their protostellar cores and the evolutionary tracks predicted for protostars of a given core mass. We reported in \cref{fig:mass vs lum} one model by \cite{duarte2013}, based on decreasing accretion rates and intermittent accretion. The accretion of high-mass protostars is taken to be strong but sporadic, as expected when gas reaches protostellar cores through mass inflows \citep{smith2009, galvanmadrid2009, csengeri2011, olguin2023}. The location of low-mass protostellar cores of nearby clouds \citep{bontemps1996, bontemps2010SI, motte2001} agrees with the protostellar number density predicted by this model. In contrast, and even when taking into account an uncertainty factor of two on the absolute value of protostellar core masses, the ALMA-IMF population of intermediate- to high-mass protostellar cores is, on average, either too luminous or not massive enough compared with predictions of \cite{duarte2013}. To fit their location in \cref{fig:mass vs lum}, the horizontal part of the evolutionary tracks tracing the main accretion phase of high-mass protostars should extend into the high-luminosity regime, with mid-lifetime (track arrows in \cref{fig:mass vs lum}) shifted toward luminosities five times higher. This implies that the mass reservoir of the high-luminosity ALMA-IMF cores would not decrease during the main accretion phase as much as assumed by the model. This mass reservoir could in fact be continuously replenished, or even increased, by external gas inflows, as ALMA-IMF studies have already suggested \citep[including companion papers, Papers~V and XIII,][]{nony2023, alvarez2024, sandoval2024}. A better model of protostellar evolution could therefore have a constant or even increasing accretion rate, as well as a core formation efficiency (CFE) greater than $100\%$ of the initial core mass reservoirs. CFE values increasing with core density have already been reported in massive protoclusters \citep[e.g.,][]{louvet2014} and used for recent protostellar evolutionary models \citep{peretto2020}. 

The $M_{\rm core}-L_{\rm proto\star}$ diagram of \cref{fig:mass vs lum} will be the subject of follow-up studies. The parameters of the accretion models must be adjusted to better describe, in ALMA-IMF environments, the variation of the accretion rate with time. Ultimately, this will help constrain star formation models suited to the star formation bursts that develop in massive protoclusters.

\section{Conclusion} \label{s:conc}

This study was carried out in the context of the ALMA-IMF Large Program \citep{motte2022}. We have made extensive use of some of the ALMA-IMF data products, and we detail them in the following list:
\begin{itemize}
    \item The currently published ALMA-IMF core catalogs are unprecedented for their core detection sensitivity and completeness while excluding spurious sources \citep{pouteau2022, nony2023, armante2024, louvet2024}. They provide high-quality measurements of the thermal dust emission flux for the vast majority of our $\sim$$2300$~au cores. A systematic search for outflows driven by ALMA-IMF cores allowed us to classify 266 protostellar cores and 617 prestellar core candidates (Nony et al. in prep., see also Sect.~\ref{s:corecat}).
    \item Recent SED analysis of ALMA-IMF protoclusters using the Bayesian technique \PPMAP has provided a rich database of dust temperature and luminosity images at the unrivalled resolution of $2.5\arcsec$      \citep{dellova2024}. Although this resolution comes close to that of ALMA-IMF images, $0.3-0.9\arcsec$, it still falls short (see Figs.~\ref{fig:lum-temp peaks G012-W43} and \ref{appendixfig:lum and temp peaks}). \PPMAP thus provides the most likely values for the background-diluted core temperature as well as the luminosity of protostellar groups 
    (see Sects.~\ref{s:lumcat} and \ref{s:data-temp}).
\end{itemize}
We have proposed a methodology for estimating the mass-averaged temperature of cores detected by ALMA-IMF as well as the luminosity of individual high-mass protostars. Our main results regarding this methodology can be summarised as follows:
\begin{itemize}
    \item We used the luminosity peak catalog of \cite{dellova2024} for the Young and Intermediate ALMA-IMF protoclusters. We augmented this work with a \getsf extraction of luminosity peaks, which are not associated with free-free emission, in the Evolved protoclusters of ALMA-IMF (see Sect.~\ref{s:lumcat}). In total, 111 out of the 255 luminosity peaks host protostellar cores; they are listed and characterized in Tables~\ref{tab:measures table}--\ref{appendixtab:measures table evolved}. We also integrated, in larger radii, the luminosity of the strongest luminosity peaks, encompassing the six brightest hot cores identified in ALMA-IMF by \cite{bonfand2024} (see Sect.~\ref{s:extremehotcore} and \cref{appendixfig:luminosity profile}). Cores that drive outflows and are associated with luminosity peaks are dubbed "luminous protostellar cores", and those that are not strong enough to be detected in the \PPMAP luminosity images are referred to as "low-luminosity protostellar cores" (see \cref{fig:temperature three profiles}).
    \item The default dust temperature image of \cite{dellova2024} is produced with the $70\,\mu$m emission corrected for the optical depth toward the protocluster and its protostellar cores. This process gives the temperature of the ALMA-IMF protoclusters as a whole and the most likely background-diluted temperature of their luminous protostellar cores (see Sect.~\ref{s:data-temp} and \cref{appendixsect:ppmap}). The background-diluted temperature of low-luminosity protostars, prestellar cores, and starless filaments is best estimated by averaging the two dust temperature images provided by \cite{dellova2024}: the original image and the one corrected for the optical depth at $70\,\mu$m (see Sect.~\ref{s:data-temp}).
    \item For luminous protostellar cores, we estimated their luminosity, using their distance to their host luminosity peak to determine their contribution to this luminosity (see Sect.~\ref{s:Tproto-InOut} and \cref{fig:spatial weight}). Lists of the luminosity of the luminous protostellar cores are provided in Tables~\ref{tab:computed table1}--\ref{appendixtab:computed table}, and they range from $50~\lsun$ to $1.5\times 10^4~\lsun$ (5th and 95th percentiles, up to $8\times 10^4~\lsun$). 
    \item For these luminous protostellar cores, the mass-averaged temperature was estimated using parametric relations derived from the radiative transfer modeling of protostellar envelopes and two approaches (see Sect.~\ref{s:Tproto} and \cref{fig:temperature three profiles}). The first approach extrapolates the background-diluted temperature of cores down to the size of cores (see Sect.~\ref{s:Tproto-OutIn} and \cref{eq:tdust-fromT}), while the second approach uses the protostellar luminosity (see Sect.~\ref{s:Tproto-InOut} and \cref{eq:tdust-fromL}). The two methods give consistent results (see Sect.~\ref{s:finalTdust} and \cref{fig:tdust variation}). Their average gives, for the 151 luminous protostellar cores, temperatures that range from 38~K to 92~K (5th and 95th percentiles) and go up to 127~K, with a median value of 54~K (see Tables~\ref{tab:computed table1}--\ref{appendixtab:computed table}).
    \item For low-luminosity protostellar cores, the mass-averaged temperature was estimated using a single approach, namely, the first approach described above for luminous protostars (see Sect.~\ref{s:Tproto-OutIn}, \cref{fig:temperature three profiles} and \cref{eq:tdust-fromT}). The resulting mass-averaged temperatures of the 115 low-luminosity protostellar cores are listed in Tables~\ref{tab:computed table1}--\ref{appendixtab:computed table}, and they range from 26~K to 51~K (5th and 95th percentiles), with a median value of 37~K. 
    \item For prestellar cores, we estimated the mass-averaged temperature, applying a parametric cooling relation from their background-diluted temperature (see Sect.~\ref{s:Tpre estimate} and \cref{fig:temperature three profiles}). The power-law index of this self-shielding has been systematically measured in published studies of observations or radiative transfer models dedicated to cloud structures following a $\rho(r)\propto r^{-2}$ law (see Sect.~\ref{s:Tpre-litterature} and \cref{eq:tdustPrestellar}). The resulting mass-averaged temperatures of the 617 ALMA-IMF prestellar cores are listed in \cref{appendixtab:prestellar table} and range from 17~K to 31~K (5th and 95th percentiles), with a median value of 22~K. 
    \item We applied our methodology to the entire sample of 883 ALMA-IMF cores detected at the original angular resolution. The core masses were computed using our final temperature estimates in \cref{eq:optically thick mass} (see Sects.~\ref{s:finalTdust}--\ref{s:mass}), and they are listed in Tables~\ref{tab:computed table1}--\ref{appendixtab:prestellar table}. These results are consistent with initial ALMA-IMF studies (see Sects.~\ref{s:consistencyPilot} and \ref{s:consistencyPcores}). However, it is worth noting that the dust-based temperature of luminous protostellar cores associated with hot cores is half that of the COM line-based temperatures (see Sect.~\ref{s:consistencyCOM}).
    \item The 151 luminous protostellar cores constitute the largest sample obtained to date, which provides both mass and luminosity, in the high-mass regime and at the core scale. We placed these protostellar cores in a $M_{\rm core}/L_{\rm proto\star}$ evolutionary diagram (see \cref{fig:mass vs lum}). A comparison with protostellar evolutionary tracks and the protostellar core number density of an accretion scenario using the initial core as the mass reservoir suggests that core mass growth during protostellar evolution would be more appropriate (see Sect.~\ref{s:M-L}).
\end{itemize}

The present ALMA-IMF catalog of 883 cores with robust characteristics, such as their size, temperature, mass, and luminosity (for 151 of them), represents a gold mine of data. It will be used in future papers studying core populations, including their CMF, CFE, and  $M_{\rm core}/L_{\rm proto\star}$ diagram, with the aim of constraining cloud and star formation models in extreme Galactic protoclusters, such as those imaged by the ALMA-IMF Large Program. Our result calls for follow-up studies to resolve the contradiction between the mass-averaged temperature we compute and their COM line-based temperatures. Among others, this will require higher angular resolution observations in the (sub)millimeter regime, surveys dedicated to chemical studies, and JWST observations of ALMA-IMF protostars. Moreover, radiative transfer modeling is needed to improve the analytical description of the self-shielding of intermediate- to high-mass prestellar cores in environments resembling the ALMA-IMF protoclusters.


\begin{acknowledgements}
This paper makes use of the following ALMA data: ADS/JAO.ALMA\#2017.1.01355.L, \#2013.1.01365.S, and \#2015.1.01273.S. ALMA is a partnership of ESO (representing its member states), NSF (USA) and NINS (Japan), together with NRC (Canada), MOST and ASIAA (Taiwan), and KASI (Republic of Korea), in cooperation with the Republic of Chile. The Joint ALMA Observatory is operated by ESO, AUI/NRAO and NAOJ.
The project leading to this publication has received support from ORP, which is funded by the European Union's Horizon 2020 research and innovation program under grant agreement No. 101004719 [ORP].
This project has received funding from the European Research Council (ERC) via the ERC Synergy Grant \textsl{ECOGAL} (grant 855130) and from the French Agence Nationale de la Recherche (ANR) through the project \textsl{COSMHIC} (ANR-20-CE31-0009).
YP and BL acknowledge funding from the European Research Council (ERC) under the European Union’s Horizon 2020 research and innovation programme, for the Project DOC, grant agreement No 741002.
TN and RGM acknowledge support from UNAM-PAPIIT project IN104319 and from CONACyT Ciencia de Frontera project ID  86372. Part of this work was performed at the high-performance computers at IRyA-UNAM.
TN acknowledges support from the Large Grant INAF 2022 YODA.
TCs and MB have received financial support from the French State in the framework of the IdEx Universit\'e de Bordeaux Investments for the future Program.
AS gratefully acknowledges support by the Fondecyt Regular (project code 1220610), and ANID BASAL project FB210003.
AG acknowledges support from the NSF under grants AST 2008101 and CAREER 2142300. 
FL acknowledges support by the Marie Curie Action of the European Union (project MagiKStar, Grant agreement number 841276). 
MB is a postdoctoral fellow in the University of Virginia’s VICO collaboration and is funded by grants from the NASA Astrophysics Theory Program (grant num- ber 80NSSC18K0558) and the NSF Astronomy \& Astrophysics program (grant number 2206516).
PS was partially supported by a Grant-in-Aid for Scientific Research (KAKENHI Number JP22H01271 and JP23H01221) of JSPS.
RA gratefully acknowledges support from ANID Beca Doctorado Nacional 21200897.
LB gratefully acknowledges support by the ANID BASAL project FB210003. 
\end{acknowledgements}

\bibliographystyle{aa}  
\bibliography{CoreDustTemperature_ProtostarLuminosity}

\begin{appendix}

\section{Background-diluted dust temperature of cores in the ALMA-IMF protoclusters}
\label{appendixsect:ppmap}


\begin{figure*}[htbp!]
\vskip -2.7cm
\hskip -0.8cm\includegraphics[width=1.85\textwidth]{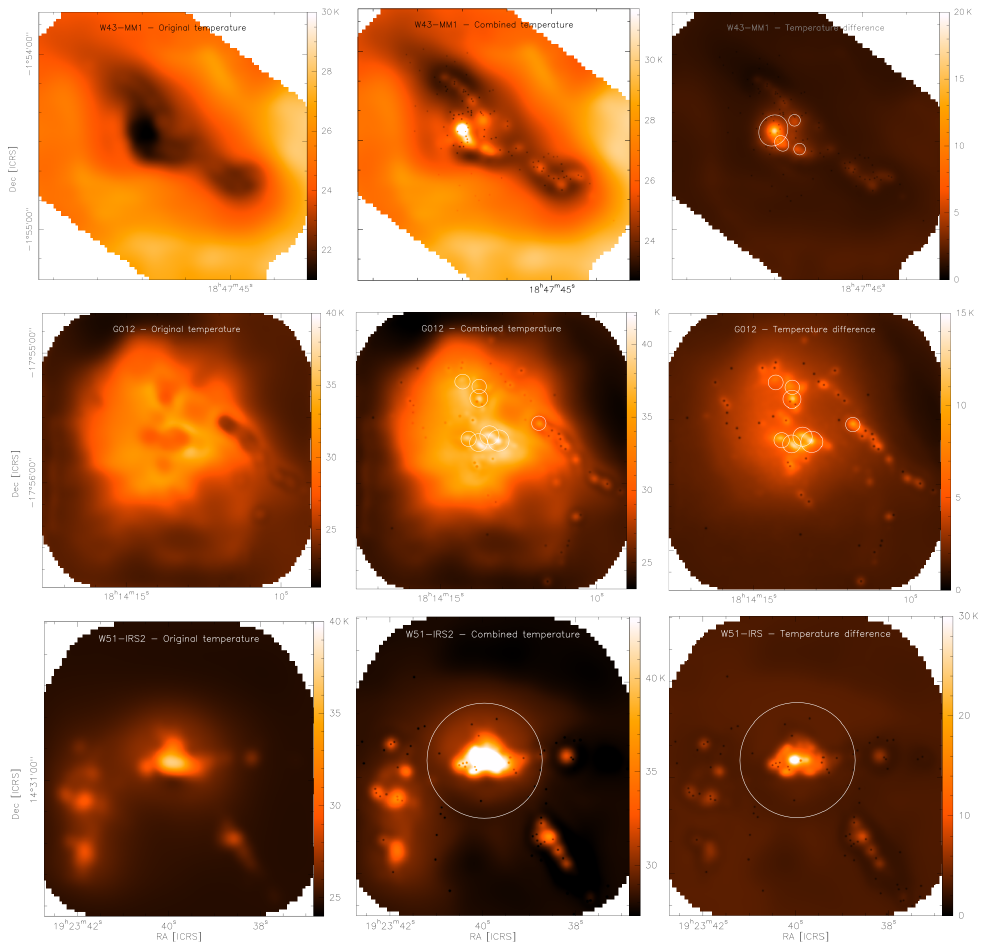}
\vskip -3cm
\caption{Original (\textit{left}) and combined (\textit{center column}) dust temperature images and their subtraction (\textit{right}) for the Young protocluster W43-MM1 (\textit{upper}) and the Evolved protoclusters G012.80 (\textit{central row}) and W51-IRS2 (\textit{lower}). In the right panels, bright areas represent gas structures, whose 70\,$\mu$m flux is extinguished by the protocluster high column density gas. Their temperature has been increased by the opacity-corrected procedure of \PPMAP. Unresolved bright and dark sources are the luminous protostellar cores and, low-luminosity protostellar or prestellar cores, respectively. White circles mark the areas heated beyond the $2.5\arcsec$ beam of \PPMAP by the brightest protostars of these regions.} 
\label{appendixfig:ppmap}
\end{figure*} 

\begin{table*}[h!]
\centering
\resizebox{1\textwidth}{!}{
\begin{threeparttable}[c]
\caption{Dust temperature in the ALMA-IMF protoclusters.}
\label{appendixtab:ppmap table}
\begin{tabular}{l|ccc|ccc|l}
\hline\hline
Region &  \multicolumn{2}{c}{$\overline{T_{\rm dust}^{\rm\PPMAP}}$} & $\Delta \overline{T_{\rm dust}^{\rm\PPMAP}}$ & \multicolumn{2}{c}{$\overline{T_{\rm dust}^{\rm\PPMAP}}$[w-$N_{\rm H_2}$]} &   $\Delta \overline{T_{\rm dust}^{\rm\PPMAP}}$[w-$N_{\rm H_2}$] & Number of\\
name & original & combined & difference & original & combined & difference & resolved \\ 
 & [K] & [K] & [K] & [K] & [K] & [K] & heating points \\
(1) & (2) & (3) & (4) & (5) & (6) & (7) & (8)\\
\hline
W43-MM1     & $27\pm5$ & $29\pm5$ & $3\pm7$ & $23\pm4$ & $27\pm5$ & $4\pm7$ & 4 \\  
W43-MM2     & $25\pm5$ & $26\pm5$ & $1\pm7$ & $24\pm5$ & $28\pm6$ & $4\pm7$ & 1 \\  
G338.93     & $25\pm5$ & $27\pm5$ & $2\pm7$ & $26\pm5$ & $28\pm6$ & $3\pm8$ & 3 \\  
G328.25     & $25\pm5$ & $27\pm5$ & $2\pm7$ & $27\pm5$ & $30\pm6$ & $3\pm8$ & 1 \\  
G337.92     & $22\pm4$ & $24\pm4$ & $2\pm6$ & $22\pm4$ & $28\pm5$ & $5\pm6$ & 2 \\  
G327.29     & $25\pm5$ & $27\pm5$ & $2\pm7$ & $26\pm5$ & $36\pm7$ & $10\pm9$ & 6 \\  
\hline
G008.67     & $21\pm4$ & $23\pm4$ & $2\pm6$ & $23\pm4$ & $26\pm5$ & $3\pm7$ & 2 \\  
W43-MM3     & $25\pm5$ & $27\pm5$ & $2\pm7$ & $24\pm5$ & $26\pm5$ & $2\pm7$ & 3 \\  
W51-E       & $23\pm4$ & $26\pm4$ & $3\pm6$ & $24\pm4$ & $38\pm7$ & $15\pm8$ & 4 \\  
G353.41     & $21\pm4$ & $23\pm4$ & $2\pm6$ & $22\pm4$ & $24\pm5$ & $3\pm6$ & 5 \\  
\hline
G010.62     & $24\pm5$ & $26\pm5$ & $2\pm7$ & $26\pm5$ & $31\pm6$ & $5\pm8$ & 1 \\  
W51-IRS2    & $29\pm5$ & $31\pm6$ & $2\pm7$ & $33\pm6$ & $50\pm9$ & $17\pm10$ & 2 \\  
G012.80     & $28\pm5$ & $30\pm5$ & $2\pm7$ & $26\pm5$ & $29\pm5$ & $3\pm7$ & 8 \\  
G333.60     & $27\pm5$ & $30\pm5$ & $3\pm7$ & $27\pm5$ & $31\pm6$ & $4\pm8$ & 3 \\  
\hline
\end{tabular}
\begin{tablenotes}[flushleft]
\item []\textbf{Notes:}
(2)--(4) Mean dust temperatures, spatially averaged over the entire protocluster area, in the original and combined images, and their difference, respectively.
(5)--(7) Mean temperatures, weighted by the column density, in the original and combined images, and their difference, respectively.
(8) Number of resolved heating points, with a heating radius of $\theta
(\overline{T_{\rm dust}^{\rm\PPMAP}}, L_{\rm proto \star}) > 2.5\arcsec$ and thus a FWHM size $>$$2.5\arcsec$, using in \cref{eq:R at Tbck} Col.~3, and Col.~3 of Tables~\ref{tab:computed table1}--\ref{appendixtab:computed table}.
\end{tablenotes}
\end{threeparttable}    }
\end{table*}

This section presents our best-estimate of the background-diluted temperature of cores, based on the two dust temperature images produced by \cite{dellova2024} using the \PPMAP technique. The "original dust temperature images" are the regular output of the \PPMAP procedure \citep{marsh2015, marsh2017}. Since ALMA-IMF protoclusters are highly embedded protoclusters, $>$$5\times 10^{22}$~cm$^{-2}$ \citep[][see their Figs.~9 and 14]{diazgonzalez2023}, \cite{dellova2024} further corrected the \PPMAP dust temperature for the opacity of their 70~$\mu$m emission. The resulting image, the "corrected dust temperature image", is better suited than the original dust temperature image for the embedded ALMA-IMF protoclusters and in particular for the line-of-sights toward their luminous protostars and \hii regions. As this correction was designed for hot, embedded sources, we must propose another background-diluted temperature image for the cold parts of the ALMA-IMF protoclusters, including low-luminosity protostellar cores and prestellar core candidates (see definitions in Sect.~\ref{s:data-cat}). We therefore created "combined dust temperature images", which consist of the opacity-corrected image except at the location of prestellar cores and low-luminosity protostars, where the average of the two is taken (see Sect.~\ref{s:Tbck}). In detail, we measured the temperatures at the location of these cores in both the original and opacity-corrected temperature images, and averaged them. Gaussians with these peak values and widths corresponding to the half power beam width of each 1.3~mm image then replace the opacity-corrected temperature for the low-luminosity protostars and prestellar cores.

\cref{appendixtab:ppmap table} lists the spatially averaged dust temperatures, $\overline{T_{\rm dust}^{\rm \PPMAP}}$, and the mean temperatures weighted by the column density, $\overline{T_{\rm dust}^{\rm\PPMAP}}$[w-$N_{\rm H_2}$], both in the original and combined images, and their differences. Figure~\ref{appendixfig:ppmap} displays the original and combined \PPMAP temperature images, and their subtraction, for the Young protocluster W43-MM1 and the Evolved protoclusters G012.80 and W51-IRS2. The comparison of the original and combined images shows that, on average, they are very close (see, e.g., \cref{appendixfig:ppmap}). Their values deviate toward protostars and \hii regions, in agreement with the purpose of the correction of the $70\,\mu$m opacity. In more detail, when spatially averaged over the full extent of the ALMA-IMF protoclusters, the mean dust temperature of the original versus combined \PPMAP images are only $1-3$~K different. As for the temperature weighted by the column density, which can meaningfully be compared to temperature derived from SED estimates in beams covering the full extent of ALMA-IMF protoclusters, they are slightly hotter for Evolved protoclusters than Young or Intermediate protoclusters. The largest increases, $\Delta \overline{T_{\rm dust}^{\rm\PPMAP}}$[w-$N_{\rm H_2}] \simeq$10~K to $\simeq$17~K, are found for the three protoclusters, which host the most luminous protostars of ALMA-IMF: G327.29, W51-E, and W51-IRS2 (see Sect.~\ref{s:extremehotcore} and \cref{tab:luminosity peaks from radial profile}). The other protoclusters globally have their mean temperature increased by $\Delta \overline{T_{\rm dust}^{\rm\PPMAP}}$[w-$N_{\rm H_2}] \simeq 2-5$~K above $24.5\pm 1.8$~K, leading to a decrease of their total mass by only $\sim$15\%. 
 
In \cref{appendixfig:ppmap}c, displaying the difference between the combined and original image of W43-MM1, we measure a small increase in temperature over the protocluster extent from a spatially averaged temperature of $\sim$27~K (see \cref{appendixtab:ppmap table}). Both the original and combined images show filaments with temperatures 2~K lower than their surroundings, consistent with an expected self-shielding of the denser gas. Moreover, in the densest parts of the W43-MM1 cloud, the combined image presents several temperature peaks up to $\sim$46~K (see \cref{appendixfig:ppmap}c). These hot areas, either point-like or extended, are in line with the heating expected from the luminosity of their protostellar cores listed in \cref{tab:computed table1}. According to \cref{eq:tdust}, the heating radius reaching the background temperature of $T_{\rm dust}^{\rm bck}$ is 
\begin{eqnarray}
     r(T_{\rm dust}^{\rm bck}, L_{\rm proto \star}) & \simeq & 2100~\text{au} \times \left(\frac{T_{\rm dust}^{\rm bck}}{\rm 43.5~K}\right)^{-1/0.36}  \times \left(\frac{L_{\rm proto \star}}{1100~\lsun}\right)^{0.5}
\label{eq:R at Tbck},
\end{eqnarray}
and for $T_{\rm dust}^{\rm bck} \simeq 23$~K in the inner part of the W43-MM1 protocluster located at 5.5~kpc, it leads to the angular radius of
\begin{eqnarray}
       \theta(T_{\rm dust}^{\rm bck}={23~\rm K}, L_{\rm proto \star}) & \simeq & 2.2\arcsec \times \left(\frac{L_{\rm proto \star}}{1100~\lsun}\right)^{0.5}.
\label{eq:R at Tbck W43}
\end{eqnarray}
W43-MM1 presents nine heating points, of which the four main ones are extended, with radii computed from \cref{eq:R at Tbck W43} to be $\theta({\rm 23~K}) \simeq 2\arcsec-5\arcsec$ (see \cref{appendixfig:ppmap}c). Prestellar cores and low-luminosity protostellar cores in the immediate proximity of these luminous protostars should be heated by their luminosity, like observed for other cores under the influence of luminous sources \citep{sanchezmonge2013, boegner2022}. 

In Figs.~\ref{appendixfig:ppmap}f and \ref{appendixfig:ppmap}i, G012.80 and W51-IRS2 display a similar increase of temperature between the original and combined images from spatially averaged temperatures of 28--29~K (see \cref{appendixtab:ppmap table}), than in Young protoclusters. Both the original and combined images display warm gas associated to their developed \hii regions. Since the combined images are constructed to correct for the opacity of the $70\,\mu$m emission, the heating of these \hii regions is better revealed in the combined images. Evolved protoclusters show a more substantial increase of their temperature weighted spatially by the column density, $\Delta \overline{T_{\rm dust}^{\rm\PPMAP}}$[w-$N_{\rm H_2}] \simeq 3-17$~K (see \cref{appendixtab:ppmap table}). The increased heating of the high-density gas is caused by their luminous protostars, as in the Young protoclusters. G012.80 and W51-IRS2 host eight and two main heating points, respectively, associated with luminous protostars or HC\hii regions (see Figs.~\ref{appendixfig:ppmap}e-f, \ref{appendixfig:ppmap}h-i and \cref{appendixtab:ppmap table}). Using the protostellar luminosities listed in Tables~\ref{tab:computed table1}--\ref{appendixtab:computed table} in \cref{eq:R at Tbck}, we computed heating radii of $\theta({\rm 26~K}) \simeq 2.5\arcsec-3.7\arcsec$ in G012.80 and $\theta({\rm 28~K}) \simeq 18\arcsec$ in W51-IRS2 (see Figs.~\ref{appendixfig:ppmap}f, \ref{appendixfig:ppmap}i).

Ultimately, the most luminous protostars largely heat the high-density parts of the ALMA-IMF protoclusters, regardless of their evolutionary status (see \cref{appendixfig:ppmap} and \cref{appendixtab:ppmap table}). The three extreme protoclusters, G327.29, W51-E, and W51-IRS2, which host protostars with luminosities larger than $10^4~\lsun$ (see Tables~\ref{tab:luminosity peaks from radial profile} and \ref{appendixtab:computed table}), have mean temperatures, measured in the combined image and weighted by column density, above 35~K (see \cref{appendixtab:ppmap table}). From \cref{eq:tdust}, the radii of heating down to their background temperature are $\theta({\rm 22.5~K}, 9000~\lsun) \simeq 15\arcsec$ in G327.29, $\theta({\rm 23.6~K}, 32\,000~\lsun) \simeq 11\arcsec$ and $\theta({\rm 23.6~K}, 90\,000~\lsun) \simeq 19\arcsec$ in W51-E and $\theta({\rm 28~K}, 200\,000~\lsun) \simeq 18\arcsec$ in W51-IRS2. These radii cover a large part of the G327.29, W51-E, and W51-IRS2 protoclusters, and in particular their high-column density parts, as illustrated in \cref{appendixfig:ppmap}h for W51-IRS2. This alone explains the overall increase of the temperature of these three extreme protoclusters when measured in their combined images. The consistency between luminosity and temperature images supports the use of the combined dust temperature images.

\begin{figure}[htbp!]
    \centering
    \includegraphics[width=1\linewidth]{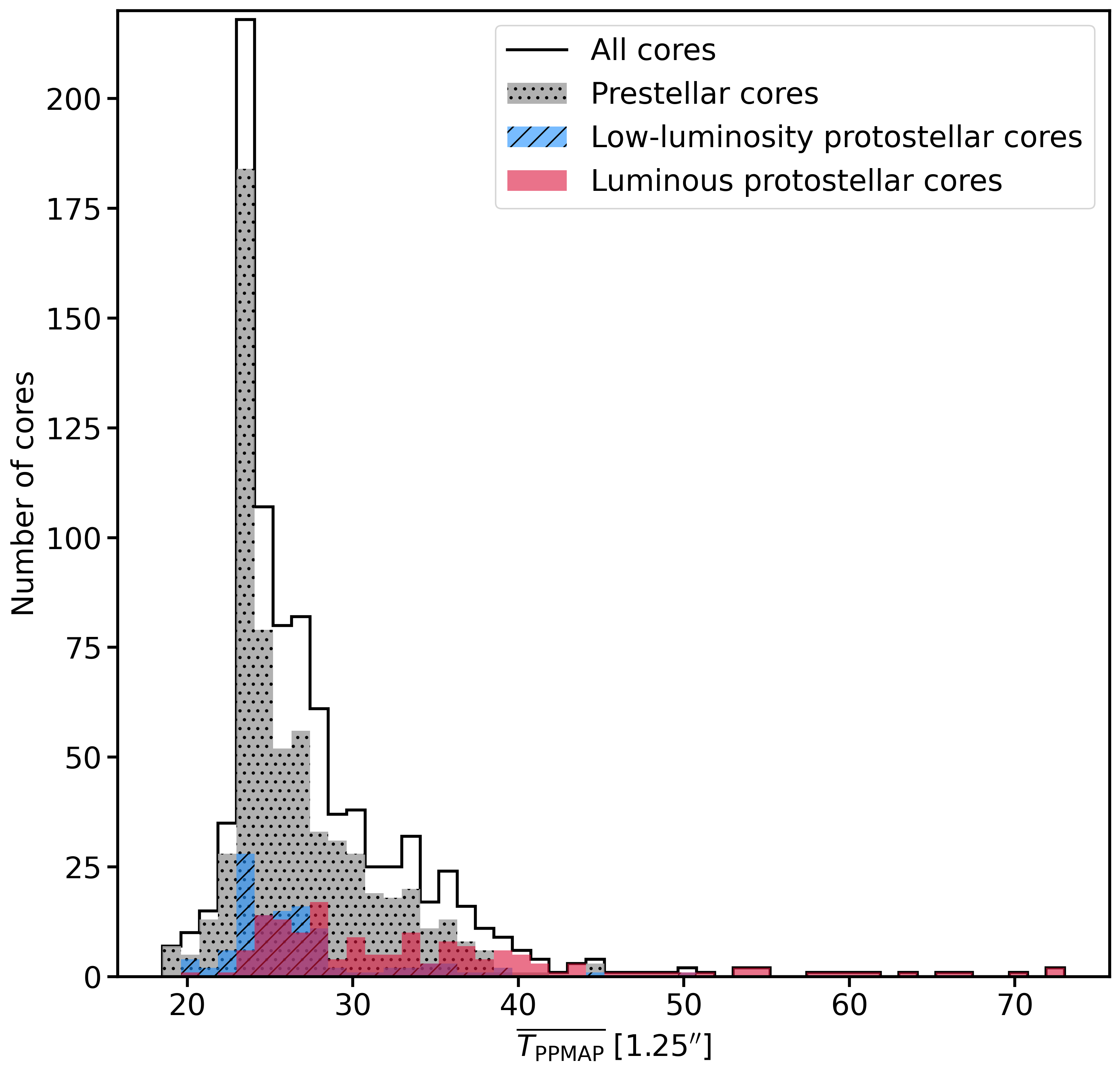}
    \vskip -0cm
    \caption{
    Distribution of the background-diluted dust temperature for protostellar cores, which are either luminous or are low-luminosity (red and blue histograms, respectively), and for prestellar cores (gray histogram). These temperatures, which approximate the background temperature of cores, are lower than those reported in \cref{fig:tdust all cores} for protostellar cores and higher for prestellar cores. While the background-diluted dust temperature of low-luminosity protostars and prestellar cores span a narrow range, $20-40$~K, the one of luminous protostars is twice larger.}
    \label{appendixfig:tbck all cores}
\end{figure}

Therefore, the combined images of dust temperature are appropriate to measure both the overall temperature of the protocluster and the background-diluted temperature of luminous protostellar cores. The background-diluted temperature of low-luminosity protostellar cores and prestellar cores is necessarily influenced by the overall cloud temperature and its local heating by luminous protostars. The original dust temperature images of, for instance, Figs.~\ref{appendixfig:ppmap}a, d, and g thus underestimate the temperature of all cores, including low-luminosity protostellar cores and prestellar cores. The opacity-corrected image, however, provides local heating within all high-density peaks, including prestellar cores \citep{dellova2024}. At the location of all prestellar cores and low-luminosity protostars, which are undetected in the \PPMAP luminosity images, we therefore averaged the temperatures measured in the original and in the opacity-corrected images. This simple average merely provides a first order estimates of their background-diluted temperature, taking into account both the underestimation of cloud heating and the overestimation of local heating. Errorbars at these locations cover the two extreme values provided by the original and opacity-corrected temperature maps.

Figure~\ref{appendixfig:tbck all cores} displays the distribution of dust temperatures measured in the combined images at the location of protostellar and prestellar cores. These temperatures are quoted as the background-diluted core temperatures and are used in Sects.~\ref{s:Tproto}--\ref{s:Tpre} (see also \cref{fig:temperature three profiles}). The background-diluted dust temperature of prestellar cores and low-luminosity protostars span similar ranges, from $\sim$20~K to $\sim$40~K. Their values are marginally increased from those measured in the original \PPMAP dust temperature images, with the notable exception of cores located close to luminous protostars (see, e.g., Figs.~\ref{appendixfig:ppmap}c and \ref{appendixfig:ppmap}i). The background-diluted temperatures of luminous protostars range from $\sim$25~K to $\sim$70~K. 

We assume in the present paper that the combined images of dust temperature are appropriate to measure both the overall temperature of the protocluster and the background-diluted temperature of cores, whatever their nature. 

\FloatBarrier

\section{Relationship between protostellar luminosity and COM line emission}
\label{appendixsect:gammaCOM}


In the ALMA-IMF pilot study, \cite{motte2018b} used the flux arising from COMs and its ratio to the continuum emission to make a first-order estimate of the luminosity of protostellar cores in the W43-MM1 protocluster. This attempt was crucial, as the only existing luminosity constraints at the time were measurements made over the $18\arcsec-36\arcsec$ beams of \textit{Herschel} in the submillimeter range \citep{nguyen2013}. Arguments in favor of a correlation between this COM intensity ratio and the protostellar luminosity were obtained by \cite{brouillet2022}, who studied this protocluster particularly rich in hot cores \citep{bonfand2024}. The molecular emission of the eight brightest hot cores of W43-MM1, when scaled to the intensities of the CH$_3$OCHO doublet lines at 216.21~GHz, turns out to be relatively homogeneous and, in detail, similar to within a factor of $2-3$ \citep{brouillet2022}.

Theoretical arguments, unfolded below, led \cite{motte2018b} to propose a proportional relationship between the protostellar luminosity of a core powering a hot core, $L_{\rm proto\star}$, and the intensity ratios of the COM line emission to continuum emission ($\gamma_{\rm COM}$, defined in \cref{appendixsect:gammaCOM def}) of this core and of the other hot cores of the protocluster. We here present their assumptions and equations to express the mass of gas heated to over 100~K, by the protostellar luminosity, as a function of $\gamma_{\rm COM}$ (see \cref{appendixsect:M100K - gammaCOM}) and $L_{\rm proto\star}$ (see \cref{appendixsect:M100K - Lproto}). By equalizing the resulting equations, \cref{appendixsect:Lproto - gammaCOM} gives a relationship between $L_{\rm proto\star}$ and $\gamma_{\rm COM}$ (see \cref{eq:lproto-gammaCOM}). The mathematical relationships derived here are not used in the present study but remain interesting. They present the methodology used in the first ALMA-IMF studies, and equations and assumptions that could be improved and discussed in future studies comparing temperature estimates based on SED and COM line analyses.

\subsection{$\gamma_{\rm COM}$ definition and measurement}
\label{appendixsect:gammaCOM def}

The intensity ratio of COM lines to continuum at 1.3~mm can be estimated by integrating the line emission measured at core center or averaged over the beam. Integrating the intensity of COM lines over a wide spectral band gives statistical significance to this intensity ratio. \cite{brouillet2022} integrated the line intensities observed over the two spectral bands of 2~GHz each  at 232.4~GHz and 233.4~GHz, from the ALMA-IMF pilot  dataset. In these two spectral bands, which cover no  molecular lines other than COMs, \cite{molet2019} identified over 100 molecular line transitions, including those from CH$_3$OH, CH$_3$OCHO, CH$_3$CN, and C$_2$H$_5$CN. The intensity ratio of COM line emission, $S_{\rm COM}$, to the continuum emission, $S_{\rm cont}$, is simply 
\begin{eqnarray}
    \gamma_{\rm COM} & = & \frac{S_{\rm COM}}{S_{\rm cont}}. \label{eq:gammaCOM}
\end{eqnarray}

While the continuum emission is assumed to represent the core mass, the brightness temperature of the line emission might be used to estimate the mass of gas radiatively heated to over $\sim$100~K (see arguments in \cref{appendixsect:M100K - gammaCOM}). If the hot core emission is not resolved by the ALMA beam, this ratio reflects the dilution of the hot core gas within the beam and hence the development of the hot core, which we assume here to be mainly heated by the protostellar luminosity. In the W43-MM1 protocluster, the intensity ratios of COM line to continuum emission range from $0.5\%$ to $50\%$ \citep{brouillet2022}. In \cref{appendixsect:M100K - gammaCOM}, we first use these $\gamma_{\rm COM}$ ratios to estimate the mass of gas, which is radiatively heated to over 100~K.

\subsection{Gas mass heated to over 100~K as a function of $\gamma_{\rm COM}$}
\label{appendixsect:M100K - gammaCOM}

The equations presented in this section are based on strong assumptions:
\begin{enumerate}
\item Firstly, we assume that COM abundances are directly and uniquely related to protostellar heating, in line with chemical models of \cite{garrod2006}. In agreement with this hypothesis, it has been observed that the abundance of certain COMs correlates with the luminosity or luminosity-to-mass ratio of high-mass protostars \citep[e.g.,][]{coletta2020}. This hypothesis rejects any other physical processes that may desorb COMs from the grain such as outflow or accretion shocks \citep[see, e.g.,][]{lefloch2017, palau2017, csengeri2019, busch2024, bouscasse2024} or form COMs in the gas phase  \citep[e.g.,][]{deSimone2020, law2021}. Furthermore, it ignores the variability of COMs production depending on their composition. For instance, is has been shown that O-bearing COMs are enriched in shocked regions and that complex cyanides are sensitive to other processes \citep{lefloch2017, palau2021, rojas2024}.
\item We also assume that continuum and COM line emissions are optically thin throughout the entire volume of the cores considered here. However, the most massive cores of ALMA-IMF are found to be partly optically thick at 1.3~mm \citep[][and see Sect.~\ref{s:mass}]{motte2018b, louvet2024}. And the hot cores in the W51-E and W51-IRS2 protoclusters also show indications of optical thickness \citep{deSimone2020, jeff2024}.
\item Finally, despite the fact that each molecular transition has its own characteristics (frequency, Einstein coefficients, energy, degeneracy, and partition function at the excitation temperature), we assume that the ensemble of all observed COM transitions in a given cloud can be statistically represented by a single line with similar median characteristics. Arguments in favor of the homogeneity of COM abundances in hot core gas are found with relatively constant abundance ratios across a wide range of physical conditions \citep{bonfand2019, brouillet2022}. In detail, however, this assumption of homogeneity is challenged when chemical differentiation is observed, particularly with O-bearing COMs and complex cyanides localized in different parts of the protostellar envelopes \citep[e.g.,][]{jimenez2012, rojas2024}. Protostars in the same protostellar group (see \cref{fig:lum-temp peaks G012-W43}) or in Young ALMA-IMF protoclusters should have formed from a relatively chemically homogeneous gas. The gas in Young protoclusters should also be pristine, that is yet unaffected by the feedback effects of stars already formed. In these environments, the relative contribution of COM lines is therefore expected to be close from one protostar to another.
\end{enumerate}

In the unlikely event that these three hypotheses hold, we would expect a correlation between the integrated flux density of COM lines, summed over a wide spectral band, and the gas mass of a protostellar envelope, assumed to be heated by the luminosity of its central protostar. Systematic biases due to optical depth effects and the different chemical origin for COMs should distort such a correlation, and at our angular resolution, could play a role that remains to be quantified. COM lines emitted by hot cores correspond to molecular transitions with excitation temperatures ranging from 50~K \citep[e.g.,][]{belloche2013, bouscasse2022} to 300~K \citep[e.g.,][]{gibb2000, ginsburg2022}, with many excitation temperatures around $\sim$100~K \citep[e.g.,][]{bonfand2017, brouillet2022}. When considering a large number of COM transitions, that is within so-called hot core line forests, the median value of the excitation temperature should therefore be statistically close to the desorption temperatures predicted for COMs in chemical models, that is $\sim$100~K \citep[e.g.,][]{garrod2006}. Moreover, since protostellar envelopes are dense regions of the interstellar medium, and if the line and continuum emissions are optically thin, the excitation temperatures correlate well with the dust temperature and can be considered equal to it.

Under these conditions, the mass of gas heated to over 100~K by the luminosity of a protostar, $M_{\rm envelope}^{\rm >100~K}$, should be statistically proportional to the integrated flux density of the COM lines, $S_{\rm COM}$, which itself is proportional to the intensity ratio between the COM line to continuum emissions, $\gamma_{\rm COM}$:
\begin{eqnarray}
M_{\rm envelope}^{\rm >100~K} \propto S_{\rm COM} = \gamma_{\rm COM} \times S_{\rm cont} .
\label{eq:mass-gammaCOM}
\end{eqnarray}
This equation can be compared to that describing the mass of gas heated to over 100~K, as a function of the protostellar luminosity (see \cref{appendixsect:M100K - Lproto}).

\subsection{Gas mass heated to over 100~K as a function of $L_{\rm proto \star}$}
\label{appendixsect:M100K - Lproto}

As in Sect.~\ref{s:Tproto-OutIn}, we here assume that protostellar envelopes are spherical, centrally heated, optically thin to infrared radiation, and well described by a power-law density profile. With these assumptions, protostellar envelopes follow the temperature profile of \cref{eq:theo-tdust}, which becomes \cref{eq:tdust} with $p=2$ and $\beta=1.5$ for the power-law indices of the density and dust emissivity laws. The radius at which the temperature of the protostellar envelope reaches $100$~K, $R_{100~\rm K}$, is then derived from \cref{eq:tdust}, according to
\begin{eqnarray}
    R_{100~\rm K}(L_{\rm proto \star}) \simeq 207~{\rm au}  \times  \left( \frac{L_{\rm proto \star}}{1100~\lsun} \right)^{0.5}.
    \label{eq:r100}
\end{eqnarray}

For a given protostellar envelope, the mass of heated gas above 100~K is therefore the integration, up to $R_{100~\rm K}$ (see \cref{eq:r100}), of its density profile assumed to be of the form $\rho(r) = \rho_0\, \left(\frac{r}{r_0}\right)^{-2}$, where $\rho_0$ is the envelope volumetric density at $r_0$ radius.
\begin{eqnarray}
    M_{\rm envelope}^{>100~\rm K} 
    & = & 4\pi \,  \rho_0 \, r_0^2 \times R_{100~\rm K}  \nonumber\\
    & = & 4\pi \,  \rho_0 \, r_0^2 \times 207~{\rm au}  \times  \left( \frac{L_{\rm proto \star}}{1100~\lsun} \right)^{0.5} .
         \label{eq:mass-r100} 
\end{eqnarray}
We introduced the mass $M_{\rm beam}$ of the protostellar envelope measured in a beam of radius $R_{\rm beam} = \theta_{\rm beam} \times d$, where $\theta_{\rm beam}$ is the half power beam width of the observation considered and $d$ is the distance to the Sun:
\begin{equation}
    M_{\rm beam}= 4\pi \, \rho_0 \, r_0^2~ R_{\rm beam} .
    \nonumber
\end{equation}
To simplify \cref{eq:mass-r100}, that becomes
\begin{eqnarray}
    M_{\rm envelope}^{>100~\rm K} & = & M_{\rm beam} \times  
    \left( \frac{\rm 207~au}{R_{\rm beam}} \right) \times  \left( \frac{L_{\rm proto \star}}{1100~\lsun} \right)^{0.5} .
\label{mass-lproto}
\end{eqnarray}

If we assume that the continuum emission at cont$\lambda=1.3$~mm is optically thin in the beam of our ALMA observations, masses integrated within the beam are computed from the following equation that, we here simplify using the Rayleigh-Jeans approximation and assuming that $d$ and $\kappa_\lambda$ will not vary from one protostar to another in a given region:
\begin{eqnarray}
    M_{\rm beam} \: & = \frac{S_{\rm cont\lambda}\; d^2}{ \kappa_\lambda\; B_\lambda(\overline{T_{\rm dust}}[R_{\rm beam}])} \nonumber \\
    & \propto S_{\rm cont\lambda} \times (\overline{T_{\rm dust}}[R_{\rm beam}])^{-1}. 
    \label{eq:mass}
\end{eqnarray}

Applying \cref{eq:tdust-fromL} to $R_{\rm beam}$, which is constant for all protostars of a given region, \cref{mass-lproto} gives the following proportional relationship between the mass of gas heated to over 100~K and the protostellar luminosity:     
\begin{eqnarray}
    M_{\rm envelope}^{>100~\rm K} 
    & \propto & S_{\rm cont\lambda} \times \left( L_{\rm proto \star} \right)^{0.5} \times \left( L_{\rm proto \star} \right)^{-0.18} \nonumber\\
    & \propto & S_{\rm cont\lambda} \times \left( L_{\rm proto \star} \right)^{0.32}.
\label{eq:mass-lproto-final}
\end{eqnarray}

As mentioned in Sect.~\ref{s:finalTdust}, these equations are strictly valid only in the case of spherical geometry. However, observations at very high spatial resolutions now indicate that protostellar envelopes subfragment and harbor disks and outflow cavities (Yoo et al. in prep.). This deviation from spherical geometry will modify the density and temperature laws within the protostellar envelope, which in turn will have a very strong impact on the chemistry expected in these various components.

\subsection{Relationship between the protostellar luminosity and the intensity ratio of the COM line to the continuum emission}
\label{appendixsect:Lproto - gammaCOM}

Combining the relationships found in Appendices~\ref{appendixsect:M100K - gammaCOM} and \ref{appendixsect:M100K - Lproto} (see Eqs.~\ref{eq:mass-gammaCOM} and \ref{eq:mass-lproto-final}) that describe the mass of gas heated to over 100~K gives a proportional relationship between $L_{\rm proto\star}$ and $\gamma_{\rm COM}$,
\begin{eqnarray}
\gamma_{\rm COM} \times S_{\rm cont} \propto S_{\rm cont\lambda} \times \left( L_{\rm proto \star} \right)^{0.32},
\nonumber
\end{eqnarray}
that leads to
\begin{eqnarray}
 L_{\rm proto \star} \propto \left(\gamma_{\rm COM}\right)^{3.1}
\label{eq:lproto-gammaCOM}
\end{eqnarray}
because the continuum flux mentioned here is the flux density of the continuum emission at 1.3~mm, as measured in a beam at the location of the protostellar core. 

This proportionality relationship is based on a very simple hot core model, whose simplistic assumptions are presented in Sect.~\ref{appendixsect:M100K - gammaCOM} and a simple protostellar envelope model, discussed in Sect.~\ref{appendixsect:M100K - Lproto}. It was interesting at the time of the ALMA-IMF pilot study of \cite{motte2018b}, and is only statistically valid for a Young region with many cores, such as W43-MM1. Its extension to all hot cores forming in ALMA-IMF protoclusters, or even more difficult in protoclusters in general, will require significant improvements to this model, notably by taking better account of some of its many limitations and even resolving some of them.


\section{Tables} \label{appendixsect:complementary tables}

\cref{appendixsect:complementary tables} presents Tables~\ref{tab:measures table}--\ref{tab:measures table evolved} and \cref{tab:computed table1}, which focus on the case study regions G012.80, W43-MM1, W43-MM2, and-MM3. Tables~\ref{appendixtab:measures table}--\ref{appendixtab:measures table evolved} and \cref{appendixtab:computed table} complement them, respectively, for the ten other ALMA-IMF regions: the Young G327.29, G328.25, G337.92, and G338.93 protoclusters, the Intermediate G008.67, G353.41, and W51-E protoclusters, and the G010.62, G333.60, and W51-IRS2 protoclusters. 

Tables~\ref{tab:measures table} and \ref{appendixtab:measures table} list, for the Young and Intermediate protoclusters, their luminosity peaks associated with protostars, characterize them and their association with temperature peaks (see Sects.~\ref{s:corecat}--\ref{s:lumcat} and \ref{s:Tpeak}, \cref{appendixfig:lum and temp peaks}). Tables~\ref{tab:measures table evolved} and \ref{appendixtab:measures table evolved} do the same for the luminosity peaks identified by the present study in Evolved protoclusters. 

As for Tables~\ref{tab:computed table1}--\ref{appendixtab:computed table}, they give the main physical properties of the protostellar cores found in the ALMA-IMF protoclusters. They list their mass-averaged temperature, mass, and luminosity, as computed in Sects.~\ref{s:Tproto} and \ref{s:finalTdust} (see also \cref{fig:temperature three profiles}), in Sect.~\ref{s:mass}, and in Sect.~\ref{s:Tproto-InOut} (see also \cref{fig:spatial weight}).

\cref{appendixtab:prestellar table} is dedicated to the prestellar core candidates found in the 14 ALMA-IMF protoclusters studied in the present study (see Sects.~\ref{s:corecat}). It focuses here on the cores more massive than $6.5~\msun$ but the complete table of 617 ALMA-IMF prestellar cores is accessible through the CDS services. The table provides the mass-averaged dust temperature and mass of prestellar cores, as computed in Sect.~\ref{s:Tpre} (see also \cref{fig:temperature three profiles}) and Sect.~\ref{s:mass}, respectively.

\begin{table*}[htbp!]
\centering
\resizebox{0.99\textwidth}{!}{
\begin{threeparttable}[c]
\caption{Luminosity peaks in Young and Intermediate W43 protoclusters and their association with temperature peaks and protostellar cores.}
\label{tab:measures table}
\begin{tabular}{lccccc|lcccc}
\hline\hline
Protocluster & Luminosity & \multirow{2}{*}{$\lbol(r<\theta_{L_{\rm bol}})$} & \multirow{2}{*}{$\theta_{L_{\rm bol}}$} & \multirow{2}{*}{FWHM$_{L_{\rm bol}}$} & Temperature & Core ID & \multirow{2}{*}{$\theta_{\rm core}$} & \multirow{2}{*}{$d_{\rm L_{\rm bol}}$} & \multirow{2}{*}{$\overline{T_{\rm dust}^{\rm PPMAP}}$[$1.25\arcsec$]} & Luminosity \\
name & peak & & & & peak & number & & & & contribution\\ 
 & & [$\times 10^3\;L_\odot$] & [$\arcsec$] & [kau] &  & & [$\arcsec$] & [$\arcsec$] & [K] & [$\%$] \\ 
(1) & (2) & (3) & (4) & (5) & (6) & (7) & (8) & (9) & (10) & (11) \\ \hline
W43-MM1     & P1     &  12.21$\pm$4.03  &  3.2 &  17.3 &   yes &    1         &   0.62 &    0.7 &   37$\pm$3   & 48     \\
            &        &                  &      &       &       &    4         &   0.82 &   0.37 &   38$\pm$3   & 52     \\ \cline{2-11}
            & P2     &   6.03$\pm$1.99  &  4.2 &  23.0 &   yes &    2         &   0.54 &   1.19 &   31$\pm$3   & 27     \\
            &        &                  &      &       &       &    5         &   0.54 &    1.4 &   28$\pm$3   & 25     \\
            &        &                  &      &       &       &   10         &   0.64 &   0.95 &   30$\pm$3   & 30     \\
            &        &                  &      &       &       &   14         &   0.47 &   1.95 &   30$\pm$3   & 19     \\ \cline{2-11}
            & P3     &   2.05$\pm$0.68  &  3.4 &  18.4 &   yes &    9         &   0.52 &   0.13 &   28$\pm$3   & 64     \\
            &        &                  &      &       &       &   12         &   0.54 &   1.51 &   26$\pm$3   & 36     \\ \cline{2-11}
            & P4     &   2.33$\pm$0.78  &  4.3 &  23.5 &   yes &    8         &   0.55 &   0.29 &   28$\pm$3   & 52     \\
            &        &                  &      &       &       &   11         &   0.50 &   0.86 &   27$\pm$3   & 48     \\ \cline{2-11}
            & P5     &   0.96$\pm$0.33  &  3.1 &  17.3 &    no &    3         &   0.62 &   1.01 &   27$\pm$3   & 50     \\
            &        &                  &      &       &       &   18         &   0.57 &   0.97 &   26$\pm$3   & 50     \\ \cline{2-11}
            & P6     &   0.64$\pm$0.22  &  3.3 &  17.9 &   yes &   15         &   0.56 &   0.47 &   25$\pm$5   & 53     \\
            &        &                  &      &       &       &   59         &   0.51 &   0.84 &   25$\pm$4   & 47     \\ \cline{2-11}
            & P7     &   0.66$\pm$0.23  &  3.8 &  20.8 &   yes &    7         &   0.51 &   0.83 &   26$\pm$3   & 100    \\ \cline{2-11}
            & P8     &   0.34$\pm$0.12  &  3.1 &  17.3 &   yes &   29         &   0.49 &   0.61 &   26$\pm$4   & 50     \\
            &        &                  &      &       &       &   51$\star$  &   0.80 &   0.63 &   26$\pm$4   & 50     \\ \cline{2-11}
            & P9     &   0.14$\pm$0.05  &  3.1 &  17.1 &    no &   67         &   0.55 &   0.26 &   26$\pm$5   & 100    \\ \cline{2-11}
            & P11    &   0.34$\pm$0.12  &  4.2 &  22.8 &   yes &   39         &   0.52 &   0.73 &   25$\pm$4   & 100    \\ \cline{2-11}
            & P21    &   0.15$\pm$0.08  &  4.3 &  23.7 &   yes &   31$\star$  &   0.50 &   0.62 &   25$\pm$4   & 66     \\
            &        &                  &      &       &       &   36$\star$  &   0.50 &   2.16 &   25$\pm$3   & 34     \\ \hline
W43-MM2     & P1     &   8.17$\pm$2.70  &  3.1 &  17.2 &   yes &    1         &   0.72 &   0.48 &   38$\pm$4   & 32     \\
            &        &                  &      &       &       &    7         &   0.78 &   0.48 &   37$\pm$4   & 32     \\
            &        &                  &      &       &       &   33         &   1.16 &   1.53 &   33$\pm$4   & 18     \\
            &        &                  &      &       &       &   78         &   0.80 &   1.56 &   33$\pm$4   & 17     \\ \cline{2-11}
            & P3     &   0.46$\pm$0.15  &  3.6 &  20.0 &   yes &    5         &   0.76 &   0.69 &   25$\pm$4   & 100    \\ \cline{2-11}
            & P4     &   0.28$\pm$0.09  &  3.0 &  16.4 &   yes &    6         &   0.57 &   0.37 &   25$\pm$4   & 100    \\ \cline{2-11}
            & P5     &   1.07$\pm$0.36  &  4.0 &  21.8 &   yes &   12         &   0.56 &   0.51 &   27$\pm$4   & 49     \\
            &        &                  &      &       &       &   28         &   0.56 &   0.11 &   27$\pm$4   & 51     \\ \cline{2-11}
            & P7     &   0.26$\pm$0.10  &  2.7 &  14.7 &   yes &   10         &   0.57 &    0.4 &   27$\pm$4   & 100    \\ \cline{2-11}
            & P8     &   0.34$\pm$0.14  &  2.9 &  15.7 &   yes &   15         &   0.53 &   0.38 &   28$\pm$4   & 100    \\ \cline{2-11}
            & P9     &   0.08$\pm$0.03  &  2.8 &  15.1 &    no &   50         &   0.56 &   0.27 &   24$\pm$4   & 100    \\ \cline{2-11}
            & P10    &   0.14$\pm$0.05  &  3.0 &  16.4 &    no &  112         &   0.51 &   0.91 &   25$\pm$4   & 46     \\
            &        &                  &      &       &       &  204         &   0.60 &   0.58 &   24$\pm$4   & 54     \\ \cline{2-11}
            & P13    &   0.02$\pm$0.01  &  2.5 &  14.0 &    no &   62         &   0.67 &   0.26 &   23$\pm$3   & 100    \\ \cline{2-11}
            & P14    &   0.04$\pm$0.01  &  2.5 &  13.7 &    no &   14         &   0.56 &   0.42 &   23$\pm$4   & 100    \\ \cline{2-11}
            & P15    &   0.05$\pm$0.02  &  2.5 &  13.8 &    no &   41         &   0.61 &   0.57 &   25$\pm$3   & 100    \\ \cline{2-11}
            & P16    &   0.05$\pm$0.02  &  2.8 &  15.4 &    no &   39$\star$  &   0.50 &   0.18 &   24$\pm$4   & 100    \\ \hline
W43-MM3     & P1$\dagger$     &   3.06$\pm$1.01  &  3.3 &  18.0 &   yes &    3         &   0.68 &   0.19 &   28$\pm$6   & 100    \\ \cline{2-11}
            & P2     &   6.12$\pm$2.03  &  3.8 &  20.8 &   yes &    2         &   0.56 &   0.57 &   32$\pm$4   & 50     \\
            &        &                  &      &       &       &    9         &   0.56 &   0.67 &   33$\pm$4   & 50     \\ \cline{2-11}
            & P6     &   2.90$\pm$0.96  &  4.0 &  22.0 &   yes &   24         &   0.61 &   1.08 &   32$\pm$4   & 100    \\ \cline{2-11}
            & P7     &   0.39$\pm$0.13  &  4.2 &  23.4 &    no &   37         &   0.52 &   0.23 &   25$\pm$6   & 59     \\
            &        &                  &      &       &       &  147         &   0.75 &   1.56 &   25$\pm$6   & 41     \\ \cline{2-11}
            & P16    &   0.07$\pm$0.03  &  3.7 &  20.6 &    no &   86$\star$  &   0.88 &   1.32 &   25$\pm$7   & 100    \\ \hline
\end{tabular}
\begin{tablenotes}[flushleft]
\item []\textbf{Notes:} (2)--(5) Luminosity peak name, integrated luminosity, and mean FWHM  size. 
Values are taken or computed from Table~B.1 of \cite{dellova2024}. 
The luminosity peak with a $\dagger$ tag is common to the W43-MM2 and W43-MM3 regions; we characterized it only once, in W43-MM3.
(6) Presence or absence of a temperature peak detected in the ellipse describing the FWHM of the luminosity peak.
(7)--(8) Protostellar cores associated with the luminosity peak of Col.~2 have their centers inside the ellipse describing the FWHM of the luminosity peak. ID numbers and FWHM angular sizes of protostellar cores are taken from the catalogs of \cite{nony2023} and \cite{pouteau2022} (see Sect.~\ref{s:corecat}). Protostellar cores with a $\star$ tag are tentative.
(9) Distance between the centers of the protostellar core and its host luminosity peak.
(10) Temperature measured in the \PPMAP dust temperature image, at core location, corresponding to the background-diluted core temperature, mass-averaged up to a $1.25\arcsec$ radius (see Sect.~\ref{s:Tbck}).
(11) Contribution level of each protostellar core to their associated luminosity peak, estimated in Sect.~\ref{s:Tproto-InOut}.
\end{tablenotes}
\end{threeparttable}}
\end{table*}

\begin{table*}[htbp!]
\centering
\resizebox{\textwidth}{!}{
\begin{threeparttable}[c]
\caption{Luminosity peaks in the Evolved G012.80 protocluster and their association with temperature peaks and protostellar cores.}
\label{tab:measures table evolved}
\begin{tabular}{lccccccc|lcccc}
\hline\hline
Protocluster & Luminosity & \multirow{2}{*}{RA} & \multirow{2}{*}{Dec} & \multirow{2}{*}{$\lbol(r<\theta_{L_{\rm bol}})$} & \multirow{2}{*}{$\theta_{L_{\rm bol}}$} & \multirow{2}{*}{FWHM$_{L_{\rm bol}}$} & Temperature & Core ID & \multirow{2}{*}{$\theta_{\rm core}$} & \multirow{2}{*}{$d_{\rm L_{\rm bol}}$} & \multirow{2}{*}{$\overline{T_{\rm dust}^{\rm PPMAP}}$[$1.25\arcsec$]} & Luminosity \\
name & peak & & & & & & peak & number & & & & contribution\\ 
 & & \multicolumn{2}{c}{[ICRS]} 
 & [$\times 10^3\;L_\odot$] & [$\arcsec$] & [kau] &  & & [$\arcsec$] & [$\arcsec$] & [K] & [$\%$] \\ 
(1) & (2a) & (2b) & (2c) & (3) & (4) & (5) & (6) & (7) & (8) & (9) & (10) & (11) \\ \hline
G012.80     & P1   & 273.54932 & -17.92576 &  0.94$\pm$0.31  &  3.1 &  7.3 & yes &   1        &  1.08 &   0.28 &   34$\pm$3   & 100 \\ \cline{2-13}
            & P2   & 273.55730 & -17.92260 &  1.37$\pm$0.46  &  3.0 &  7.3 & yes &   2        &  1.31 &   0.45 &   39$\pm$3   & 100 \\ \cline{2-13}
            & P3   & 273.55479 & -17.92796 &  1.99$\pm$0.68  &  3.4 &  8.2 & yes &   7        &  1.21 &   0.26 &   42$\pm$3   & 100 \\ \cline{2-13}
            & P5   & 273.55904 & -17.92761 &  0.99$\pm$0.33  &  3.2 &  7.6 & yes &  27$\star$ &  1.29 &   1.09 &   39$\pm$3   & 100 \\ \cline{2-13}
            & P6   & 273.54202 & -17.93270 &  0.19$\pm$0.06  &  3.0 &  7.2 & yes &  24        &  1.40 &   0.11 &   28$\pm$6   & 100 \\ \cline{2-13}
            & P7   & 273.54439 & -17.93756 &  0.14$\pm$0.05  &  2.9 &  6.9 & yes &   3        &  0.92 &   0.15 &   28$\pm$6   & 100 \\ \cline{2-13}
            & P8   & 273.55600 & -17.92715 &  1.57$\pm$0.54  &  3.5 &  8.5 & yes &  53$\star$ &  1.16 &   0.31 &   41$\pm$3   & 100 \\ \cline{2-13}
            & P9   & 273.55717 & -17.92849 &  1.44$\pm$0.51  &  3.3 &  8.0 & yes &  28        &  1.06 &   1.46 &   40$\pm$4   & 100 \\ \cline{2-13}
            & P10  & 273.54846 & -17.92645 &  0.35$\pm$0.12  &  3.0 &  7.2 &  no &  13        &  1.42 &   1.02 &   31$\pm$3   & 100 \\ \cline{2-13}
            & P11  & 273.54032 & -17.93339 &  0.23$\pm$0.08  &  4.3 & 10.4 & yes &  77        &  1.33 &   0.31 &   26$\pm$6   & 100 \\ \cline{2-13}
            & P12  & 273.55785 & -17.92095 &  0.91$\pm$0.31  &  4.0 &  9.6 & yes &  94$\star$ &  1.17 &   1.78 &   35$\pm$3   & 100 \\ \cline{2-13}
            & P14  & 273.55612 & -17.92125 &  0.46$\pm$0.16  &  3.3 &  7.9 & yes &  12$\star$ &  1.17 &   0.37 &   35$\pm$3   & 100 \\ \cline{2-13}
            & P15  & 273.55685 & -17.92379 &  0.42$\pm$0.15  &  3.0 &  7.2 & yes &  40        &  1.33 &   0.42 &   38$\pm$3   & 100 \\ \cline{2-13}
            & P16  & 273.55970 & -17.92007 &  0.93$\pm$0.33  &  4.2 & 10.1 & yes &  15$\star$ &  2.05 &   1.21 &   37$\pm$3   & 100 \\ \cline{2-13}
            & P17  & 273.55758 & -17.93096 &  0.53$\pm$0.19  &  3.4 &  8.1 & yes &  59$\star$ &  1.35 &   1.55 &   34$\pm$3   & 100 \\ \cline{2-13}
            & P19  & 273.55795 & -17.92009 &  0.39$\pm$0.15  &  3.4 &  8.3 &  no & 101$\star$ &  1.15 &   0.23 &   36$\pm$3   & 100 \\ \cline{2-13}
            & P20  & 273.55321 & -17.92079 &  0.29$\pm$0.11  &  3.4 &  8.2 & yes &   5        &  1.11 &   0.17 &   30$\pm$4   & 100 \\ \cline{2-13}
            & P21  & 273.55568 & -17.91504 &  0.07$\pm$0.03  &  2.8 &  6.7 & yes &   6        &  1.11 &   0.37 &   28$\pm$6   & 100 \\ \cline{2-13}
            & P23  & 273.54701 & -17.93328 &  0.08$\pm$0.03  &  2.9 &  7.0 & yes &  19        &  1.26 &   0.47 &   29$\pm$6   & 100 \\ \cline{2-13}
            & P24  & 273.55333 & -17.92807 &  0.50$\pm$0.18  &  3.6 &  8.7 & yes &  17$\star$ &  1.69 &   0.31 &   38$\pm$4   & 100 \\ \cline{2-13}
            & P25  & 273.55260 & -17.92854 &  0.29$\pm$0.11  &  2.7 &  6.6 & yes &  11$\star$ &  1.04 &   0.31 &   39$\pm$4   & 100 \\ \cline{2-13}
            & P28  & 273.54595 & -17.92846 &  0.23$\pm$0.09  &  4.1 &  9.8 & yes &  41        &  1.07 &   0.55 &   30$\pm$4   & 55  \\
            &      &           &           &                 &      &      &     &  51$\star$ &  1.22 &   1.19 &   30$\pm$4   & 45  \\ \cline{2-13}
            & P29  & 273.55422 & -17.91412 &  0.03$\pm$0.01  &  2.8 &  6.6 &  no &   9$\star$ &  1.00 &   0.18 &   26$\pm$6   & 100 \\ \cline{2-13}
            & P30  & 273.54443 & -17.93034 &  0.11$\pm$0.04  &  3.7 &  8.9 & yes &   8$\star$ &  1.26 &   0.24 &   30$\pm$5   & 100 \\ \cline{2-13}
            & P34  & 273.56901 & -17.92472 &  0.03$\pm$0.02  &  2.7 &  6.4 &  no &  22        &  1.24 &    0.1 &   28$\pm$6   & 100 \\ \hline
\end{tabular}
\begin{tablenotes}
\item []\textbf{Notes:} (2)--(5) Luminosity peak name, RA and Dec peak coordinates, integrated luminosity, and mean FWHM angular size. Values are taken and computed from the catalog obtained in Sect.~\ref{s:lumcat}. (7)--(8) Protostellar cores associated with the luminosity peak of Col.~2 have their centers inside the ellipse describing the FWHM of the
luminosity peak. ID numbers and FWHM angular sizes of protostellar cores are taken from the catalog of \cite{armante2024} (see Sect.~\ref{s:corecat}). Protostellar cores with a $\star$ tag are tentative.
For other columns see \cref{tab:measures table}.
\end{tablenotes}
\end{threeparttable}}
\end{table*}

\begin{table*}[htbp!]
\centering
\resizebox{\textwidth}{!}{
\begin{threeparttable}[c]
\caption{Luminosity peaks and their association with temperature peaks and protostellar cores in the Young G327.29, G328.25, G337.92, and G338.93 protoclusters and in the Intermediate G008.67, G353.41, and W51-E protoclusters.}
\label{appendixtab:measures table}
\begin{tabular}{lccccc|lccccc}
\hline\hline
Protocluster & Luminosity & \multirow{2}{*}{$\lbol(r<\theta_{L_{\rm bol}})$} & \multirow{2}{*}{$\theta_{L_{\rm bol}}$} & \multirow{2}{*}{FWHM$_{L_{\rm bol}}$} & Temperature & Core ID & \multirow{2}{*}{$\theta_{\rm core}$} & \multirow{2}{*}{$d_{\rm L_{\rm bol}}$} & \multirow{2}{*}{$\overline{T_{\rm dust}^{\rm PPMAP}}$[$1.25\arcsec$]} & Luminosity \\
name & peak & & & & peak & number & & & & contribution\\ 
 & & [$\times 10^3\;L_\odot$] & [$\arcsec$] & [kau] &  & & [$\arcsec$] & [$\arcsec$] & [K] & [$\%$] \\ 
(1) & (2) & (3) & (4) & (5) & (6) & (7) & (8) & (9) & (10) & (11) \\ \hline
G008.67     & P1     &   5.69$\pm$1.88  &  3.5 &  11.8 &   yes &    1         &   1.27 &   0.53 &   36$\pm$3   & 100    \\ \cline{2-11}
            & P2     &   1.86$\pm$0.61  &  3.4 &  11.7 &   yes &    2         &   0.75 &   0.65 &   33$\pm$11  & 100    \\ \cline{2-11}
            & P7     &   0.19$\pm$0.07  &  3.2 &  11.0 &    no &    7         &   0.86 &   0.46 &   25$\pm$9   & 100    \\ \hline
G327.29     & P1     &   5.34$\pm$1.76  &  3.7 &   9.1 &   yes &    1         &   1.33 &   0.48 &   67$\pm$6   & 29     \\
            &        &                  &      &       &       &    2$\star$  &   0.98 &   0.36 &   66$\pm$6   & 30     \\
            &        &                  &      &       &       &    6$\star$  &   1.46 &   1.47 &   55$\pm$6   & 20     \\
            &        &                  &      &       &       &    9$\star$  &   1.31 &   1.27 &   54$\pm$6   & 22     \\ \cline{2-11}
            & P2     &   0.39$\pm$0.13  &  3.4 &   8.4 &   yes &   11         &   0.86 &   0.29 &   27$\pm$5   & 100    \\ \cline{2-11}
            & P3     &   0.63$\pm$0.21  &  4.1 &  10.2 &   yes &    7         &   1.13 &   0.77 &   28$\pm$4   & 58     \\
            &        &                  &      &       &       &   12$\star$  &   0.90 &   1.58 &   26$\pm$4   & 42     \\ \cline{2-11}
            & P4     &   1.37$\pm$0.46  &  3.7 &   9.3 &   yes &   16         &   0.93 &   0.35 &   36$\pm$5   & 100    \\ \cline{2-11}
            & P5     &   0.22$\pm$0.07  &  3.1 &   7.7 &   yes &    5$\star$  &   0.77 &   0.35 &   28$\pm$4   & 100    \\ \cline{2-11}
            & P6     &   0.19$\pm$0.06  &  3.1 &   7.6 &   yes &   19         &   0.95 &   0.13 &   27$\pm$5   & 100    \\ \cline{2-11}
            & P7     &   0.60$\pm$0.21  &  3.5 &   8.7 &   yes &    4         &   1.39 &   0.61 &   32$\pm$4   & 100    \\ \cline{2-11}
            & P9     &   0.43$\pm$0.16  &  3.2 &   8.1 &    no &   23         &   0.70 &   0.27 &   33$\pm$4   & 100    \\ \cline{2-11}
            & P11    &   0.07$\pm$0.02  &  3.6 &   8.9 &   yes &   25$\star$  &   0.76 &   0.29 &   26$\pm$5   & 100    \\ \cline{2-11}
            & P13    &   0.12$\pm$0.04  &  3.3 &   8.2 &   yes &   15$\star$  &   0.90 &   0.65 &   27$\pm$4   & 100    \\ \cline{2-11}
            & P16    &   0.04$\pm$0.02  &  3.1 &   7.8 &   yes &   24         &   0.82 &   1.17 &   25$\pm$4   & 100    \\ \cline{2-11}
            & P19    &   0.02$\pm$0.01  &  2.8 &   7.1 &   yes &   14$\star$  &   0.77 &   0.71 &   26$\pm$4   & 100    \\ \hline
G328.25     & P1     &   3.62$\pm$1.20  &  3.5 &   8.8 &   yes &    1         &   0.86 &   0.54 &   44$\pm$8   & 100    \\ \hline
G337.92     & P1     &   8.42$\pm$2.78  &  3.6 &  13.0 &   yes &    1         &   0.85 &   0.94 &   51$\pm$4   & 50     \\
            &        &                  &      &       &       &    2         &   0.80 &   0.98 &   49$\pm$4   & 50     \\ \cline{2-11}
            & P2     &   5.12$\pm$1.70  &  5.1 &  18.2 &    no &    5$\star$  &   0.77 &   0.69 &   39$\pm$4   & 36     \\
            &        &                  &      &       &       &   14         &   0.63 &    1.6 &   39$\pm$4   & 28     \\
            &        &                  &      &       &       &   31$\star$  &   1.04 &   0.56 &   39$\pm$4   & 36     \\ \cline{2-11}
            & P6     &   0.05$\pm$0.02  &  2.8 &   9.9 &   yes &    3         &   0.63 &   0.38 &   22$\pm$5   & 100    \\ \cline{2-11}
            & P9     &   0.10$\pm$0.04  &  3.5 &  12.7 &   yes &   26         &   0.61 &   0.94 &   26$\pm$4   & 100    \\ \hline
G338.93     & P1     &   6.69$\pm$2.21  &  3.5 &  13.5 &   yes &    3         &   0.65 &   0.75 &   40$\pm$7   & 37     \\
            &        &                  &      &       &       &    6$\star$  &   0.83 &   0.32 &   40$\pm$8   & 42     \\
            &        &                  &      &       &       &   20$\star$  &   0.90 &   1.73 &   37$\pm$8   & 21     \\ \cline{2-11}
            & P2     &   2.10$\pm$0.69  &  2.9 &  11.4 &   yes &    1         &   0.67 &   0.34 &   30$\pm$8   & 54     \\
            &        &                  &      &       &       &   11$\star$  &   0.61 &    0.8 &   28$\pm$8   & 46     \\ \cline{2-11}
            & P3     &   3.99$\pm$1.32  &  3.8 &  14.9 &    no &   10         &   0.78 &   0.46 &   37$\pm$8   & 100    \\ \cline{2-11}
            & P4     &   1.57$\pm$0.52  &  3.6 &  14.1 &    no &    4         &   0.66 &   0.55 &   34$\pm$8   & 100    \\ \cline{2-11}
            & P10    &   0.08$\pm$0.03  &  4.6 &  17.8 &    no &   25         &   1.02 &   1.78 &   24$\pm$7   & 100    \\ \hline
G353.41     & P1     &   0.30$\pm$0.10  &  3.1 &   6.2 &   yes &    4         &   1.22 &   0.72 &   28$\pm$6   & 100    \\ \cline{2-11}
            & P2     &   0.82$\pm$0.27  &  4.3 &   8.5 &   yes &    8$\star$  &   1.33 &   0.49 &   37$\pm$4   & 100    \\ \cline{2-11}
            & P3     &   0.20$\pm$0.07  &  3.2 &   6.5 &   yes &    2$\star$  &   1.04 &    0.4 &   29$\pm$5   & 58     \\
            &        &                  &      &       &       &   14$\star$  &   0.98 &   1.15 &   28$\pm$5   & 42     \\ \cline{2-11}
            & P4     &   0.40$\pm$0.13  &  3.8 &   7.6 &   yes &    1         &   1.03 &   1.13 &   32$\pm$4   & 100    \\ \cline{2-11}
            & P5     &   0.17$\pm$0.06  &  3.7 &   7.4 &    no &   27$\star$  &   1.40 &   0.54 &   30$\pm$5   & 100    \\ \cline{2-11}
            & P7     &   0.14$\pm$0.05  &  3.5 &   7.0 &   yes &    5         &   0.95 &    0.3 &   26$\pm$6   & 100    \\ \cline{2-11}
            & P8     &   0.08$\pm$0.03  &  3.0 &   6.0 &   yes &    3         &   0.94 &    0.5 &   30$\pm$6   & 100    \\ \hline
W51-E       & P1     &  74.84$\pm$24.70 &  3.5 &  19.0 &   yes &    4         &   0.56 &   0.52 &   61$\pm$8   & 60     \\
            &        &                  &      &       &       &   25$\star$  &   0.42 &   1.43 &   54$\pm$8   & 40     \\ \cline{2-11}
            & P2     &  73.74$\pm$24.34 &  3.3 &  17.9 &   yes &    3         &   0.67 &   0.66 &   64$\pm$8   & 42     \\
            &        &                  &      &       &       &    6$\star$  &   0.45 &   1.29 &   58$\pm$8   & 31     \\
            &        &                  &      &       &       &   10$\star$  &   0.59 &   1.52 &   55$\pm$8   & 26     \\ \cline{2-11}
            & P4     &   3.87$\pm$1.30  &  4.0 &  21.7 &   yes &   26$\star$  &   0.78 &   0.17 &   25$\pm$8   & 100    \\ \cline{2-11}
            & P6     &   7.84$\pm$2.63  &  5.4 &  28.9 &    no &   11         &   0.46 &   1.47 &   32$\pm$9   & 34     \\
            &        &                  &      &       &       &   13         &   0.42 &   0.59 &   34$\pm$9   & 41     \\
            &        &                  &      &       &       &   21         &   0.50 &    2.3 &   37$\pm$8   & 25     \\ \hline
\end{tabular}
\begin{tablenotes}[flushleft]
\item []\textbf{Notes:} (2)--(5) Luminosity peak name, integrated luminosity, and mean FWHM angular size. Values are taken or computed from Table~B.1 of \cite{dellova2024}. 
(6)  Presence or absence of a temperature peak detected in the ellipse describing the FWHM of the luminosity peak.
(7)--(8) Protostellar cores associated with the luminosity peak of Col.~2 have their centers inside the ellipse describing the FWHM of the luminosity peak. ID numbers and FWHM angular sizes of protostellar cores are taken from the catalog of \cite{louvet2024} (see Sect.~\ref{s:corecat}). Protostellar cores with a $\star$ tag are tentative.
(9) Distance between the centers of the protostellar core and its host luminosity peak.
(10) Temperature measured in the \PPMAP dust temperature image, at core location, corresponding to the background-diluted temperature of cores, mass-averaged up to a $1.25\arcsec$ radius (see Sect.~\ref{s:Tbck}).
(11) Contribution level of each protostellar core to their associated luminosity peak, estimated in Sect.~\ref{s:Tproto-InOut}.
\end{tablenotes}
\end{threeparttable}}
\end{table*}

\begin{table*}[htbp!]
\centering
\resizebox{\textwidth}{!}{
\begin{threeparttable}[c]
\caption{Luminosity peaks and their association with temperature peaks and protostellar cores in the Evolved G010.62, G333.60, and W51-IRS2 protoclusters.
}
\label{appendixtab:measures table evolved}
\begin{tabular}{lccccccc|lcccc}
\hline\hline
Protocluster & Luminosity & \multirow{2}{*}{RA} & \multirow{2}{*}{Dec} & \multirow{2}{*}{$\lbol(r<\theta_{L_{\rm bol}})$} & \multirow{2}{*}{$\theta_{L_{\rm bol}}$} & \multirow{2}{*}{FWHM$_{L_{\rm bol}}$} & Temperature & Core ID & \multirow{2}{*}{$\theta_{\rm core}$} & \multirow{2}{*}{$d_{\rm L_{\rm bol}}$} & \multirow{2}{*}{$\overline{T_{\rm dust}^{\rm PPMAP}}$[$1.25\arcsec$]} & Luminosity \\
name & peak & & & & & & peak & number & & & & contribution\\ 
 & & \multicolumn{2}{c}{[ICRS]}
 & [$\times 10^3\;L_\odot$] & [$\arcsec$] & [kau] &  & & [$\arcsec$] & [$\arcsec$] & [K] & [$\%$] \\ 
(1) & (2a) & (2b) & (2c) & (3) & (4) & (5) & (6) & (7) & (8) & (9) & (10) & (11) \\ \hline
G010.62     & P1   & 272.61950 & -19.93054 & 18.04$\pm$5.97  &  4.3 & 21.1 & yes &   1        &  1.06 &   0.51 &   48$\pm$7   & 40  \\
            &      &           &           &                 &      &      &     &  19        &  0.71 &   0.27 &   47$\pm$7   & 41  \\
            &      &           &           &                 &      &      &     &  22        &  0.70 &    2.3 &   46$\pm$7   & 19  \\ \cline{2-13}
            & P2   & 272.63033 & -19.92245 &  0.31$\pm$0.10  &  3.1 & 15.4 &  no &  15        &  0.50 &   0.23 &   20$\pm$5   & 100 \\ \cline{2-13}
            & P4   & 272.62108 & -19.92943 &  6.01$\pm$2.15  &  4.1 & 20.5 & yes &  18        &  0.55 &   1.28 &   41$\pm$5   & 32  \\
            &      &           &           &                 &      &      &     &  21        &  0.66 &   1.22 &   40$\pm$5   & 32  \\
            &      &           &           &                 &      &      &     &  23        &  0.48 &   0.91 &   40$\pm$5   & 36  \\ \cline{2-13}
            & P5   & 272.62184 & -19.92800 &  1.37$\pm$0.60  &  2.9 & 14.4 & yes &   3        &  0.51 &   0.15 &   36$\pm$5   & 100 \\ \cline{2-13}
            & P6   & 272.62233 & -19.93008 &  0.76$\pm$0.49  &  2.6 & 12.8 &  no &  29$\star$ &  0.70 &   0.24 &   36$\pm$5   & 100 \\ \hline
G333.60     & P1   & 245.54694 & -50.09874 & 12.55$\pm$4.16  &  3.2 & 13.5 & yes &   1$\star$ &  1.05 &   0.44 &   50$\pm$12  & 100 \\ \cline{2-13}
            & P3   & 245.51733 & -50.10861 &  1.07$\pm$0.36  &  3.5 & 14.9 & yes &  24        &  0.69 &   1.56 &   27$\pm$12  & 40  \\
            &      &           &           &                 &      &      &     &  89$\star$ &  0.69 &   0.73 &   28$\pm$12  & 60  \\ \cline{2-13}
            & P4   & 245.52836 & -50.10492 &  2.73$\pm$0.94  &  3.0 & 12.7 & yes &   4        &  0.72 &   0.52 &   36$\pm$13  & 62  \\
            &      &           &           &                 &      &      &     &   5$\star$ &  0.63 &   1.37 &   36$\pm$13  & 38  \\ \cline{2-13}
            & P5   & 245.53612 & -50.10325 &  3.64$\pm$1.26  &  3.7 & 15.5 & yes &   8        &  1.15 &   1.14 &   43$\pm$12  & 100 \\ \cline{2-13}
            & P6   & 245.53540 & -50.10487 &  2.80$\pm$1.02  &  3.9 & 16.5 & yes &   2        &  0.75 &   0.53 &   41$\pm$13  & 100 \\ \hline
W51-IRS2    & P1   & 290.91684 &  14.51820 & 29.63$\pm$9.79  &  3.4 & 18.6 & yes &   1        &  0.66 &   0.13 &   73$\pm$10  & 39  \\
            &      &           &           &                 &      &      &     &   3        &  0.59 &   1.28 &   70$\pm$10  & 27  \\
            &      &           &           &                 &      &      &     &   6        &  0.72 &   0.76 &   73$\pm$10  & 34  \\ \cline{2-13}
            & P2   & 290.91558 &  14.51820 & 22.66$\pm$7.49  &  3.6 & 19.3 & yes &   5        &  0.68 &   1.29 &   60$\pm$9   & 41  \\
            &      &           &           &                 &      &      &     &   8        &  0.60 &   0.26 &   59$\pm$9   & 59  \\ \cline{2-13}
            & P4   & 290.91076 &  14.51156 &  4.83$\pm$1.61  &  3.4 & 18.2 & yes &   4        &  0.64 &    0.2 &   36$\pm$8   & 100 \\ \cline{2-13}
            & P5   & 290.92418 &  14.51484 &  8.94$\pm$2.96  &  4.3 & 23.5 & yes &  25        &  0.56 &   1.48 &   35$\pm$7   & 100 \\ \cline{2-13}
            & P6   & 290.90913 &  14.50917 &  2.29$\pm$0.77  &  3.2 & 17.3 & yes &  20        &  0.60 &   0.25 &   31$\pm$8   & 100 \\ \cline{2-13}
            & P7   & 290.91443 &  14.51766 &  6.83$\pm$2.28  &  3.1 & 16.9 &  no &  34        &  0.58 &   0.84 &   43$\pm$6   & 100 \\ \cline{2-13}
            & P8   & 290.90985 &  14.51033 &  2.57$\pm$0.87  &  3.4 & 18.5 & yes &  17        &  0.82 &   0.79 &   31$\pm$8   & 100 \\ \cline{2-13}
            & P9   & 290.91128 &  14.51264 &  2.41$\pm$0.81  &  3.1 & 16.6 & yes &  11        &  0.55 &   0.34 &   31$\pm$8   & 52  \\
            &      &           &           &                 &      &      &     &  51$\star$ &  0.75 &   0.59 &   29$\pm$9   & 48  \\ \cline{2-13}
            & P12  & 290.90507 &  14.51843 &  0.82$\pm$0.27  &  3.6 & 19.7 & yes &  98        &  0.59 &   1.23 &   27$\pm$8   & 100 \\ \cline{2-13}
            & P13  & 290.92359 &  14.51721 &  3.12$\pm$1.05  &  3.3 & 17.9 & yes &  41        &  0.76 &   0.31 &   34$\pm$9   & 100 \\ \cline{2-13}
            & P18  & 290.90829 &  14.50836 &  0.83$\pm$0.28  &  3.4 & 18.2 &  no &  65$\star$ &  0.63 &   1.68 &   28$\pm$8   & 34  \\
            &      &           &           &                 &      &      &     & 106$\star$ &  0.80 &   0.29 &   28$\pm$8   & 66  \\ \hline
\end{tabular}
\begin{tablenotes}
\item []\textbf{Notes:} (2)--(5) Luminosity peak name, RA and Dec peak coordinates, integrated luminosity, and mean FWHM angular size. Values are taken and computed from the catalog obtained in Sect.~\ref{s:lumcat}. For other columns see \cref{appendixtab:measures table}.
\end{tablenotes}
\end{threeparttable}}
\end{table*}

\begin{table*}[hbtp!]
\centering
\resizebox{\textwidth}{!}{
\begin{threeparttable}[c]
\caption{Luminosity and mass-averaged temperature of the protostellar cores driving the six brightest hot cores of ALMA-IMF.}
\label{tab:luminosity peaks from radial profile}
\begin{tabular}{l|ccc|ccc|ccc|ccc}
    \hline\hline
    Protocluster & \multicolumn{3}{c|}{Profile center} &   \multicolumn{3}{c|}{Hot core luminosity estimates} &     \multicolumn{3}{c|}{Protostellar core} &     \multicolumn{3}{c}{$\overline{T_{\rm dust}}$ estimates from} \\
    name & name & RA & Dec &     $L_{\rm MF}({\rm peak})$ &     $L_{\rm MF}({\rm pow})$ & $L_{\rm MF}({\rm tot})$ &      ID &  $L_{\rm MF}$ & $L_{\rm proto\star}(\mathrm{pow})$ & $L_{\rm proto\star}[R_{\rm out}]$     &  $L_{\rm proto\star}(\mathrm{pow})$ & $L_{\rm proto\star}(\mathrm{tot})$ \\ 
     & & \multicolumn{2}{c|}{[ICRS]}  & \multicolumn{3}{c|}{[$\times 10^3~\lsun$]} &     number &  contrib & [$\times 10^3~\lsun$] & \multicolumn{3}{c}{[K]} \\
    (1) & (2a) & (2b) & (2c) & (3a) & (3b) & (3c) & (4) & (5) & (6) & (7a)  & (7b) & (7c)\\ \hline
    G327.29     & MF1     & 15:53:07.79 & $-$54:37:06.40 & $3.1\pm0.5$ & $9\pm3$ & $46\pm8$ &  1  & $50\%$ & $4.5\pm1.5$ & $64\pm4$   & $78\pm5$ & $104\pm3$ \\ 
                & & & & & & & 2$\star$ & $50\%$ &  $4.5\pm1.5$ & $77\pm5$ & $93\pm6$ & $124\pm4$ \\ \hline
    W51-E       & MF1=e2     & 19:23:43.97 & +14:30:34.50   & $31\pm10$ & $32 \pm 6$ & $290\pm50$ & 3    & $100\%$ & $32 \pm 6$ & $106\pm6$   & $107\pm4$ & $158\pm5$ \\ 
                & MF2\&3  & 19:23:43.87 & +14:30:27.60   & $75\pm25$ & $90 \pm 30$ & $230\pm40$ &      4   & $60\%$ &   $54\pm 18$ &$123\pm7$ & $127\pm8$  & $151\pm5$ \\
                & & & & & & & 25$\star$  & $40\%$ & $36\pm12$ & $137\pm8$  & $142\pm9$ & $168\pm5$                \\  \hline
    W51-IRS2    & MF1\&3  & 19:23:39.91 & +14:31:05:02   & $30\pm10$ & $200\pm30$ & $470\pm80$ &1  & $39\%$ & $80\pm10$ & $97\pm6$  & $137\pm4$ & $160\pm5$ \\
               & & & & & & & 3  & $27\%$ & $54\pm8$ & $101\pm6$  & $143\pm4$ & $167\pm5$                \\ 
               & & & & & & & 6  & $34\%$ & $68\pm10$ & $90\pm5$  & $127\pm3$ & $148\pm5$ \\ \hline

\end{tabular}
\begin{tablenotes}
    \item []\textbf{Notes:}
(2a)--(2c) Name, RA and Dec coordinates of the four locations around which the luminosity profiles, presented in \cref{appendixfig:luminosity profile}, are built to characterize the six brightest methyl formate (MF) sources of the ALMA-IMF survey \citep{bonfand2024}, encompassing eight protostellar cores. 
(3a)--(3c) Three luminosity estimates for these extreme hot cores, derived from the luminosity peak (lower limit), integration of the power-law component of the profile, and total integration of the profile (upper limit).
(4)--(6) The eight protostellar cores, located within the extent of the six brightest MF sources of ALMA-IMF, their contribution to the hot core luminosity, and their luminosity computed from Cols.~3b and 5 in \cref{eq:spatial luminosity}. Protostellar cores with a $\star$ tag are tentative.
(7a)--(7c) Dust temperature of the protostellar cores, which is estimated from Cols.~3a--3c, Col.~5 and the core size of \cref{appendixtab:computed table}.
\end{tablenotes}
\end{threeparttable}}
\end{table*}

\begin{table*}[htbp!]
\centering
\resizebox{0.9\linewidth}{!}{
\begin{threeparttable}[c]
\caption{Main characteristics of protostellar cores in the G012.80, W43-MM1, W43-MM2, and W43-MM3 protoclusters.}
\label{tab:computed table1}
\begin{tabular}{lccccc|ccc}
    \hline\hline
    Protocluster & Core ID & \multirow{2}{*}{$R_{\rm out}$} & \multirow{2}{*}{$L_{\rm proto \star}$[$R_{\rm out}$]} &
    \multicolumn{2}{c|}{$\overline{T_{\rm dust}}(L_{\rm proto \star})$[$R_{\rm out}$]}
    & \multirow{2}{*}{$\overline{T_{\rm dust}}\pm 1\sigma$} & \multirow{2}{*}{$M_{\rm core}$[$\overline{T_{\rm dust}}$]} & \multirow{2}{*}{Comments} \\
    name & number & & & from \PPMAP \tdust & from $L_{\rm proto\star}$& & & \\ 
     & & [au] & [$\times 10^3$~\lsun] & [K] & [K] & [K] & [\msun] & \\
     (1) & (2) & (3) & (4) & (5) & (6) & (7) & (8) &  (9)\\ \hline
G012.80     &    1         & 1520  &     0.94$\pm$0.31  &    43$\pm$4   &   74$\pm$4   &    59$\pm$15  &     9.5$\pm$2.9  & O, hot-core \\
            &    2         & 2320  &     1.37$\pm$0.46  &    43$\pm$3   &   68$\pm$4   &    55$\pm$13  &     9.1$\pm$0.5  & O, hot-core \\
            &    3         & 1050  &     0.14$\pm$0.05  &    41$\pm$9   &   60$\pm$4   &    50$\pm$10  &     1.4$\pm$0.1  & O \\
            &    5         & 1640  &     0.29$\pm$0.11  &    37$\pm$5   &   58$\pm$4   &    48$\pm$11  &     2.7$\pm$0.2  & O \\
            &    6         & 1640  &     0.07$\pm$0.03  &    35$\pm$7   &   45$\pm$3   &    40$\pm$5   &     1.1$\pm$0.1  & O \\
            &    7         & 2000  &     1.99$\pm$0.68  &    49$\pm$3   &   77$\pm$5   &    63$\pm$14  &     3.4$\pm$0.4  & O, hot-core, FF corrected \\
            &    8$\star$  & 2170  &     0.11$\pm$0.04  &    34$\pm$6   &   44$\pm$3   &    39$\pm$5   &     1.7$\pm$0.2  & O \\
            &    9$\star$  & 1140  &     0.03$\pm$0.01  &    37$\pm$9   &   44$\pm$4   &    41$\pm$4   &     0.6$\pm$0.1  & O \\
            &   11$\star$  & 1340  &     0.29$\pm$0.11  &    52$\pm$5   &   63$\pm$4   &    57$\pm$5   &     0.9$\pm$0.1  & O \\
            &   12$\star$  & 1880  &     0.46$\pm$0.16  &    41$\pm$4   &   60$\pm$4   &    51$\pm$9   &     1.5$\pm$0.1  & O \\
            &   13         & 2690  &     0.35$\pm$0.12  &    32$\pm$3   &   50$\pm$3   &    41$\pm$9   &     5.2$\pm$0.6  & O, hot-core \\
            &   15$\star$  & 4450  &     0.93$\pm$0.33  &    32$\pm$3   &   50$\pm$3   &    41$\pm$9   &     4.4$\pm$0.5  & O \\
            &   17$\star$  & 3480  &     0.50$\pm$0.18  &    36$\pm$4   &   49$\pm$3   &    43$\pm$7   &     1.0$\pm$0.2  & O \\
            &   19         & 2180  &     0.08$\pm$0.03  &    33$\pm$7   &   42$\pm$3   &    37$\pm$5   &     0.8$\pm$0.1  & O \\
            &   22         & 2110  &     0.03$\pm$0.02  &    32$\pm$7   &   35$\pm$3   &    34$\pm$3   &     0.8$\pm$0.1  & O \\
            &   24         & 2630  &     0.19$\pm$0.06  &    29$\pm$6   &   46$\pm$3   &    37$\pm$8   &     1.4$\pm$0.2  & O \\
            &   27$\star$  & 2280  &     0.99$\pm$0.33  &    43$\pm$3   &   65$\pm$4   &    54$\pm$11  &     1.9$\pm$0.3  & O, FF corrected \\
            &   28         & 1430  &     1.44$\pm$0.51  &    52$\pm$5   &   82$\pm$5   &    67$\pm$15  &     0.7$\pm$0.2  & O, FF corrected \\
            &   36$\star$  & 1340  &                    &    43$\pm$4   &              &    43$\pm$4   &     1.2$\pm$0.2  & O \\
            &   40         & 2390  &     0.42$\pm$0.15  &    41$\pm$3   &   54$\pm$4   &    48$\pm$7   &     1.2$\pm$0.2  & O \\
            &   41         & 1470  &     0.13$\pm$0.05  &    39$\pm$5   &   52$\pm$3   &    45$\pm$7   &     0.9$\pm$0.2  & O \\
            &   43$\star$  & 2600  &                    &    29$\pm$4   &              &    29$\pm$4   &     2.0$\pm$0.4  & O \\
            &   51$\star$  & 2050  &     0.10$\pm$0.04  &    34$\pm$5   &   45$\pm$3   &    40$\pm$5   &     1.1$\pm$0.2  & O \\
            &   53$\star$  & 1840  &     1.57$\pm$0.54  &    49$\pm$4   &   76$\pm$5   &    62$\pm$13  &     0.9$\pm$0.2  & O, FF corrected \\
            &   59$\star$  & 2470  &     0.53$\pm$0.19  &    36$\pm$3   &   56$\pm$4   &    46$\pm$10  &     1.4$\pm$0.2  & O \\
            &   77         & 2390  &     0.23$\pm$0.08  &    28$\pm$7   &   49$\pm$3   &    39$\pm$10  &     0.4$\pm$0.1  & O \\
            &   79         & 1900  &                    &    38$\pm$3   &              &    38$\pm$3   &     1.0$\pm$0.2  & O \\
            &   84$\star$  & 1050  &                    &    40$\pm$5   &              &    40$\pm$5   &     0.3$\pm$0.2  & O \\
            &   94$\star$  & 1850  &     0.91$\pm$0.31  &    42$\pm$4   &   69$\pm$4   &    55$\pm$13  &     0.5$\pm$0.2  & O \\
            &  101$\star$  & 1800  &     0.39$\pm$0.15  &    43$\pm$4   &   59$\pm$4   &    51$\pm$8   &     0.3$\pm$0.1  & O \\ \hline
W43-MM1     &    1         & 2520  &     5.80$\pm$1.91  &    53$\pm$4   &   86$\pm$5   &    69$\pm$16  &   115.2$\pm$49.6 & O, hot-core \\
            &    2         & 1830  &     1.63$\pm$0.54  &    50$\pm$5   &   77$\pm$5   &    63$\pm$13  &    50.8$\pm$2.6  & O, hot-core \\
            &    3         & 2520  &     0.48$\pm$0.16  &    39$\pm$4   &   55$\pm$3   &    47$\pm$8   &    49.1$\pm$3.0  & O, hot-core \\
            &    4         & 3900  &     6.41$\pm$2.12  &    47$\pm$4   &   75$\pm$4   &    61$\pm$14  &    54.4$\pm$2.4  & O, hot-core \\
            &    5         & 1850  &     1.50$\pm$0.49  &    45$\pm$5   &   75$\pm$4   &    60$\pm$15  &    11.6$\pm$0.7  & hot-core \\
            &    7         & 1630  &     0.66$\pm$0.23  &    44$\pm$5   &   68$\pm$4   &    56$\pm$12  &     9.8$\pm$0.5  & O \\
            &    8         & 1920  &     1.22$\pm$0.41  &    44$\pm$5   &   71$\pm$4   &    58$\pm$14  &    11.5$\pm$0.6  & O, hot-core \\
            &    9         & 1710  &     1.30$\pm$0.43  &    46$\pm$5   &   75$\pm$5   &    61$\pm$15  &     6.9$\pm$0.6  & O, hot-core \\
            &   10         & 2670  &     1.78$\pm$0.59  &    42$\pm$4   &   68$\pm$4   &    55$\pm$13  &     7.5$\pm$0.8  & O, hot-core \\
            &   11         & 1510  &     1.11$\pm$0.37  &    47$\pm$5   &   76$\pm$5   &    62$\pm$15  &     2.4$\pm$0.2  & O, hot-core \\
            &   12         & 1820  &     0.75$\pm$0.25  &    42$\pm$5   &   67$\pm$4   &    54$\pm$12  &     4.8$\pm$0.6  & O \\
            &   13$\star$  & 1160  &                    &    50$\pm$4   &              &    50$\pm$4   &     3.5$\pm$0.7  & O \\
            &   14         & 1170  &     1.12$\pm$0.37  &    57$\pm$6   &   84$\pm$5   &    70$\pm$14  &     2.5$\pm$0.5  & O, hot-core \\
            &   15         & 2070  &     0.34$\pm$0.11  &    39$\pm$8   &   55$\pm$3   &    47$\pm$8   &     3.8$\pm$0.3  & O, hot-core \\
            &   16         & 2680  &                    &    34$\pm$3   &              &    34$\pm$3   &    12.1$\pm$1.7  & O, hot-core \\
            &   18         & 2100  &     0.48$\pm$0.17  &    40$\pm$5   &   59$\pm$4   &    49$\pm$9   &     4.6$\pm$0.4  & O, hot-core \\
            &   19         & 1410  &                    &    46$\pm$4   &              &    46$\pm$4   &     3.5$\pm$0.8  & O, hot-core \\
            &   22         & 1160  &                    &    45$\pm$4   &              &    45$\pm$4   &     2.4$\pm$0.5  & O \\
            &   23$\star$  & 1660  &                    &    41$\pm$4   &              &    41$\pm$4   &     3.8$\pm$0.6  & O \\
            &   26         & 1920  &                    &    37$\pm$6   &              &    37$\pm$6   &     2.9$\pm$0.3  & O \\
            &   29         & 1410  &     0.17$\pm$0.06  &    46$\pm$7   &   56$\pm$3   &    51$\pm$5   &     1.3$\pm$0.1  & O \\
            &   31$\star$  & 1460  &     0.10$\pm$0.05  &    44$\pm$7   &   50$\pm$5   &    47$\pm$4   &     1.2$\pm$0.1  & O \\
            &   36$\star$  & 1470  &     0.05$\pm$0.03  &    44$\pm$5   &   44$\pm$4   &    44$\pm$3   &     1.1$\pm$0.2  & O \\
            &   39         & 1660  &     0.34$\pm$0.12  &    42$\pm$7   &   60$\pm$4   &    51$\pm$9   &     0.7$\pm$0.1  & O \\
            &   44         & 2080  &                    &    36$\pm$6   &              &    36$\pm$6   &     1.2$\pm$0.2  & O \\
            &   49         & 1450  &                    &    42$\pm$5   &              &    42$\pm$5   &     1.3$\pm$0.2  & O \\
            &   51$\star$  & 3730  &     0.17$\pm$0.06  &    32$\pm$5   &   39$\pm$2   &    36$\pm$3   &     2.8$\pm$0.3  & O \\
            &   59         & 1550  &     0.30$\pm$0.10  &    43$\pm$7   &   60$\pm$4   &    51$\pm$9   &     0.5$\pm$0.2  & O \\
            &   67         & 1900  &     0.14$\pm$0.05  &    41$\pm$8   &   49$\pm$3   &    45$\pm$4   &     0.7$\pm$0.1  & O \\
            &  174$\star$  & 3090  &                    &    32$\pm$5   &              &    32$\pm$5   &     0.3$\pm$0.1  & O \\ \hline
\end{tabular}
\begin{tablenotes}[flushleft]
\item []\textbf{Notes:} (2) Protostellar core ID numbers. A $\star$ tag indicates tentative protostellar cores. (3) Outer radius of protostellar cores, assumed to be equal to their FWHM (Col.~8 of Tables~\ref{tab:measures table}--\ref{tab:measures table evolved}), deconvolved by the beam and set at their distance to the Sun \citep{motte2022}.
(4) Protostellar luminosity estimated from their luminosity contribution to their host \PPMAP luminosity peak (Cols.~3 and 11 of Tables~\ref{tab:measures table}--\ref{tab:measures table evolved}) in \cref{eq:spatial luminosity}. 
(5) Mass-averaged temperature of protostellar cores estimated from the \PPMAP temperature (Col.~10 of Tables~\ref{tab:measures table}--\ref{tab:measures table evolved}) and outer radius (Col.~3) in \cref{eq:tdust-fromT}. 
(6) Mass-averaged dust temperature of protostellar cores estimated from their outer radius and protostellar luminosity (Cols.~3--4) in \cref{eq:tdust-fromL}. 
(7) Average value of the dust temperatures in Cols.~5--6. 
(8) Mass of protostellar cores computed from its 1.3~mm (peak and integrated) fluxes measured in the catalogs of \cite{pouteau2022}, \cite{nony2023}, and \cite{armante2024}, and the temperature of Col.~7 in \cref{eq:optically thick mass}. When necessary, fluxes have been corrected for contamination by free-free and line emission.
(9) Protostellar cores driving an outflow (tag "O", \citealt{nony2020, nony2023}, in prep.; \citealt{valeille2024}), coincident with a hot core candidate \citep[tag "hot-core",][]{bonfand2024}, and/or whose flux is corrected for contamination by free-free emission \citep[tag "FF corrected",][]{armante2024}. Masses of Col.~8 are uncertain when the core flux is corrected for free-free contamination.
\end{tablenotes}
\end{threeparttable}}
\end{table*}

\setcounter{table}{5}
\begin{table*}[htbp!]
\centering
\resizebox{0.9\linewidth}{!}{
\begin{threeparttable}[c]
\caption{\textbf{continued.} Main characteristics of protostellar cores in the G012.80, W43-MM1, W43-MM2, and W43-MM3 protoclusters.}
\begin{tabular}{lccccc|ccc}
    \hline\hline
    Protocluster & Core ID & \multirow{2}{*}{$R_{\rm out}$} & \multirow{2}{*}{$L_{\rm proto \star}$[$R_{\rm out}$]} & 
    \multicolumn{2}{c|}{$\overline{T_{\rm dust}}(L_{\rm proto \star})$[$R_{\rm out}$]}
    & \multirow{2}{*}{$\overline{T_{\rm dust}}\pm 1\sigma$} & \multirow{2}{*}{$M_{\rm core}$[$\overline{T_{\rm dust}}$]} & \multirow{2}{*}{Comments} \\
        name & number & & & from \PPMAP \tdust & from $L_{\rm proto\star}$ & & & \\ 
     & & [au] & [$\times 10^3$~\lsun] & [K] & [K] & [K] & [\msun] & \\ 
     (1) & (2) & (3) & (4) & (5) & (6) & (7) & (8) &  (9)\\ \hline
W43-MM2     &    1         & 2770  &     2.65$\pm$0.87  &    53$\pm$6   &   72$\pm$4   &    62$\pm$10  &    66.7$\pm$13.2 & O, hot-core \\
            &    5         & 3270  &     0.46$\pm$0.15  &    33$\pm$5   &   49$\pm$3   &    41$\pm$8   &    10.0$\pm$0.6  & O \\
            &    6         & 1780  &     0.28$\pm$0.09  &    41$\pm$7   &   56$\pm$3   &    48$\pm$8   &     4.0$\pm$0.2  & O \\
            &    7         & 3660  &     2.65$\pm$0.87  &    46$\pm$5   &   65$\pm$4   &    56$\pm$9   &    14.2$\pm$0.7  & O \\
            &   10         & 1960  &     0.26$\pm$0.10  &    42$\pm$6   &   54$\pm$4   &    48$\pm$6   &     3.1$\pm$0.2  & O, hot-core \\
            &   12         & 1740  &     0.52$\pm$0.18  &    44$\pm$7   &   63$\pm$4   &    54$\pm$10  &     4.6$\pm$0.3  & O, hot-core \\
            &   14         & 1700  &     0.04$\pm$0.01  &    38$\pm$7   &   40$\pm$3   &    39$\pm$2   &     2.6$\pm$0.2  & O \\
            &   15         & 1400  &     0.34$\pm$0.14  &    50$\pm$7   &   64$\pm$5   &    57$\pm$7   &     1.9$\pm$0.1  & O \\
            &   20$\star$  & 2480  &                    &    35$\pm$4   &              &    35$\pm$4   &     2.6$\pm$0.2  & O \\
            &   28         & 1750  &     0.55$\pm$0.18  &    44$\pm$7   &   64$\pm$4   &    54$\pm$10  &     2.6$\pm$0.2  & O \\
            &   33         & 5860  &     1.45$\pm$0.48  &    35$\pm$4   &   49$\pm$3   &    42$\pm$7   &    15.8$\pm$1.0  & O \\
            &   39$\star$  & 1280  &     0.05$\pm$0.02  &    44$\pm$7   &   46$\pm$4   &    45$\pm$3   &     0.9$\pm$0.1  & O \\
            &   41         & 2160  &     0.05$\pm$0.02  &    38$\pm$5   &   38$\pm$3   &    38$\pm$2   &     1.3$\pm$0.2  & O \\
            &   44$\star$  & 4540  &                    &    34$\pm$4   &              &    34$\pm$4   &     6.4$\pm$0.6  & O \\
            &   50         & 1690  &     0.08$\pm$0.03  &    40$\pm$7   &   46$\pm$3   &    43$\pm$3   &     0.7$\pm$0.1  & O \\
            &   51         & 1970  &                    &    41$\pm$5   &              &    41$\pm$5   &     3.3$\pm$0.4  & O \\
            &   56         & 4280  &                    &    27$\pm$4   &              &    27$\pm$4   &     0.5$\pm$0.1  & O \\
            &   62         & 2680  &     0.02$\pm$0.01  &    32$\pm$4   &   30$\pm$2   &    31$\pm$2   &     0.9$\pm$0.1  & O \\
            &   71         & 3960  &                    &    29$\pm$4   &              &    29$\pm$4   &     0.9$\pm$0.1  & O \\
            &   75         & 2320  &                    &    35$\pm$5   &              &    35$\pm$5   &     0.9$\pm$0.1  & O \\
            &   78         & 3580  &     1.42$\pm$0.47  &    42$\pm$5   &   59$\pm$3   &    50$\pm$8   &     3.7$\pm$0.3  & O \\
            &   85         & 1930  &                    &    37$\pm$4   &              &    37$\pm$4   &     0.1$\pm$0.1  & O \\
            &   88         & 7270  &                    &    23$\pm$3   &              &    23$\pm$3   &     0.8$\pm$0.2  & O \\
            &   97$\star$  & 1340  &                    &    41$\pm$5   &              &    41$\pm$5   &     0.4$\pm$0.1  & O \\
            &   99         & 3540  &                    &    29$\pm$4   &              &    29$\pm$4   &     0.7$\pm$0.1  & O \\
            &  112         & 1280  &     0.06$\pm$0.02  &    46$\pm$7   &   49$\pm$3   &    47$\pm$3   &     0.6$\pm$0.1  & O \\
            &  133         & 1280  &                    &    42$\pm$6   &              &    42$\pm$6   &     0.2$\pm$0.1  & O \\
            &  140         & 1280  &                    &    42$\pm$5   &              &    42$\pm$5   &     0.1$\pm$0.1  & O \\
            &  182$\star$  & 5060  &                    &    25$\pm$3   &              &    25$\pm$3   &     0.5$\pm$0.1  & O \\
            &  192         & 2950  &                    &    34$\pm$4   &              &    34$\pm$4   &     0.9$\pm$0.2  & O \\
            &  204         & 2120  &     0.08$\pm$0.03  &    37$\pm$6   &   42$\pm$3   &    39$\pm$3   &     0.5$\pm$0.2  & O \\
            &  230         & 3670  &                    &    29$\pm$4   &              &    29$\pm$4   &     0.2$\pm$0.1  & O \\
            &  236         & 2800  &                    &    36$\pm$5   &              &    36$\pm$5   &     0.4$\pm$0.1  & O \\
            &  239         & 3780  &                    &    30$\pm$4   &              &    30$\pm$4   &     0.7$\pm$0.2  & O \\
            &  262         & 4760  &                    &    27$\pm$4   &              &    27$\pm$4   &     0.3$\pm$0.1  & O \\
            &  265$\star$  & 4530  &                    &    27$\pm$4   &              &    27$\pm$4   &     0.2$\pm$0.1  & O \\
            &  287         & 1280  &                    &    43$\pm$7   &              &    43$\pm$7   &     0.1$\pm$0.1  & O \\ \hline
W43-MM3     &    2         & 1560  &     3.09$\pm$1.02  &    55$\pm$7   &   91$\pm$5   &    73$\pm$18  &    12.8$\pm$0.5  & O, hot-core \\
            &    3         & 2530  &     3.06$\pm$1.01  &    40$\pm$9   &   76$\pm$5   &    58$\pm$18  &     6.7$\pm$0.3  & O, hot-core \\
            &    9         & 1510  &     3.03$\pm$1.00  &    57$\pm$7   &   92$\pm$5   &    74$\pm$17  &     4.8$\pm$0.2  & O, hot-core \\
            &   24         & 2030  &     2.90$\pm$0.96  &    50$\pm$6   &   82$\pm$5   &    66$\pm$16  &     2.7$\pm$0.3  & O \\
            &   25$\star$  & 4000  &                    &    35$\pm$4   &              &    35$\pm$4   &    10.4$\pm$0.9  & O \\
            &   37         & 1330  &     0.23$\pm$0.08  &    45$\pm$11  &   60$\pm$4   &    53$\pm$8   &     0.9$\pm$0.1  & O \\
            &   86$\star$  & 4040  &     0.07$\pm$0.03  &    30$\pm$8   &   33$\pm$2   &    31$\pm$2   &     0.2$\pm$0.1  & O \\
            &  113         & 3360  &                    &    34$\pm$9   &              &    34$\pm$9   &     0.5$\pm$0.1  & O \\
            &  147         & 3180  &     0.16$\pm$0.05  &    33$\pm$8   &   41$\pm$2   &    37$\pm$4   &     0.6$\pm$0.1  & O \\
            &  157$\star$  & 4460  &                    &    28$\pm$7   &              &    28$\pm$7   &     0.6$\pm$0.1  & O \\
            &  187         & 1330  &                    &    44$\pm$11  &              &    44$\pm$11  &     0.2$\pm$0.1  & O \\
            &  237         & 2560  &                    &    34$\pm$9   &              &    34$\pm$9   &     0.3$\pm$0.1  & O \\
            &  238         & 1770  &                    &    38$\pm$9   &              &    38$\pm$9   &     0.5$\pm$0.1  & O \\
            &  244         & 3110  &                    &    33$\pm$7   &              &    33$\pm$7   &     0.4$\pm$0.1  & O \\ \hline
\end{tabular}
\end{threeparttable}}
\end{table*}

\begin{table*}[htbp!]
\centering
\resizebox{0.8\linewidth}{!}{
\begin{threeparttable}[c]
\caption{Main characteristics of protostellar cores in the Young G327.29, G328.25, G337.92, and G338.93 protoclusters}
\label{appendixtab:computed table}
\begin{tabular}{lccccc|ccc}
    \hline\hline
    Protocluster & Core ID & \multirow{2}{*}{$R_{\rm out}$} & \multirow{2}{*}{$L_{\rm proto \star}$[$R_{\rm out}$]} &
    \multicolumn{2}{c|}{$\overline{T_{\rm dust}}(L_{\rm proto \star})$[$R_{\rm out}$]}
    & \multirow{2}{*}{$\overline{T_{\rm dust}}\pm 1\sigma$} & \multirow{2}{*}{$M_{\rm core}$[$\overline{T_{\rm dust}}$]} & \multirow{2}{*}{Comments} \\
    name & number & & & from \PPMAP \tdust & from $L_{\rm proto\star}$& & & \\ 
     & & [au] & [$\times 10^3$~\lsun] & [K] & [K] & [K] & [\msun] & \\
     (1) & (2) & (3) & (4) & (5) & (6) & (7) & (8) &  (9)\\ \hline
G327.29     &    1         & 2900  &     4.50$\pm$1.50  &    69$\pm$6   &   78$\pm$5   &    73$\pm$5   &   111.9$\pm$11.2 & O, hot-core \\
            &    2$\star$  & 1800  &     4.50$\pm$1.50  &    80$\pm$7   &   93$\pm$6   &    87$\pm$6   &    37.0$\pm$1.3  & O, hot-core \\
            &    4         & 3060  &     0.60$\pm$0.21  &    32$\pm$4   &   53$\pm$3   &    43$\pm$10  &    14.8$\pm$0.9  & O, hot-core \\
            &    5$\star$  & 970   &     0.22$\pm$0.07  &    43$\pm$6   &   67$\pm$4   &    55$\pm$12  & <   1.9$\pm$0.1  & O, free-free \\
            &    6$\star$  & 3260  &     1.04$\pm$0.34  &    54$\pm$6   &   57$\pm$3   &    56$\pm$3   &    31.4$\pm$1.6  & O \\
            &    7         & 2300  &     0.36$\pm$0.12  &    31$\pm$4   &   54$\pm$3   &    43$\pm$11  &     4.7$\pm$0.4  & hot-core \\
            &    9$\star$  & 2830  &     1.16$\pm$0.38  &    56$\pm$6   &   62$\pm$4   &    59$\pm$3   &    16.8$\pm$0.8  & O \\
            &   10         & 1780  &                    &    44$\pm$6   &              &    44$\pm$6   &     5.7$\pm$0.6  & O \\
            &   11         & 1370  &     0.39$\pm$0.13  &    36$\pm$7   &   66$\pm$4   &    51$\pm$15  &     0.9$\pm$0.1  & O \\
            &   12$\star$  & 1520  &     0.27$\pm$0.09  &    34$\pm$5   &   59$\pm$3   &    46$\pm$13  & <   2.1$\pm$0.2  & O, free-free \\
            &   13$\star$  & 1280  &                    &    36$\pm$5   &              &    36$\pm$5   &     1.3$\pm$0.2  & O \\
            &   14$\star$  & 970   &     0.02$\pm$0.01  &    40$\pm$6   &   44$\pm$2   &    42$\pm$2   &     0.8$\pm$0.1  & O \\
            &   15$\star$  & 1510  &     0.12$\pm$0.04  &    35$\pm$5   &   51$\pm$3   &    43$\pm$8   &     1.6$\pm$0.2  & O \\
            &   16         & 1630  &     1.37$\pm$0.46  &    46$\pm$6   &   77$\pm$5   &    61$\pm$16  &     2.2$\pm$0.2  & O \\
            &   19         & 1720  &     0.19$\pm$0.06  &    33$\pm$6   &   53$\pm$3   &    43$\pm$10  &     1.0$\pm$0.1  & O \\
            &   21$\star$  & 830   &                    &    43$\pm$6   &              &    43$\pm$6   & <   0.8$\pm$0.2  & O, free-free \\
            &   23         & 830   &     0.43$\pm$0.16  &    53$\pm$6   &   80$\pm$5   &    67$\pm$13  & <   0.7$\pm$0.1  & O, free-free \\
            &   24         & 1220  &     0.04$\pm$0.02  &    35$\pm$6   &   45$\pm$4   &    40$\pm$5   &     0.6$\pm$0.1  & O \\
            &   25$\star$  & 930   &     0.07$\pm$0.02  &    40$\pm$8   &   55$\pm$3   &    48$\pm$8   &     0.3$\pm$0.1  & O \\
            &   26         & 1360  &                    &    36$\pm$5   &              &    36$\pm$5   &     0.8$\pm$0.1  & O \\
            &   27         & 1110  &                    &    36$\pm$5   &              &    36$\pm$5   &     0.7$\pm$0.2  & O \\
            &   28         & 830   &                    &    39$\pm$6   &              &    39$\pm$6   &     0.7$\pm$0.2  & O \\
            &   31$\star$  & 2160  &                    &    31$\pm$5   &              &    31$\pm$5   &     0.9$\pm$0.1  & O \\
            &   35         & 1300  &                    &    34$\pm$5   &              &    34$\pm$5   &     0.8$\pm$0.2  & O \\
            &   45$\star$  & 1240  &                    &    32$\pm$6   &              &    32$\pm$6   &     0.3$\pm$0.1  & O \\ \hline
G328.25     &    1         & 1650  &     3.62$\pm$1.20  &    55$\pm$10  &   92$\pm$5   &    74$\pm$18  &    11.1$\pm$3.3  & O, hot-core \\
            &    6$\star$  & 830   &                    &    38$\pm$13  &              &    38$\pm$13  &     0.3$\pm$0.1  & O \\
            &   18$\star$  & 680   &                    &    40$\pm$13  &              &    40$\pm$13  &     0.1$\pm$0.1  & O \\ \hline
G337.92     &    1         & 2340  &     4.24$\pm$1.40  &    65$\pm$5   &   83$\pm$5   &    74$\pm$9   &    33.3$\pm$5.5  & O, hot-core \\
            &    2         & 2130  &     4.18$\pm$1.38  &    64$\pm$5   &   86$\pm$5   &    75$\pm$11  &     8.1$\pm$1.1  & O, hot-core \\
            &    3         & 1170  &     0.05$\pm$0.02  &    36$\pm$8   &   48$\pm$4   &    42$\pm$6   & <   0.9$\pm$0.1  & O, free-free \\
            &    4         & 1030  &                    &    37$\pm$7   &              &    37$\pm$7   &     1.2$\pm$0.1  & O \\
            &    5$\star$  & 1960  &     1.82$\pm$0.60  &    53$\pm$5   &   76$\pm$5   &    64$\pm$12  &     3.2$\pm$0.4  & O \\
            &   10         & 1750  &                    &    49$\pm$5   &              &    49$\pm$5   &     2.2$\pm$0.4  & O \\
            &   14         & 1140  &     1.45$\pm$0.48  &    64$\pm$7   &   89$\pm$5   &    76$\pm$12  &     0.8$\pm$0.2  & hot-core \\
            &   15         & 1430  &                    &    31$\pm$7   &              &    31$\pm$7   &     0.5$\pm$0.1  & O \\
            &   26         & 1000  &     0.10$\pm$0.04  &    45$\pm$7   &   58$\pm$5   &    51$\pm$6   &     0.3$\pm$0.1  & O \\
            &   31$\star$  & 3170  &     1.85$\pm$0.61  &    44$\pm$5   &   64$\pm$4   &    54$\pm$10  &     1.2$\pm$0.3  & O \\
            &   32$\star$  & 2000  &                    &    45$\pm$4   &              &    45$\pm$4   &     0.8$\pm$0.2  & O \\
            &   40$\star$  & 980   &                    &    44$\pm$7   &              &    44$\pm$7   &     0.2$\pm$0.1  & O \\ \hline
G338.93     &    1         & 1560  &     1.14$\pm$0.38  &    45$\pm$12  &   76$\pm$4   &    61$\pm$15  &    19.7$\pm$6.2  & O, hot-core \\
            &    2         & 1410  &                    &    43$\pm$12  &              &    43$\pm$12  &    36.7$\pm$2.4  & O, hot-core \\
            &    3         & 1390  &     2.49$\pm$0.82  &    63$\pm$11  &   91$\pm$5   &    77$\pm$14  &     7.3$\pm$0.3  & O, hot-core \\
            &    4         & 1500  &     1.57$\pm$0.52  &    52$\pm$12  &   82$\pm$5   &    67$\pm$15  &    11.1$\pm$0.5  & O, hot-core \\
            &    6$\star$  & 2480  &     2.78$\pm$0.92  &    51$\pm$10  &   75$\pm$4   &    63$\pm$12  &     6.9$\pm$0.3  & O \\
            &    9         & 1790  &                    &    37$\pm$11  &              &    37$\pm$11  &     5.4$\pm$0.4  & O \\
            &   10         & 2220  &     3.99$\pm$1.32  &    49$\pm$11  &   84$\pm$5   &    66$\pm$17  &     2.8$\pm$0.2  & O, hot-core \\
            &   11$\star$  & 1150  &     0.96$\pm$0.32  &    47$\pm$13  &   82$\pm$5   &    65$\pm$18  &     2.7$\pm$0.2  & O \\
            &   13$\star$  & 3440  &                    &    30$\pm$9   &              &    30$\pm$9   &     9.6$\pm$0.9  & O \\
            &   14$\star$  & 4320  &                    &    28$\pm$9   &              &    28$\pm$9   &     1.2$\pm$0.2  & O \\
            &   20$\star$  & 2840  &     1.42$\pm$0.47  &    45$\pm$10  &   64$\pm$4   &    54$\pm$9   &     1.8$\pm$0.2  & O \\
            &   23$\star$  & 2630  &                    &    39$\pm$9   &              &    39$\pm$9   &     0.6$\pm$0.2  & O \\
            &   25         & 3390  &     0.08$\pm$0.03  &    27$\pm$8   &   36$\pm$2   &    31$\pm$4   &     0.9$\pm$0.2  & O \\ \hline
\end{tabular}
\begin{tablenotes}[flushleft]
\item []\textbf{Notes:} 
(2) Protostellar core ID numbers. A $\star$ tag indicates tentative protostellar cores.
(3) Outer radius of protostellar cores, assumed to be equal to their FWHM (see Tables~\ref{appendixtab:measures table}--\ref{appendixtab:measures table evolved}), deconvolved by the beam and set at their distance to the Sun \citep{motte2022}.
(4) Protostellar luminosity estimated from their luminosity contribution to their host \PPMAP luminosity peak (Cols.~3 and 11 of Tables~\ref{appendixtab:measures table}--\ref{appendixtab:measures table evolved}) in \cref{eq:spatial luminosity}. A $\diamond$ tag indicates protostellar luminosities taken from \cref{tab:luminosity peaks from radial profile} (Col.~8).
(5) Mass-averaged temperature of protostellar cores estimated from the \PPMAP temperature (Col.~10 of Tables~\ref{appendixtab:measures table}--\ref{appendixtab:measures table evolved}) and outer radius (Col.~3) in \cref{eq:tdust-fromT}. 
(6) Mass-averaged dust temperature of protostellar cores estimated from their outer radius (Col.~3) and protostellar luminosity (Col.~4 or \cref{tab:luminosity peaks from radial profile}) in \cref{eq:tdust-fromL}. 
(7) Average value of the dust temperatures in Cols.~5--6. 
(8) Mass of protostellar cores computed from their 1.3~mm (peak and integrated) fluxes measured in the catalog at the original resolution of \cite{louvet2024} and the temperature of Col.~7 in \cref{eq:mass}. Fluxes should be free of line contamination and are sometimes corrected for contamination by free-free emission.
(9) Protostellar cores driving an outflow (tag "O", \citealt{nony2020, nony2023}, in prep.; \citealt{valeille2024}), coincident with a hot core candidate \citep[tag "hot-core",][]{bonfand2024}, and/or whose flux is either contaminated by free-free emission \citep[tag "free-free",][]{louvet2024} or corrected for free-free contamination \citep[tag "FF corrected",][]{bonfand2024}. Masses of Col.~8 are upper limits when the core flux is contaminated by free-free emission, and are uncertain when corrected for free-free contamination.
\end{tablenotes}
\end{threeparttable}}
\end{table*}

\setcounter{table}{6}
\begin{table*}[htbp!]
\centering
\resizebox{0.9\linewidth}{!}{
\begin{threeparttable}[c]
\caption{\textbf{continued.} Main characteristics of protostellar cores in the Intermediate G008.67, G353.41, and W51-E protoclusters.}
\begin{tabular}{lccccc|ccc}
    \hline\hline
    Protocluster & Core ID & \multirow{2}{*}{$R_{\rm out}$} & \multirow{2}{*}{$L_{\rm proto \star}$[$R_{\rm out}$]} & 
    \multicolumn{2}{c|}{$\overline{T_{\rm dust}}(L_{\rm proto \star})$[$R_{\rm out}$]}
    & \multirow{2}{*}{$\overline{T_{\rm dust}}\pm 1\sigma$} & \multirow{2}{*}{$M_{\rm core}$[$\overline{T_{\rm dust}}$]} & \multirow{2}{*}{Comments} \\
        name & number & & & from \PPMAP \tdust & from $L_{\rm proto\star}$ & & & \\ 
     & & [au] & [$\times 10^3$~\lsun] & [K] & [K] & [K] & [\msun] & \\ 
     (1) & (2) & (3) & (4) & (5) & (6) & (7) & (8) &  (9)\\ \hline
G008.67     &    1         & 3690  &     5.69$\pm$1.88  &    38$\pm$3   &   74$\pm$4   &    56$\pm$18  &    16.0$\pm$6.0  & hot-core, FF corrected \\
            &    2         & 1180  &     1.86$\pm$0.61  &    52$\pm$17  &   92$\pm$5   &    72$\pm$20  &     4.0$\pm$0.2  & O, hot-core \\
            &    4         & 1130  &                    &    49$\pm$4   &              &    49$\pm$4   &     5.5$\pm$0.5  & O \\
            &    7         & 1860  &     0.19$\pm$0.07  &    34$\pm$12  &   52$\pm$3   &    43$\pm$9   &     2.5$\pm$0.2  & O \\
            &   13         & 2020  &                    &    27$\pm$12  &              &    27$\pm$12  &     1.1$\pm$0.2  & O \\
            &   14$\star$  & 3040  &                    &    31$\pm$7   &              &    31$\pm$7   &     3.1$\pm$0.4  & O \\ \hline
G353.41     &    1         & 1340  &     0.40$\pm$0.13  &    40$\pm$5   &   66$\pm$4   &    53$\pm$13  & <   2.8$\pm$0.9  & O, free-free \\
            &    2$\star$  & 1360  &     0.12$\pm$0.04  &    36$\pm$6   &   53$\pm$3   &    44$\pm$8   &     4.4$\pm$0.3  & O \\
            &    3         & 1020  &     0.08$\pm$0.03  &    41$\pm$8   &   55$\pm$4   &    48$\pm$7   &     1.7$\pm$0.1  & O \\
            &    4         & 1870  &     0.30$\pm$0.10  &    31$\pm$7   &   56$\pm$3   &    44$\pm$12  &     3.7$\pm$0.3  & O, hot-core \\
            &    5         & 1040  &     0.14$\pm$0.05  &    36$\pm$8   &   60$\pm$4   &    48$\pm$12  &     1.4$\pm$0.1  & O \\
            &    8$\star$  & 2140  &     0.82$\pm$0.27  &    39$\pm$4   &   64$\pm$4   &    52$\pm$12  &     3.7$\pm$0.3  & O \\
            &   10$\star$  & 1520  &                    &    30$\pm$7   &              &    30$\pm$7   &     2.6$\pm$0.3  & O \\
            &   11$\star$  & 3610  &                    &    17$\pm$6   &              &    17$\pm$6   &     0.7$\pm$0.2  & O \\
            &   14$\star$  & 1150  &     0.08$\pm$0.03  &    37$\pm$7   &   53$\pm$3   &    45$\pm$8   &     1.3$\pm$0.2  & O \\
            &   20         & 990   &                    &    32$\pm$9   &              &    32$\pm$9   &     0.4$\pm$0.1  & O \\
            &   21         & 1680  &                    &    32$\pm$6   &              &    32$\pm$6   &     0.8$\pm$0.1  & O \\
            &   26$\star$  & 1040  &                    &    36$\pm$8   &              &    36$\pm$8   &     0.3$\pm$0.1  & O \\
            &   27$\star$  & 2320  &     0.17$\pm$0.06  &    31$\pm$5   &   47$\pm$3   &    39$\pm$8   &     1.7$\pm$0.3  & O \\
            &   33$\star$  & 980   &                    &    28$\pm$9   &              &    28$\pm$9   &     0.4$\pm$0.1  & O \\
            &   40         & 1550  &                    &    30$\pm$6   &              &    30$\pm$6   &     1.0$\pm$0.2  & O \\
            &   43         & 1600  &                    &    28$\pm$7   &              &    28$\pm$7   &     0.4$\pm$0.1  & O \\
            &   44$\star$  & 1780  &                    &    29$\pm$5   &              &    29$\pm$5   &     0.5$\pm$0.1  & O \\
            &   45$\star$  & 1920  &                    &    26$\pm$7   &              &    26$\pm$7   &     0.5$\pm$0.1  & O \\ \hline
W51-E       &    3         & 3230  &    32.00$\pm$6.00  &    83$\pm$10  &  107$\pm$4   &    95$\pm$12  &   262.0$\pm$78.0 & O, hot-core, FF corrected \\
            &    4         & 2560  &    54.00$\pm$18.00 &    86$\pm$11  &  127$\pm$8   &   107$\pm$20  &    73.3$\pm$3.8  & O, hot-core, FF corrected \\
            &    6$\star$  & 1780  &    23.01$\pm$7.60  &    94$\pm$13  &  124$\pm$7   &   109$\pm$15  & <  13.8$\pm$1.1  & O, free-free \\
            &   10$\star$  & 2720  &    19.54$\pm$6.45  &    76$\pm$11  &  104$\pm$6   &    90$\pm$14  &    12.3$\pm$1.5  & O \\
            &   11         & 1870  &     2.68$\pm$0.90  &    51$\pm$14  &   83$\pm$5   &    67$\pm$16  &     4.1$\pm$0.5  & O, hot-core \\
            &   12         & 940   &                    &    50$\pm$15  &              &    50$\pm$15  &     2.5$\pm$0.3  & O \\
            &   13         & 1520  &     3.19$\pm$1.07  &    58$\pm$15  &   92$\pm$6   &    75$\pm$17  &     3.5$\pm$0.6  & O, hot-core \\
            &   14$\star$  & 4050  &                    &    31$\pm$9   &              &    31$\pm$9   &    11.6$\pm$1.1  & O \\
            &   21         & 2150  &     1.97$\pm$0.66  &    56$\pm$12  &   75$\pm$5   &    65$\pm$9   &     6.1$\pm$1.0  & hot-core \\
            &   25$\star$  & 1550  &    36.00$\pm$12.00 &    92$\pm$14  &  142$\pm$9   &   117$\pm$25  &     3.8$\pm$0.9  & O, hot-core, FF corrected \\
            &   34$\star$  & 1450  &                    &    38$\pm$12  &              &    38$\pm$12  & <   0.8$\pm$0.2  & O, free-free \\
            &   35         & 4280  &                    &    26$\pm$9   &              &    26$\pm$9   &     4.1$\pm$0.5  & O \\
            &   47         & 1110  &                    &    42$\pm$15  &              &    42$\pm$15  &     0.5$\pm$0.1  & O \\ \hline
\end{tabular}
\end{threeparttable}}
\end{table*}

\setcounter{table}{6}
\begin{table*}[htbp!]
\centering
\resizebox{0.9\linewidth}{!}{
\begin{threeparttable}[c]
\caption{\textbf{continued.} Main characteristics of protostellar cores in the Evolved G010.62, G333.60 and W51-IRS2 protoclusters.}
\begin{tabular}{lccccc|ccc}
    \hline\hline
    Protocluster & Core ID & \multirow{2}{*}{$R_{\rm out}$} & \multirow{2}{*}{$L_{\rm proto \star}$[$R_{\rm out}$]} & 
    \multicolumn{2}{c|}{$\overline{T_{\rm dust}}(L_{\rm proto \star})$[$R_{\rm out}$]}
    & \multirow{2}{*}{$\overline{T_{\rm dust}}\pm 1\sigma$} & \multirow{2}{*}{$M_{\rm core}$[$\overline{T_{\rm dust}}$]} & \multirow{2}{*}{Comments} \\
        name & number & & & from \PPMAP \tdust & from $L_{\rm proto\star}$ & & & \\ 
     & & [au] & [$\times 10^3$~\lsun] & [K] & [K] & [K] & [\msun] & \\ 
     (1) & (2) & (3) & (4) & (5) & (6) & (7) & (8) &  (9)\\ \hline
G010.62     &    1         & 4740  &     7.25$\pm$2.40  &    53$\pm$8   &   71$\pm$4   &    62$\pm$9   &     --  & O, hot-core, FF corrected \\
            &    3         & 1150  &     1.37$\pm$0.60  &    66$\pm$9   &   88$\pm$7   &    77$\pm$11  &     1.9$\pm$0.1  & O, hot-core \\
            &   10         & 1970  &                    &    33$\pm$6   &              &    33$\pm$6   &     1.5$\pm$0.2  & O \\
            &   15         & 1150  &     0.31$\pm$0.10  &    37$\pm$9   &   67$\pm$4   &    52$\pm$15  &     0.5$\pm$0.1  & O \\
            &   17$\star$  & 2160  &                    &    54$\pm$9   &              &    54$\pm$9   & <   3.6$\pm$0.6  & O, free-free \\
            &   18         & 1500  &     1.90$\pm$0.68  &    68$\pm$8   &   84$\pm$5   &    76$\pm$8   & <   2.1$\pm$0.3  & O, free-free \\
            &   19         & 2690  &     7.45$\pm$2.47  &    63$\pm$9   &   88$\pm$5   &    75$\pm$12  &     --  & O, hot-core, FF corrected \\
            &   21         & 2360  &     1.95$\pm$0.70  &    57$\pm$7   &   72$\pm$5   &    64$\pm$8   &     2.6$\pm$0.3  & O \\
            &   22         & 2580  &     3.36$\pm$1.11  &    63$\pm$10  &   77$\pm$5   &    70$\pm$7   &     4.2$\pm$0.7  & hot-core, FF corrected \\
            &   23         & 1150  &     2.16$\pm$0.78  &    73$\pm$9   &   95$\pm$6   &    84$\pm$11  &     1.6$\pm$0.3  & O, hot-core \\
            &   26         & 1950  &                    &    44$\pm$6   &              &    44$\pm$6   &     1.2$\pm$0.1  & O \\
            &   29$\star$  & 2570  &     0.76$\pm$0.49  &    49$\pm$7   &   59$\pm$7   &    54$\pm$5   &     2.3$\pm$0.3  & O \\
            &   42$\star$  & 1650  &                    &    57$\pm$7   &              &    57$\pm$7   &     1.4$\pm$0.4  & O \\
            &   52$\star$  & 3980  &                    &    40$\pm$5   &              &    40$\pm$5   &     1.4$\pm$0.3  & O \\ \hline
G333.60     &    1$\star$  & 3750  &    12.55$\pm$4.16  &    56$\pm$14  &   85$\pm$5   &    71$\pm$14  &    24.0$\pm$5.7  & O \\
            &    2         & 2100  &     2.80$\pm$1.02  &    57$\pm$18  &   80$\pm$5   &    69$\pm$12  &     7.4$\pm$0.3  & O \\
            &    3         & 1280  &                    &    64$\pm$19  &              &    64$\pm$19  &     6.6$\pm$0.5  & O, hot-core \\
            &    4         & 1930  &     1.69$\pm$0.58  &    52$\pm$19  &   76$\pm$5   &    64$\pm$12  &     3.2$\pm$0.2  & O \\
            &    5$\star$  & 1290  &     1.04$\pm$0.36  &    60$\pm$22  &   80$\pm$5   &    70$\pm$10  &     2.1$\pm$0.2  & O \\
            &    6$\star$  & 3530  &                    &    46$\pm$13  &              &    46$\pm$13  &    16.6$\pm$1.1  & O \\
            &    8         & 4210  &     3.64$\pm$1.26  &    47$\pm$13  &   65$\pm$4   &    56$\pm$9   &     8.3$\pm$0.6  & O, hot-core, FF corrected \\
            &   18$\star$  & 4550  &                    &    38$\pm$13  &              &    38$\pm$13  & <   9.4$\pm$1.0  & O, free-free \\
            &   24         & 1700  &     0.42$\pm$0.14  &    41$\pm$18  &   62$\pm$4   &    51$\pm$11  &     0.7$\pm$0.1  & O \\
            &   32$\star$  & 6170  &                    &    37$\pm$10  &              &    37$\pm$10  &    10.9$\pm$0.9  & O \\
            &   47$\star$  & 5080  &                    &    35$\pm$11  &              &    35$\pm$11  & <   5.2$\pm$0.8  & O, free-free \\
            &   60$\star$  & 1400  &                    &    40$\pm$16  &              &    40$\pm$16  &     0.2$\pm$0.1  & O \\
            &   73$\star$  & 4730  &                    &    35$\pm$13  &              &    35$\pm$13  &     1.5$\pm$0.2  & O \\
            &   87$\star$  & 4850  &                    &    24$\pm$10  &              &    24$\pm$10  &     1.0$\pm$0.2  & O \\
            &   89$\star$  & 1700  &     0.65$\pm$0.22  &    42$\pm$18  &   67$\pm$4   &    54$\pm$12  &     0.3$\pm$0.1  & O \\ \hline
W51-IRS2    &    1         & 2510  &    80.00$\pm$10.00 &   104$\pm$14  &  137$\pm$4   &   121$\pm$16  &   165.0$\pm$40.8 & O, hot-core, FF corrected \\
            &    3         & 1840  &    54.00$\pm$8.00  &   112$\pm$16  &  143$\pm$4   &   127$\pm$16  &    28.4$\pm$0.9  & O, hot-core, FF corrected \\
            &    4         & 2310  &     4.83$\pm$1.61  &    53$\pm$12  &   86$\pm$5   &    69$\pm$16  &    24.7$\pm$0.9  & O, hot-core \\
            &    5         & 2640  &     9.36$\pm$3.09  &    84$\pm$13  &   92$\pm$5   &    88$\pm$5   &    33.9$\pm$1.5  & hot-core, FF corrected \\
            &    6         & 2910  &    68.00$\pm$10.00 &    99$\pm$14  &  127$\pm$3   &   113$\pm$14  &    42.6$\pm$1.5  & hot-core \\
            &    8         & 2000  &    13.30$\pm$4.40  &    91$\pm$14  &  108$\pm$6   &   100$\pm$8   &    17.7$\pm$1.1  & O, hot-core \\
            &   11         & 1490  &     1.25$\pm$0.42  &    53$\pm$14  &   78$\pm$5   &    66$\pm$13  & <   3.2$\pm$0.2  & O, free-free \\
            &   15         & 1320  &                    &    49$\pm$15  &              &    49$\pm$15  & <   1.3$\pm$0.1  & O, free-free \\
            &   17         & 3640  &     2.57$\pm$0.87  &    39$\pm$10  &   65$\pm$4   &    52$\pm$13  &     5.4$\pm$0.4  & O, hot-core \\
            &   18         & 2180  &                    &    39$\pm$9   &              &    39$\pm$9   & <   1.2$\pm$0.1  & O, free-free \\
            &   20         & 1940  &     2.29$\pm$0.77  &    49$\pm$13  &   80$\pm$5   &    64$\pm$16  & <   2.8$\pm$0.2  & O, free-free \\
            &   25         & 1590  &     8.94$\pm$2.96  &    59$\pm$12  &  109$\pm$7   &    84$\pm$25  & <   1.2$\pm$0.1  & O, hot-core, free-free \\
            &   30$\star$  & 3270  &                    &    36$\pm$11  &              &    36$\pm$11  &     2.0$\pm$0.2  & O \\
            &   33         & 1280  &                    &    51$\pm$13  &              &    51$\pm$13  &     1.5$\pm$0.2  & O \\
            &   34         & 1830  &     6.83$\pm$2.28  &    69$\pm$10  &   99$\pm$6   &    84$\pm$15  &     3.5$\pm$0.4  & O, hot-core \\
            &   41         & 3180  &     3.12$\pm$1.05  &    45$\pm$12  &   70$\pm$4   &    58$\pm$13  &     2.1$\pm$0.2  & O \\
            &   45         & 2210  &                    &    41$\pm$12  &              &    41$\pm$12  &     1.0$\pm$0.1  & O \\
            &   47         & 1950  &                    &    43$\pm$13  &              &    43$\pm$13  &     1.5$\pm$0.2  & O \\
            &   48$\star$  & 2950  &                    &    67$\pm$12  &              &    67$\pm$12  &    11.5$\pm$1.6  & O \\
            &   49         & 1280  &                    &    50$\pm$15  &              &    50$\pm$15  &     0.8$\pm$0.1  & O \\
            &   51$\star$  & 3160  &     1.16$\pm$0.39  &    38$\pm$12  &   59$\pm$4   &    49$\pm$11  &     2.1$\pm$0.2  & O \\
            &   53         & 1660  &                    &    44$\pm$14  &              &    44$\pm$14  &     0.9$\pm$0.1  & O \\
            &   61         & 4800  &                    &    32$\pm$8   &              &    32$\pm$8   &     4.5$\pm$0.6  & O \\
            &   63         & 1280  &                    &    51$\pm$16  &              &    51$\pm$16  &     0.4$\pm$0.1  & O \\
            &   65$\star$  & 2270  &     0.28$\pm$0.10  &    41$\pm$12  &   52$\pm$3   &    47$\pm$5   &     0.7$\pm$0.1  & O \\
            &   67         & 1570  &                    &    46$\pm$14  &              &    46$\pm$14  &     0.5$\pm$0.1  & O \\
            &   68$\star$  & 2640  &                    &    63$\pm$8   &              &    63$\pm$8   & <   2.5$\pm$0.4  & O, free-free \\
            &   71$\star$  & 1280  &                    &    48$\pm$10  &              &    48$\pm$10  &     0.2$\pm$0.1  & O \\
            &   73$\star$  & 3130  &                    &    35$\pm$11  &              &    35$\pm$11  &     0.9$\pm$0.1  & O \\
            &   75$\star$  & 2170  &                    &    41$\pm$12  &              &    41$\pm$12  &     0.5$\pm$0.1  & O \\
            &   81$\star$  & 1850  &                    &    42$\pm$11  &              &    42$\pm$11  &     1.0$\pm$0.2  & O \\
            &   90         & 2310  &                    &    39$\pm$9   &              &    39$\pm$9   &     0.6$\pm$0.1  & O \\
            &   92$\star$  & 2870  &                    &    35$\pm$10  &              &    35$\pm$10  &     0.5$\pm$0.1  & O \\
            &   98         & 1890  &     0.82$\pm$0.27  &    43$\pm$13  &   67$\pm$4   &    55$\pm$12  &     0.2$\pm$0.1  & O \\
            &  106$\star$  & 3500  &     0.55$\pm$0.19  &    35$\pm$10  &   50$\pm$3   &    43$\pm$7   &     0.9$\pm$0.2  & O \\
            &  117$\star$  & 1710  &                    &    46$\pm$11  &              &    46$\pm$11  &     0.7$\pm$0.2  & O \\ \hline
\end{tabular}
\end{threeparttable}}
\end{table*}

\begin{table*}[htbp!]
\centering
\resizebox{0.9\linewidth}{!}{
\begin{threeparttable}[c]
\caption{Main characteristics of the 615 prestellar cores in the ALMA-IMF protoclusters.}
\label{appendixtab:prestellar table}
\begin{tabular}{lcccccc}
    \hline\hline
    Protocluster & Core ID & \multirow{2}{*}{$R_{\rm out}$} & \multirow{2}{*}{$\overline{T_{\rm dust}^{\rm PPMAP}}$[$1.25\arcsec$]} & \multirow{2}{*}{$\overline{\tdust({\rm prestellar~core})}[R_{\rm out}]$} & \multirow{2}{*}{$M_{\rm core}$[$\overline{T_{\rm dust}}$]} & \multirow{2}{*}{Comments}\\
    name & number & & & & & \\ 
     & & [au] & [K] & [K] & [\msun] & \\
     (1) & (2) & (3) & (4) & (5) & (6) & (7)\\ \hline
G008.67     &    3 & 2780  &   30$\pm$3    &   28$\pm$2    &   10.8$\pm$1.2  &           \\
            &    5 & 3900  &   27$\pm$10   &   27$\pm$9    &    6.3$\pm$0.2  &           \\
            &    6 & 2650  &   29$\pm$3    &   27$\pm$2    &    5.2$\pm$0.9  &           \\
            &    8 & 2680  &   30$\pm$11   &   27$\pm$9    &    4.9$\pm$0.3  &           \\
            &   10 & 4120  &   30$\pm$9    &   30$\pm$9    &    4.5$\pm$0.4  &           \\
G010.62     &    6 & 1800  &   33$\pm$5    &   26$\pm$3    &   10.5$\pm$0.6  &           \\
G327.29     &    8 & 2340  &   41$\pm$6    &   39$\pm$5    &   26.4$\pm$0.6  &           \\
G333.60     &    7 & 2400  &   29$\pm$13   &   25$\pm$11   &    6.7$\pm$0.3  &           \\
            &   69 & 1640  &   39$\pm$12   &   31$\pm$9    &    5.7$\pm$1.2  &           \\
            &   20 & 2170  &   36$\pm$13   &   31$\pm$10   &    5.4$\pm$0.5  &           \\
            &   30 & 5240  &   37$\pm$12   &   37$\pm$11   &    5.3$\pm$0.3  &           \\
            &   53 & 4380  &   35$\pm$12   &   34$\pm$11   &    4.3$\pm$0.5  &           \\
            &   34 & 2620  &   36$\pm$12   &   31$\pm$10   &    4.0$\pm$0.5  &           \\
G337.92     &    7 & 1890  &   35$\pm$4    &   29$\pm$3    &    7.0$\pm$0.7  &           \\
            &    8 & 1680  &   36$\pm$4    &   29$\pm$3    &    6.9$\pm$1.0  &           \\
            &   12 & 2670  &   33$\pm$4    &   30$\pm$3    &    5.1$\pm$0.9  &           \\
G338.93     &    5 & 1600  &   35$\pm$8    &   28$\pm$6    &   14.1$\pm$0.3  &           \\
            &   17 & 3750  &   32$\pm$8    &   30$\pm$7    &    4.3$\pm$0.2  &           \\
G353.41     &    6 & 2230  &   26$\pm$6    &   25$\pm$5    &    8.6$\pm$0.3  &           \\
W43-MM1     &    6 & 1930  &   24$\pm$3    &   18$\pm$2    &   45.2$\pm$1.1  &           \\
            &  134 & 2480  &   27$\pm$3    &   22$\pm$2    &   21.1$\pm$2.0  &           \\
            &   17 & 1820  &   27$\pm$3    &   20$\pm$2    &   18.2$\pm$1.7  &           \\
            &  136 & 1850  &   28$\pm$3    &   21$\pm$2    &   12.2$\pm$1.9  &           \\
            &   21 & 1380  &   24$\pm$4    &   17$\pm$2    &   12.2$\pm$0.5  &           \\
            &   20 & 1410  &   24$\pm$4    &   17$\pm$3    &   10.3$\pm$0.3  &           \\
            &   37 & 3570  &   23$\pm$4    &   21$\pm$3    &    6.6$\pm$0.3  &           \\
W43-MM2     &   22 & 3120  &   28$\pm$4    &   24$\pm$3    &   14.2$\pm$0.8  &           \\
            &   13 & 1280  &   24$\pm$4    &   17$\pm$3    &    9.3$\pm$0.3  &           \\
W43-MM3     &   11 & 2050  &   26$\pm$7    &   20$\pm$5    &    7.9$\pm$0.2  &           \\
            &   43 & 5350  &   26$\pm$6    &   24$\pm$6    &    5.5$\pm$0.2  &           \\
W51-E       &   19 & 3470  &   45$\pm$8    &   40$\pm$6    &   65.5$\pm$5.2  &           \\
            &   37 & 2870  &   45$\pm$8    &   38$\pm$6    &   43.6$\pm$4.8  &           \\
            &   23 & 3300  &   44$\pm$8    &   38$\pm$6    &   33.3$\pm$3.3  &           \\
            &   26 & 3890  &   25$\pm$8    &   22$\pm$7    &   13.2$\pm$0.9  &           \\
            &   29 & 1890  &   38$\pm$8    &   30$\pm$6    &   11.5$\pm$2.0  &           \\
            &   43 & 2820  &   34$\pm$8    &   28$\pm$6    &    5.3$\pm$0.8  &           \\
W51-IRS2    &   12 & 2710  &   35$\pm$7    &   29$\pm$6    &   16.0$\pm$0.7  &           \\
            &   10 & 2360  &   31$\pm$8    &   25$\pm$6    &   13.8$\pm$0.4  &           \\
            &   13 & 3640  &   33$\pm$7    &   29$\pm$6    &   12.7$\pm$0.4  &           \\
            &   24 & 2930  &   33$\pm$7    &   28$\pm$5    &    9.3$\pm$0.5  &           \\
            &   39 & 2730  &   36$\pm$6    &   30$\pm$5    &    7.8$\pm$0.6  &           \\
            &   31 & 3140  &   28$\pm$9    &   24$\pm$7    &    7.3$\pm$0.4  &           \\
            &   19 & 4000  &   29$\pm$9    &   26$\pm$7    &    6.3$\pm$0.2  &           \\
            &   80 & 3720  &   30$\pm$8    &   27$\pm$7    &    5.5$\pm$0.4  &           \\
            &   ... & ...  &   ...    &   ...    &    ...  &           \\
\hline
\end{tabular}
\begin{tablenotes}[flushleft]
\item []\textbf{Notes:} 
(3) Outer radius of cores, assumed to be equal to their FWHM (see Tables~\ref{tab:measures table}--\ref{tab:measures table evolved}), deconvolved by the beam and set at their distance to the Sun \citep{motte2022}.
(4) Temperature measured in the \PPMAP dust temperature image at core location. 
(5) Mass-averaged dust temperature of prestellar cores estimated from \cref{eq:tdustPrestellar} using Cols.~3--4. (6) Core mass computed from the fluxes measured in the catalogs listed in Sect.~\ref{s:corecat} and the temperature of Col.~7 in \cref{eq:mass}.
\end{tablenotes}
\end{threeparttable}}
\end{table*}


\newpage
\section{Complementary figures} \label{appendixsect:complementary figures}

\cref{appendixsect:complementary figures} complements \cref{fig:lum-temp peaks G012-W43} by presenting in \cref{appendixfig:lum and temp peaks} all the other coincidences of luminosity peaks with the ALMA-IMF protostellar cores (\citealt{nony2020, nony2023}, in prep.; \citealt{armante2024, valeille2024}). In addition, \cref{appendixfig:luminosity profile} displays the luminosity profiles at four locations encompassing the six brightest hot cores of ALMA-IMF \citep{bonfand2024}. These profiles are used to estimate the total luminosity of their eight associated protostellar cores, which are listed in \cref{tab:luminosity peaks from radial profile}. And \cref{appendixfig:tdust prestellar cores} compares the mass-averaged dust temperature of prestellar cores, whether they are located in luminosity peaks or not.

\begin{figure*}[!hb]
    \centering\includegraphics[width=1.\linewidth]{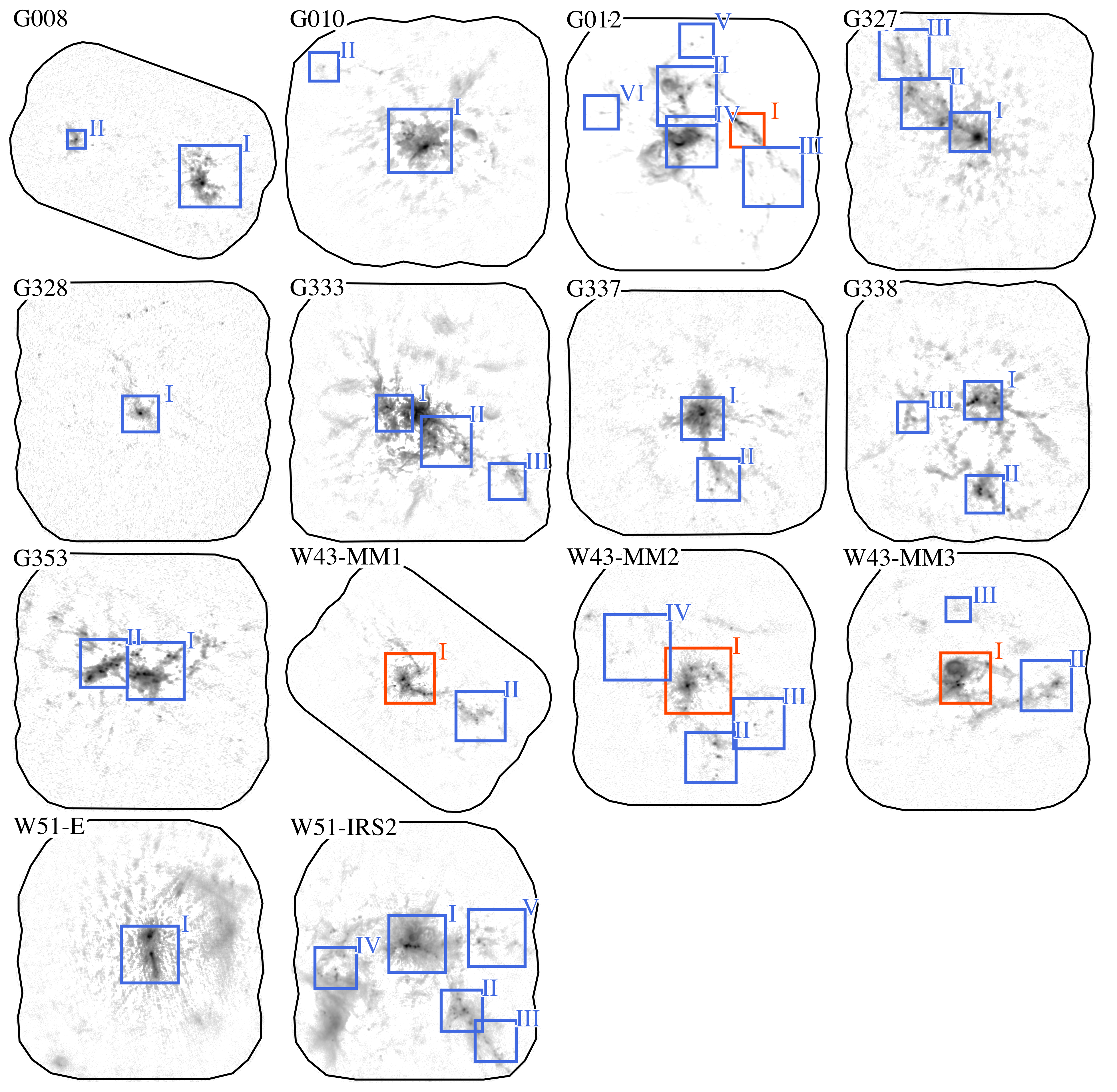}
    \caption{Zoom into the luminosity and temperature peaks, characterized at $2.5\arcsec$, to identify their associated protostellar and prestellar cores, detected at $0.3\arcsec-0.9\arcsec$ resolution. \textit{Fourteen first panels:} Location of the different zooms on the 1.3~mm image of each ALMA-IMF protocluster. Red squares outline the zooms of \cref{fig:lum-temp peaks G012-W43} and blue squares the complementary zooms shown here.}
    \label{appendixfig:zoom positions}
\end{figure*}

\setcounter{figure}{0}
\begin{figure*}
    \centering\includegraphics[width=0.85\linewidth]{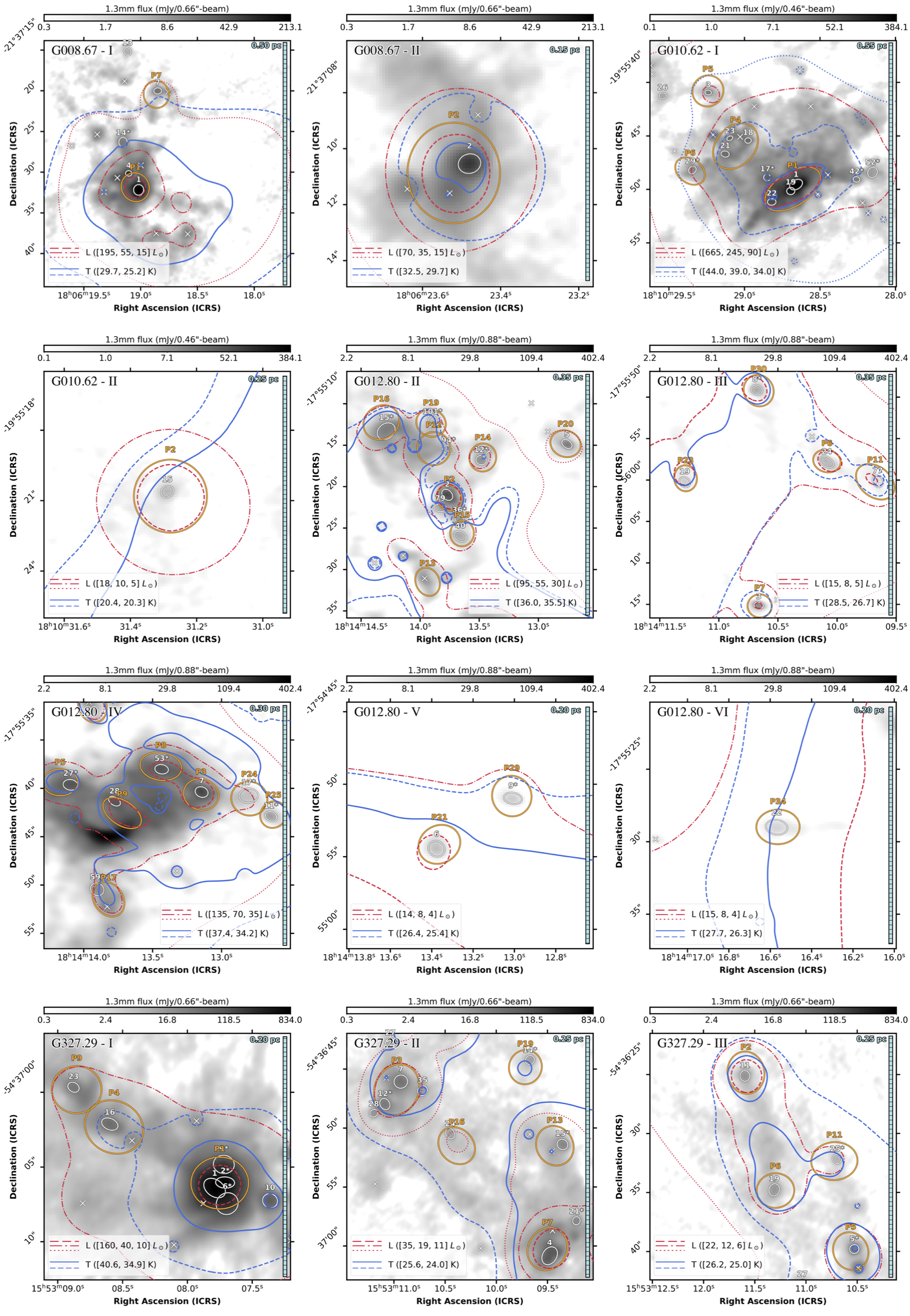}
    \caption{\textbf{continued.} Zoom-in of regions in ALMA-IMF protoclusters: G008.67 (two panels), G10.62 (two panels), G012.80 (five panels completing \cref{fig:lum-temp peaks G012-W43}a), and G327.29 (three panels). Red and blue contours display the \PPMAP luminosity and dust temperature map values, respectively, overlaid on the 1.3~mm continuum map shown in the grayscale background. Orange ellipses outline the FWHM size of the luminosity peaks associated with at least one protostellar core \citep[see][Tables~\ref{tab:measures table evolved} and \ref{appendixtab:measures table evolved}]{dellova2024}. White ellipses and crosses locate the protostellar and prestellar cores, respectively, identified by (\citealt{nony2020, nony2023}, in prep.), \cite{pouteau2022}, \cite{armante2024}, and \cite{louvet2024}. A scale bar is shown in the right-hand side of each panel. Some luminosity (and temperature) peaks host two and up to four protostellar cores of 1900~au typical size (see, e.g., P1 and P4 in G10.62).}
    \label{appendixfig:lum and temp peaks}
\end{figure*}

\setcounter{figure}{0}
\begin{figure*}
    \centering\includegraphics[width=0.85\linewidth]{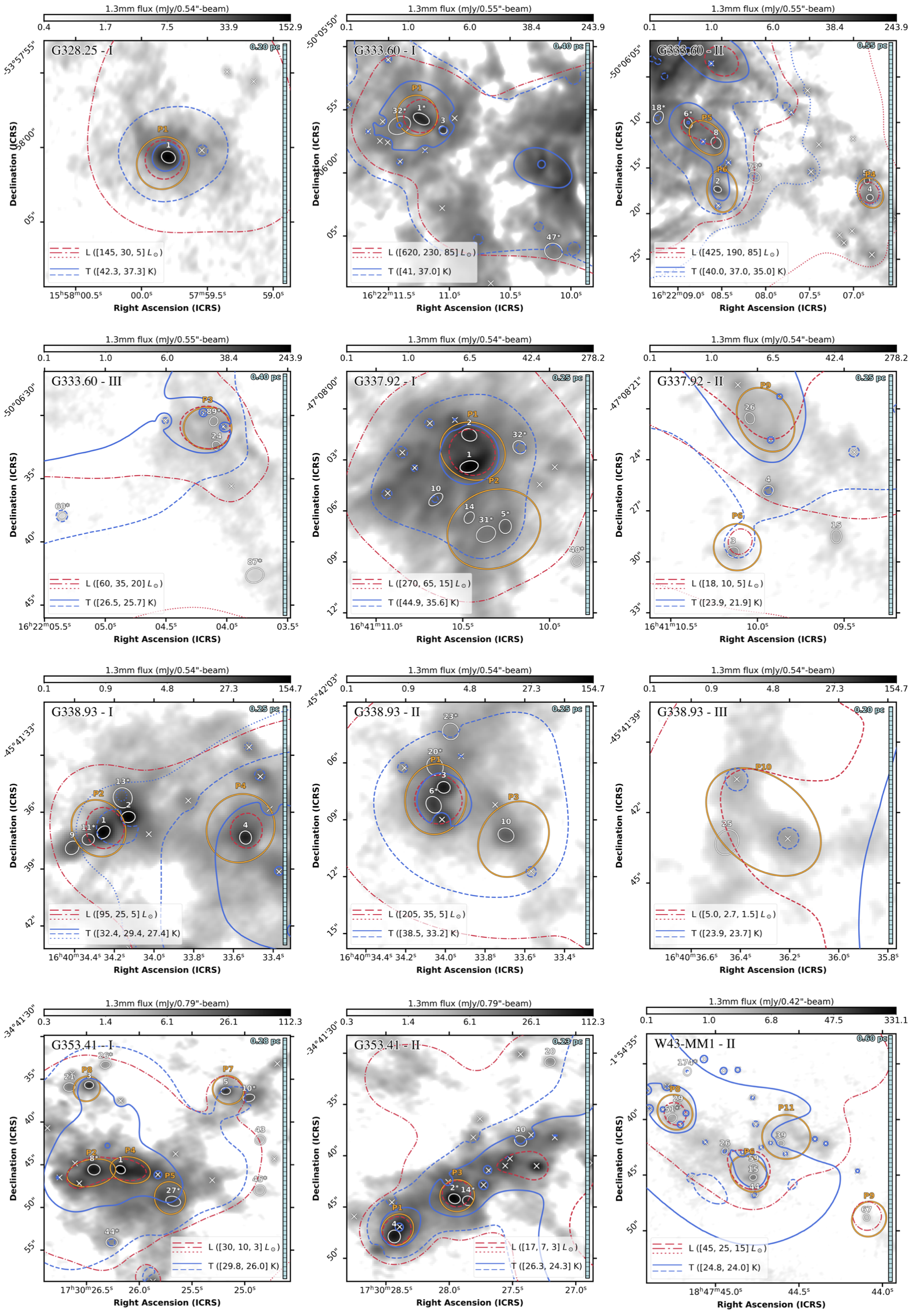}
    \caption{\textbf{continued.} Zoom-in of regions in ALMA-IMF protoclusters: G328.25 (one panel), G333.60 (three panels), G337.92.80 (two panels), G338.93 (three panels), G353.41 (two panels), and W43-MM1 (one panel completing \cref{fig:lum-temp peaks G012-W43}b). Some luminosity (and temperature) peaks host two and up to four protostellar cores of 1900~au typical size (see, e.g., P4 in G333.60, P2 in G337.92, and P3 and P10 in G338.93).}
\end{figure*}

\setcounter{figure}{0}
\begin{figure*}
    \centering\includegraphics[width=0.85\linewidth]{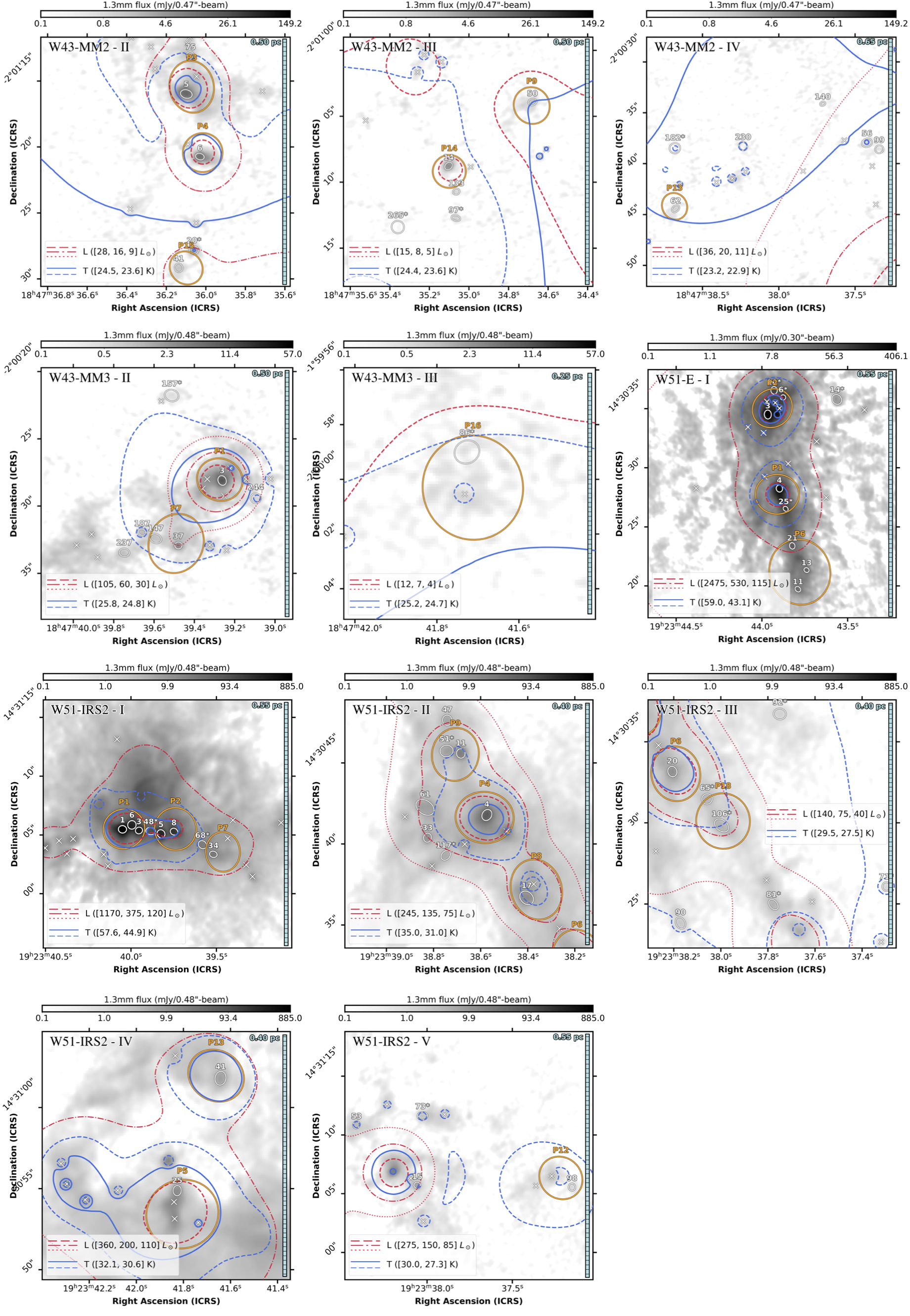}
    \caption{\textbf{continued.} Zoom-in of regions in ALMA-IMF protoclusters: W43-MM2 (three panels completing \cref{fig:lum-temp peaks G012-W43}c), W43-MM3 (two panels completing \cref{fig:lum-temp peaks G012-W43}d), W51-E (two panels), and W51-IRS2 (five panels). Some luminosity (and temperature) peaks host two and up to four protostellar cores of 1900~au typical size (see, e.g., P1 and P2 in W51-E, P1 and P2 in W51-IRS2).}  
\end{figure*}

\begin{figure*}[htbp!]
\centering
\begin{minipage}{1\textwidth}
    \centering
    \includegraphics[width=0.49\textwidth]{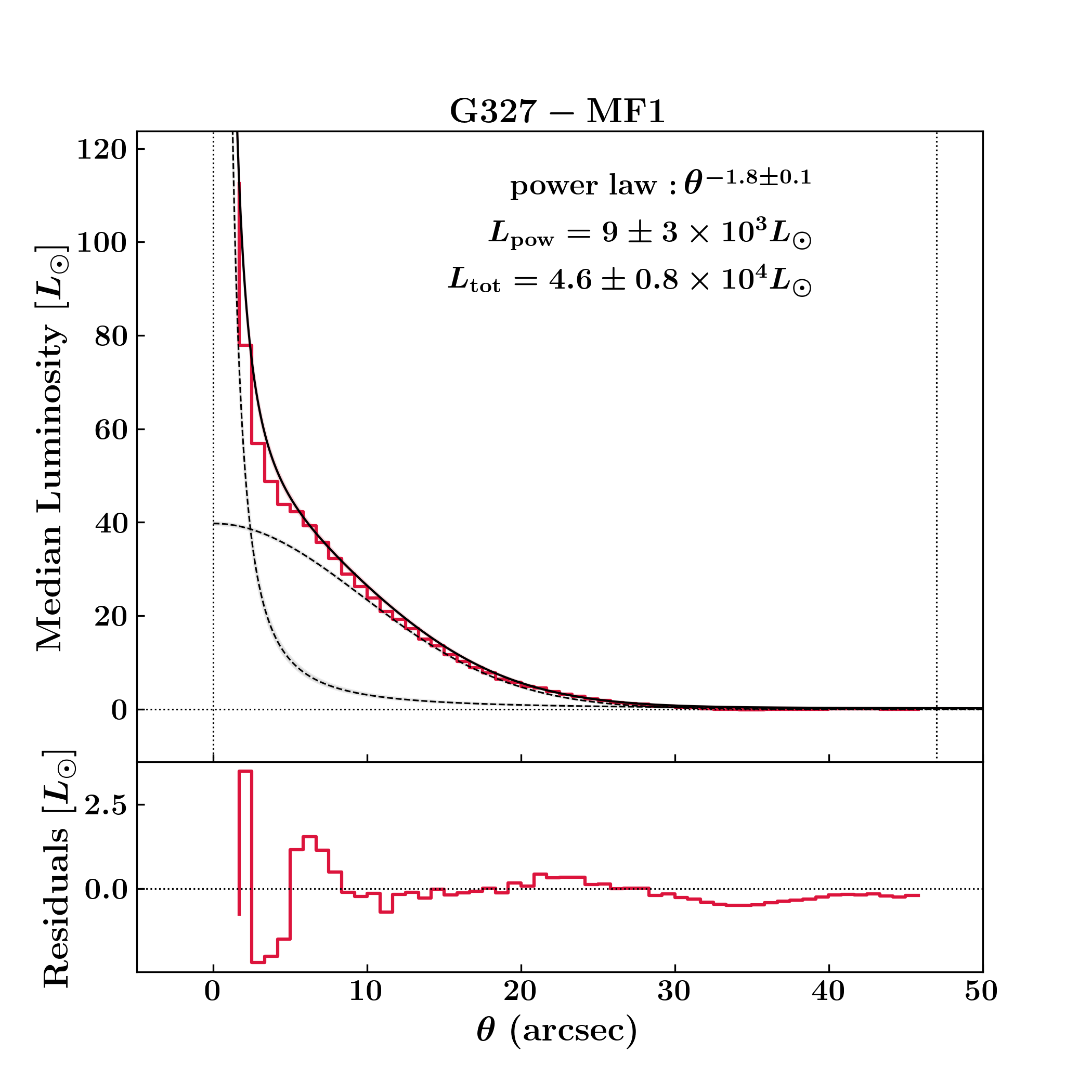} 
    \includegraphics[width=0.49\textwidth]{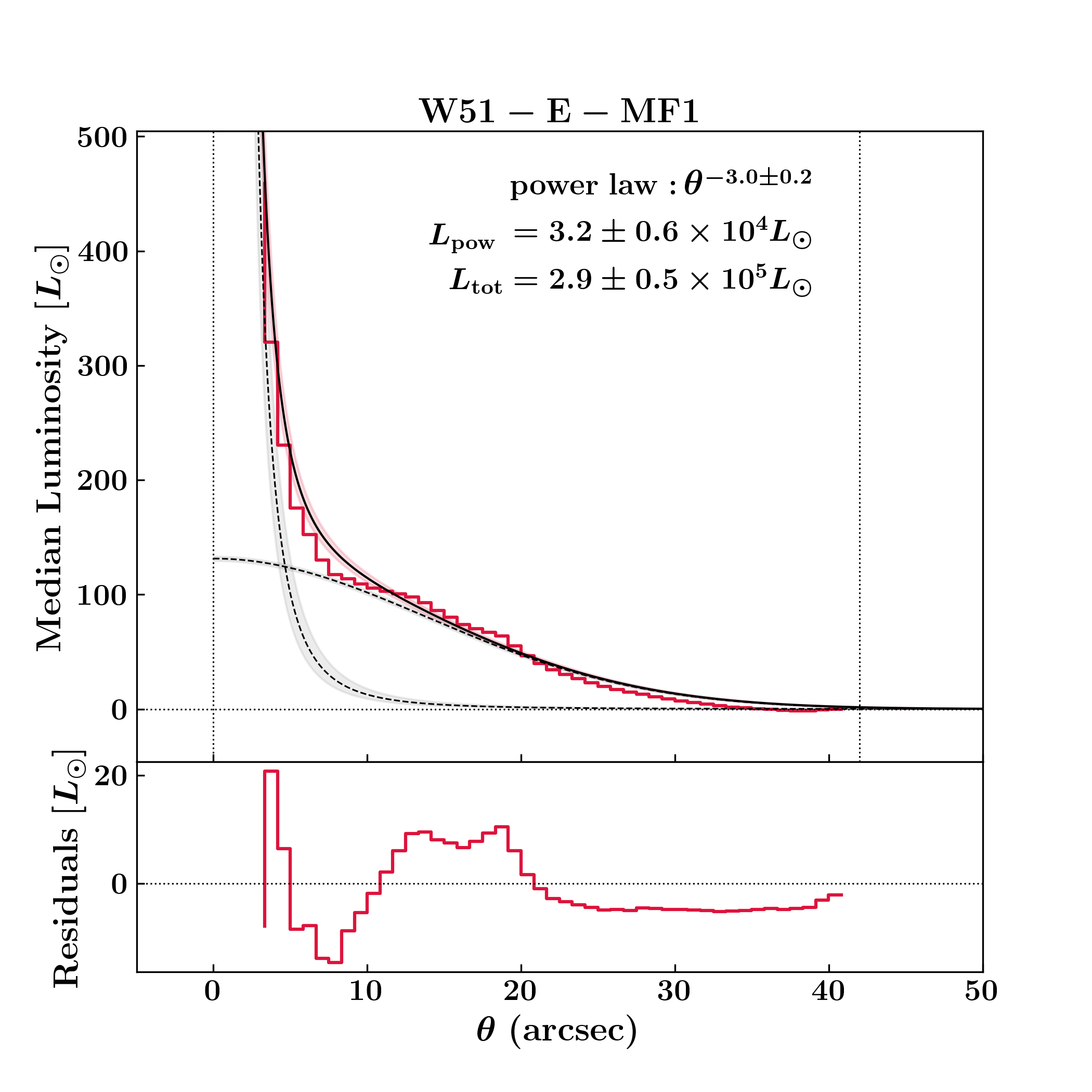}
    \vskip 10pt
    \includegraphics[width=0.49\textwidth]{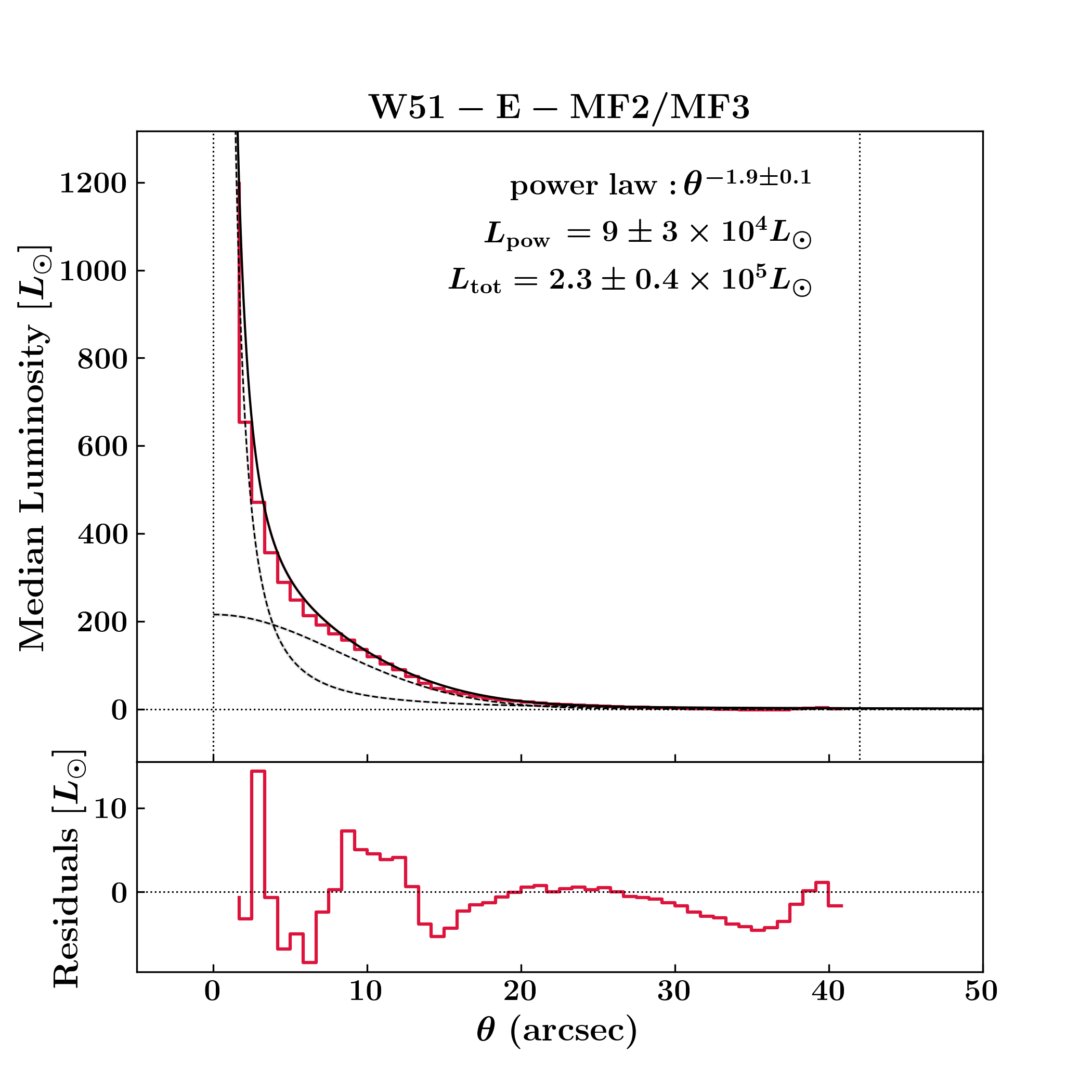}
    \includegraphics[width=0.49\textwidth]{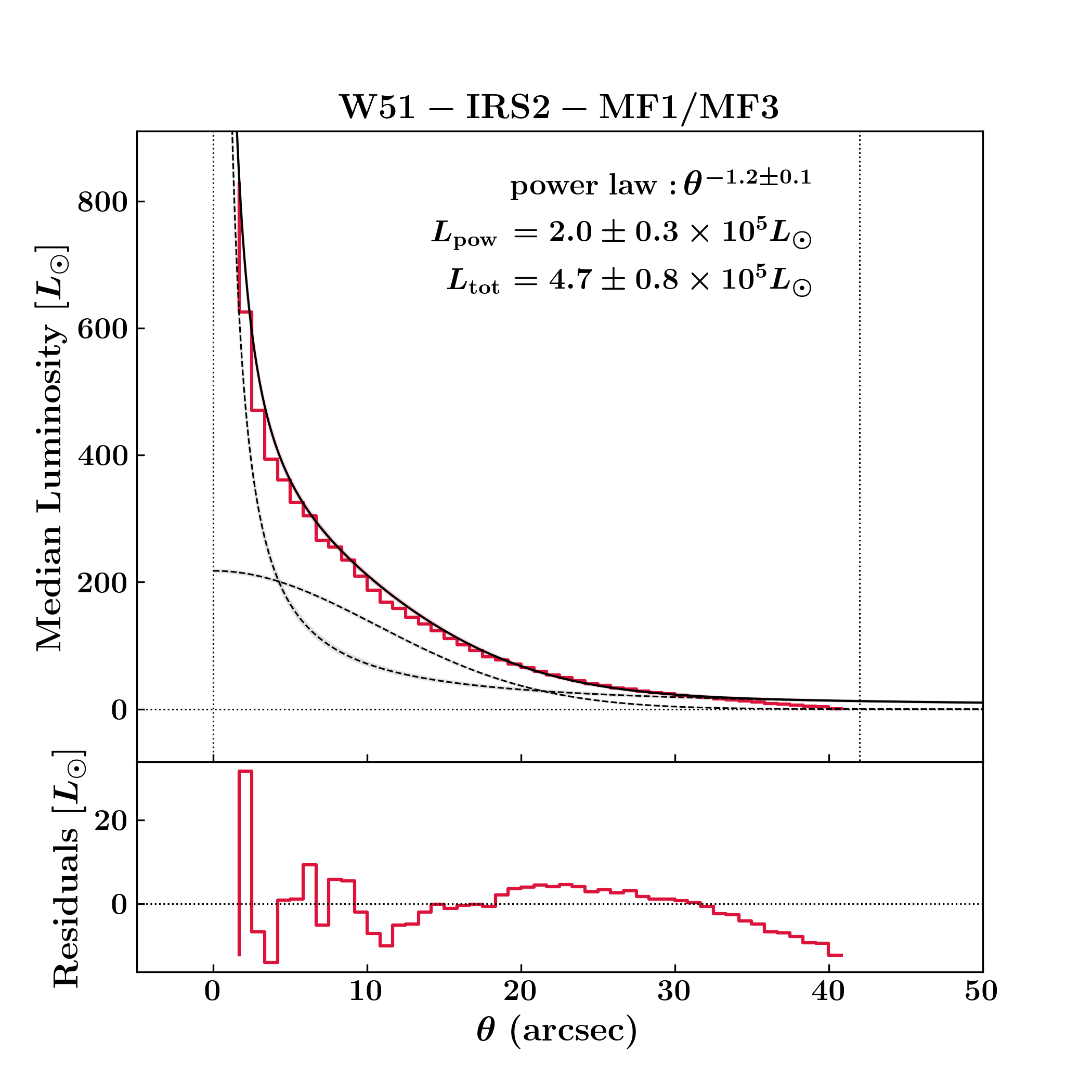}
\end{minipage}
\caption{Luminosity profiles of the four positions at $2.5\arcsec$ resolution encompassing the six brightest ALMA-IMF hot cores \citep{bonfand2024}. The luminosity is azimuthally averaged at four locations encompassing these hot cores using the median in rings spaced by $1.25\arcsec$ in radius from these locations. Profiles are fitted by the combination of a power-law and a Gaussian.} 
\label{appendixfig:luminosity profile}
\end{figure*}


\begin{figure}[htbp!]
    \centering
    \includegraphics[width=1\linewidth]{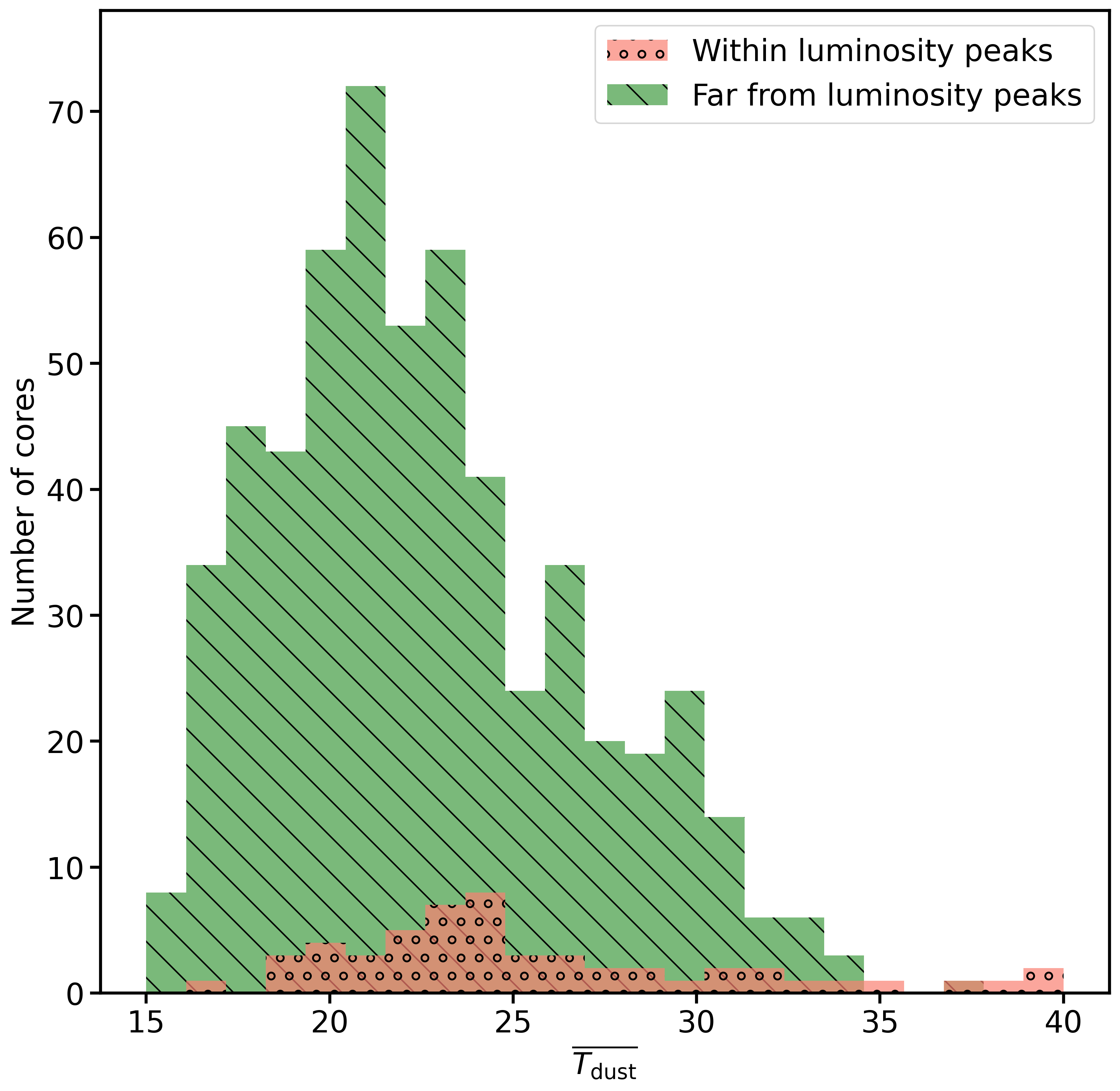}
    \vskip -0.cm
    \caption{Distribution of the mass-averaged dust temperature of prestellar cores located within luminosity peaks (orange histogram) and far away from them (green histogram), as indicated by their correspondence in Figs.~\ref{fig:lum-temp peaks G012-W43} and  \ref{appendixfig:lum and temp peaks}. The prestellar cores in the immediate vicinity of luminous protostellar cores have higher dust temperatures.}
    \label{appendixfig:tdust prestellar cores}
\end{figure}

\end{appendix}

\end{document}